%% file: clementini_DR2_revised2_no_bf.tex
\documentclass{aa}  

\usepackage{graphicx}
\usepackage{txfonts}
\usepackage{longtable}
\usepackage{upquote}
\usepackage{bm}

\begin{document}

\title{ \textit{\textbf{Gaia}} Data Release 2}

\subtitle{Specific characterisation and validation of all-sky Cepheids and RR Lyrae stars}

\titlerunning{{\it Gaia} Data Release 2, Cepheids and RR Lyrae stars}
\author{
G. Clementini\inst{\ref{inst1}},
V. Ripepi\inst{\ref{inst2}}, 
R. Molinaro\inst{\ref{inst2}},
A. Garofalo\inst{\ref{inst1},\ref{inst3}},
T. Muraveva\inst{\ref{inst1}},
L. Rimoldini\inst{\ref{inst4}},
L.~P. Guy\inst{\ref{inst4}}, 
G. Jevardat de Fombelle\inst{\ref{inst5}},
K. Nienartowicz\inst{\ref{inst5}},
O. Marchal\inst{\ref{inst4},\ref{inst6}},
M. Audard\inst{\ref{inst4}}, 
B. Holl\inst{\ref{inst4}}, 
S. Leccia\inst{\ref{inst2}}, 
M. Marconi\inst{\ref{inst2}},
I. Musella\inst{\ref{inst2}},
N. Mowlavi\inst{\ref{inst4}}, 
I. Lecoeur-Taibi\inst{\ref{inst4}},
L. Eyer\inst{\ref{inst4}},  
J. De Ridder\inst{\ref{inst7}}, 
S. Regibo\inst{\ref{inst7}}, 
L.~M. Sarro\inst{\ref{inst8}}, 
L. Szabados\inst{\ref{inst9}},
D.~W. Evans\inst{\ref{inst10}},  
and M. Riello\inst{\ref{inst10}}
}
\institute{INAF-Osservatorio di Astrofisica e Scienza dello Spazio di Bologna, Via Gobetti 93/3, I - 40129 Bologna, Italy\label{inst1}\\
\email{gisella.clementini@inaf.it}
\and
INAF-Osservatorio Astronomico di Capodimonte, Salita Moiariello 16, I - 80131 Napoli, Italy\label{inst2}
\and
Dipartimento di Fisica e Astronomia, Universit\`a di Bologna,  Via Gobetti 93/2, I - 40129 Bologna, Italy\label{inst3}
\and
Department of Astronomy, University of Geneva, Ch. d'Ecogia 16, CH-1290 Versoix, Switzerland\label{inst4}
\and
SixSq, Rue du Bois-du-Lan 8, CH-1217 Geneva, Switzerland\label{inst5}
\and
GEPI, Observatoire de Paris, Universit\'e PSL, CNRS, 5 Place Jules 
Janssen, 92190 Meudon, France\label{inst6}
\and
Institute of Astronomy, KU Leuven, Celestijnenlaan 200D, 3001 Leuven, Belgium\label{inst7}
\and
Dpto. Inteligencia Artificial, UNED, c/ Juan del Rosal 16, 28040 Madrid, Spain\label{inst8}
\and 
Konkoly Observatory, Research Centre for Astronomy and Earth Sciences, Hungarian Academy of Sciences, H-1121 Budapest, Konkoly Thege M. ut 15-17, Hungary\label{inst9}
\and
Institute of Astronomy, University of Cambridge, Madingley Road, Cambridge CB3 0HA, UK\label{inst10}
}
\authorrunning{Clementini et al.}
   \date{Received ... ; accepted .....}

  \abstract
 {The {\it Gaia} second Data Release (DR2) presents a first mapping of full-sky RR Lyrae stars and Cepheids observed by the spacecraft during the initial  22 months of science operations.}
 {The Specific Object Study (SOS) pipeline, developed to validate and fully characterise Cepheids and RR Lyrae stars (SOS Cep\&RRL) observed by {\it Gaia}, has been  
presented in the documentation and 
papers  accompanying the {\it Gaia} first Data Release. Here we describe how the SOS pipeline  was modified to allow for  processing the {\it Gaia} multiband ($G$, $G_{\rm BP}$ and $G_{\rm RP}$) time series photometry of all-sky candidate variables and produce specific results for confirmed RR Lyrae stars and Cepheids that are published in the  DR2 catalogue.}
{The SOS Cep\&RRL  processing uses  tools such as the period-amplitude and the period-luminosity relations in the $G$ band.
For the analysis of  the {\it Gaia} DR2 candidates  we also used  tools based on the $G_{\rm BP}$ and $G_{\rm RP}$ photometry, such as the period-Wesenheit 
relation in ($G$,$G_{\rm RP}$).} 
{
Multiband time series photometry and characterisation by the 
SOS Cep\&RRL pipeline 
are published in {\it Gaia} DR2   for 
150,359 such variables (9,575 classified as Cepheids and  140,784 as RR Lyrae stars) distributed all over the sky. The sample includes variables in 87 globular clusters and 14 dwarf galaxies (the  Magellanic Clouds, 5 classical and  7 ultra-faint dwarfs). To the best of our knowledge, as of 25 April 2018,  variability of 50,570 of these sources (350 Cepheids and 50,220 RR Lyrae stars)  is not  known in the literature, hence likely they are new discoveries by {\it Gaia}. 
An estimate of the interstellar absorption is  published for 54,272 fundamental-mode RR Lyrae stars from a relation based on the $G$-band amplitude and the pulsation period.
Metallicities derived from the Fourier parameters of the light curves are  also released for 64,932 RR Lyrae stars and 3,738 fundamental-mode classical Cepheids with period shorter than 6.3 days.}
{}


   \maketitle
%

\section{Introduction} 
The \textit{Gaia} mission has been  repeatedly monitoring the  celestial sphere since the start of scientific operation on 25 July 2014. The spacecraft 
 is collecting multi-band photometry and astrometric  measurements 
 of sources crossing its field of view  (FoV)  down to a limiting magnitude $G  \sim 21$ mag, along with  low-resolution spectroscopy for sources brighter than  $G  \sim 16$ mag. 
A description of the Gaia mission (spacecraft, instruments, survey and measurement principles) as well as the structure and activities  of the {\it Gaia} Data Processing and Analysis Consortium (DPAC) 
can be found in \citet{gaiacol-prusti}. 

On 25 April 2018 \textit{Gaia} Data Release 2  (DR2)  has published photometry in three pass-bands (\textit{Gaia} $G$, $G_{\rm BP}$ and $G_{\rm RP}$), five-parameter astrometry 
and radial velocities collected over the initial  22 months of observations.  
 A summary of the  {\it Gaia}  DR2 contents and survey properties is  provided in  \citet{gaiacol-brown18}.  The photometric dataset and the processing of the  $G$, $G_{\rm BP}$ and $G_{\rm RP}$-band (time series) photometry 
used for the analysis in the present paper  are described in \citet{riello2018} and \citet{evans2018}, whereas a detailed description of the astrometric dataset and processing is provided in \citet{lindegren2018}.

The multi-epoch operating procedure makes \textit{Gaia} a very powerful tool  to identify and characterise stellar variability phenomena across the whole HR diagram (e.g. \citealt{gaiacol-eyer}). 
In \textit{Gaia}  Data Relase 1 (DR1; \citealt{gaiacol-brown16}) 
$G$-band time series photometry  and characteristic parameters were released only for  a small number of pulsating variables, comprising 599 Cepheids and  2595 RR Lyrae stars (\citealt{clementini2016}, hereafter Paper~1) in a region of the Large Magellanic Cloud (LMC)  that the spacecraft  
observed at high cadence during the first 28 days of scientific operation in  Ecliptic Poles Scanning  Law. 
The catalogue  of variables released in \textit{Gaia} DR2 \citep{holl2018} contains  thousands of Cepheids and hundred of thousands RR Lyrae stars in the Milky Way (MW) and its nearest neighbours. They represent a first census of full-sky RR Lyrae stars and Cepheids and provide a flavour of \textit{Gaia} capabilities by recovering most of the known MW Cepheids,  identifying a few bona fide new ones and increasing the number of known Galactic RR Lyrae stars well beyond the current value of more than a hundred of thousands. 

The general approach of variability analysis and classification developed within 
 the {\it Gaia}  DPAC  was presented in \citet{eyer2017b}.   For DR2 an additional  fully statistical approach was  developed to classify  all-sky high-amplitude pulsating stars. This approach is extensively described  in \citet{rimoldini2018}.  The Specific Objects Study (SOS)  pipeline  (hereafter referred to as 
SOS Cep\&RRL pipeline), which is specifically designed to  validate and fully characterise Cepheids and RR Lyrae stars observed by {\it Gaia}, is described in detail in 
Paper~1.
The general properties
of the whole sample of variable sources released in  \textit{Gaia} DR2 are described in  \citet{holl2018}, which also briefly summarises steps of the 
general variability analysis  
prior the  SOS Cep\&RRL processing. 

In this paper we describe how the SOS Cep\&RRL pipeline was modified and further developed
 to process the DR2 multi-band time-series photometry of candidate Cepheids and RR Lyrae stars identified by the general variable star classification pipelines  (\citealt{eyer2017b}, \citealt{rimoldini2018}).
We describe our  validation procedures and briefly present results from the SOS Cep\&RRL  processing  of Cepheids and RR Lyrae stars confirmed by the pipeline,  that are  released  in the \textit{Gaia} DR2 variability catalogue.
 We recall that conforming to the  strict DPAC policy,  only a very limited interpretative overview and no 
scientific exploitation of the data is presented in this paper.  We also remind the reader that   
characterisation and classification of any RR Lyrae stars and Cepheids released in DR2  is 
purely and exclusively based on the {\it Gaia} data of the sources. That is we do not complement {\it Gaia}'s time series with external non-{\it Gaia} data.
Literature published data are only used as training sets for the classification tools and for the final validation of the results.

 The paper is organised as follows. Section~\ref{s2} provides a brief summary of the SOS Cep\&RRL pipeline specifically highlighting  changes and improvements with respect to the DR1 processing.
 Section~\ref{s3new}  presents the datasets and selections of all-sky sources on which the SOS Cep\&RRL  pipeline was run.
Sections ~\ref{validation} and \ref{results2} present 
the SOS Cep\&RRL analysis describing extensively our procedures to validate the SOS results that we publish in {\it Gaia} DR2 along with a comparison with the literature and the DR1 results.  In those sections we specifically acknowledge  
the limitation of the current analysis and results and  warn the reader about oversimplifications and possible biases of the SOS Cep\&RRL 
processing.   We also discuss reasons why well-known  sources are missing in DR2 as well as a few misclassifications of the  SOS Cep\&RRL processing for DR1 (Section~\ref{DR2limitation}). 
After  DR2 we become aware of a number of misclassifications in the released DR2 Cepheid and RR Lyrae samples. A partial listing of these misclassifications 
is provided in Appendix C.
   Finally, main results  and future developments of the SOS Cep\&RRL pipeline are summarised in Section~\ref{conclusions}.

\section{ SOS Cep\&RRL pipeline applied to the DR2 data: general overview}\label{s2}

  \begin{figure*}
   \centering
   \includegraphics[trim= 0 -30 0 -20,width=16 cm,clip]{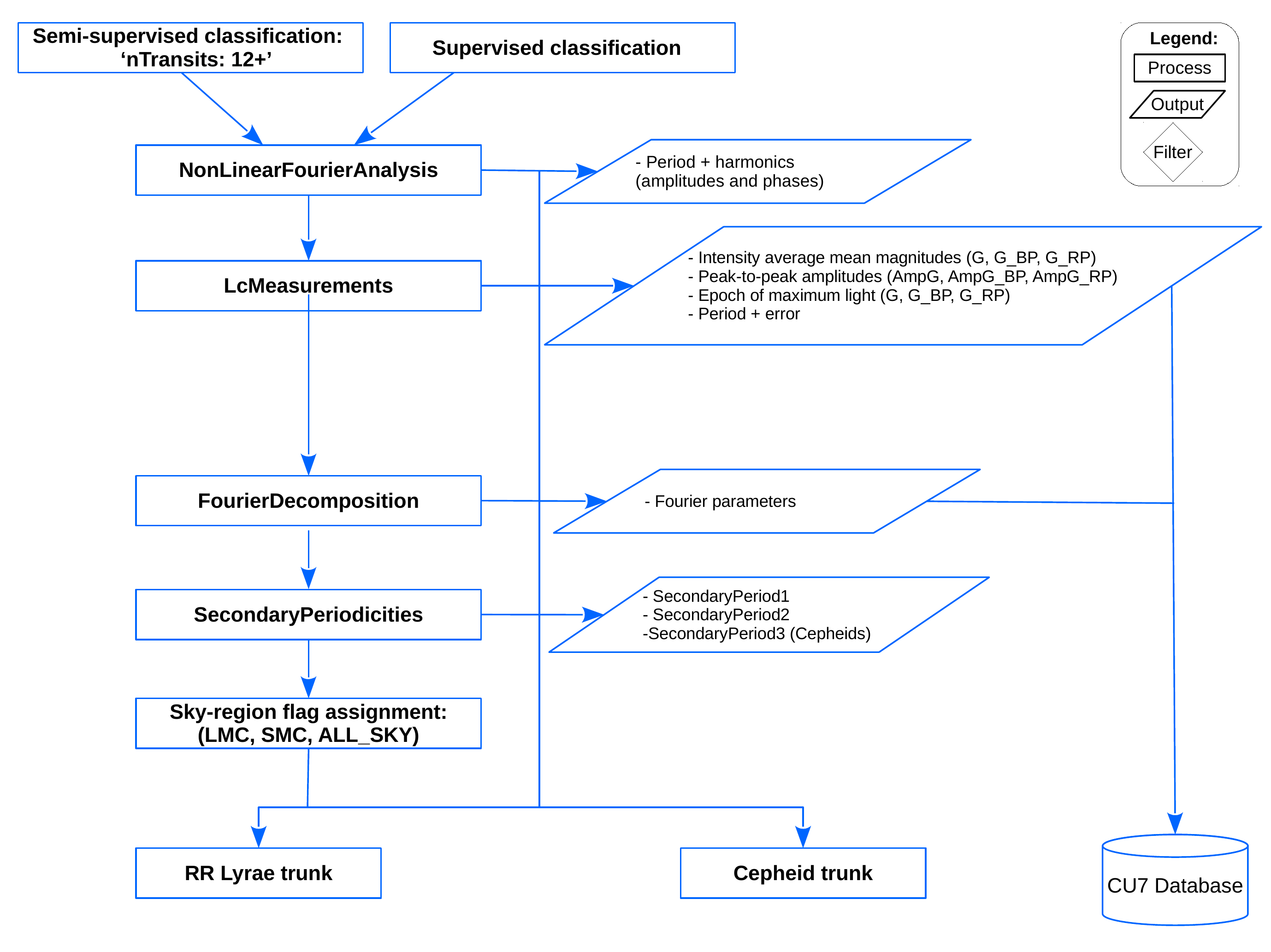}
   \caption{Overview of the SOS Cep\&RRL processing that is common to both Cepheid  and RR Lyrae stars.  Rectangles show 
   different processing modules of the general trunk, with names highlighted in boldface. Their outputs  are indicated within rhombs. The figure is updated from fig.~1 of  Paper~1
 to include the processing 
   of the $G_{\rm BP}$ and $G_{\rm RP}$ time-series and the search for a third periodicity for Cepheids. A further addition for DR2 is the selection of the region of the sky where the sources are 
   located, whether LMC, SMC or All-Sky (see text for details).}
              \label{decomposition1}
    \end{figure*}

\begin{figure*}[h!]
   \centering
    \includegraphics[trim= 0 0 0 0, width=16 cm,clip]{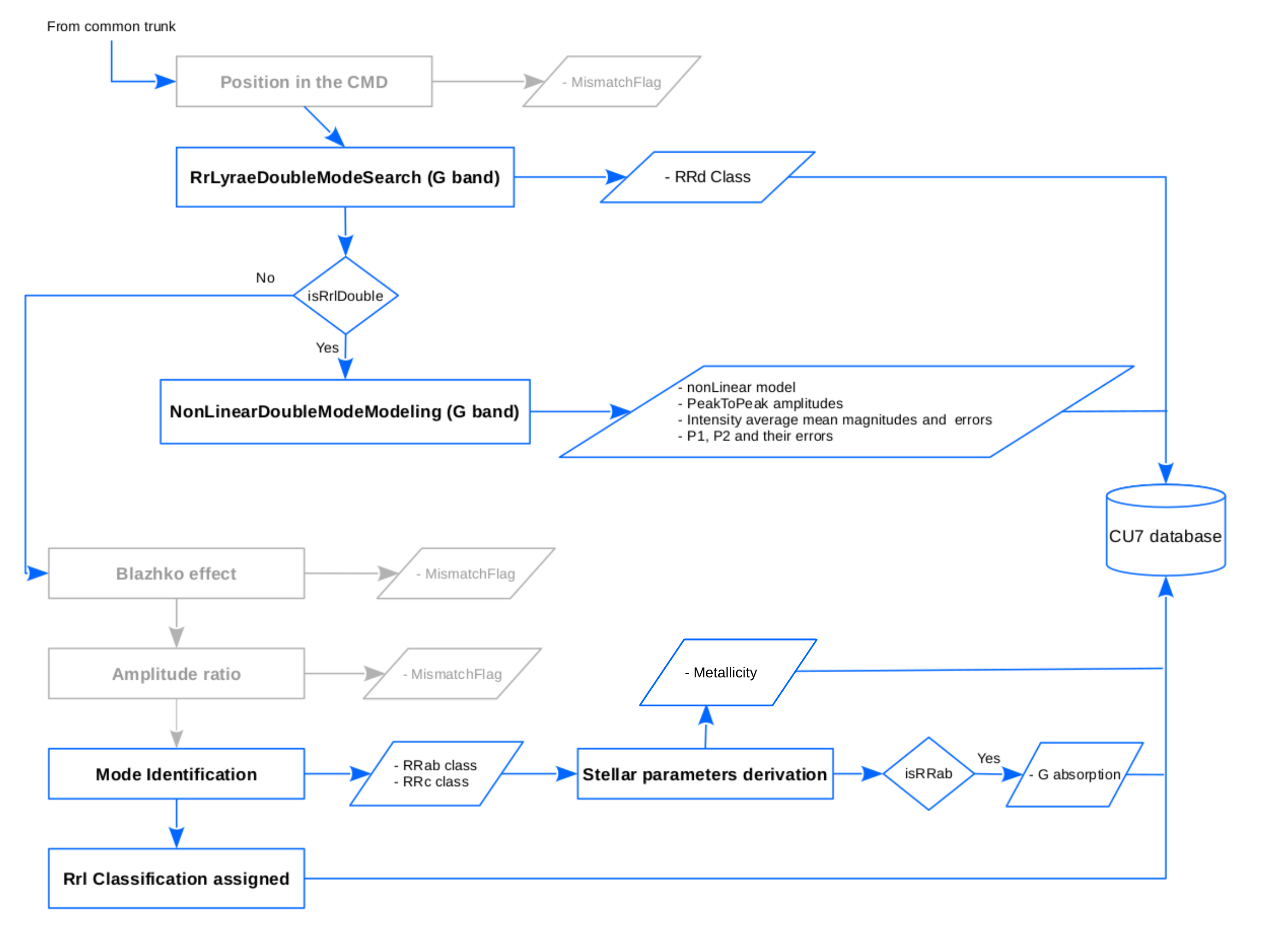}
   \caption{Flow chart of the RR Lyrae branch in  the SOS Cep\&RRL pipeline.  As in Fig.~\ref{decomposition1} rectangles show the different processing modules (with names highlighted in boldface) of this  branch. Their outputs  are indicated within rhombs.  Diamonds indicate filters leading to different processing options. We have marked in light grey modules not operational for the {\it Gaia} DR2 processing. The figure is updated from fig.~2 of  
   Paper~1  to include modifications implemented  to process the DR2 photometry. 
  }
              \label{decomposition2}
    \end{figure*}

\begin{figure*}[h!]
   \centering
   \includegraphics[trim= 0 0 0 0,  width=16 cm,clip]{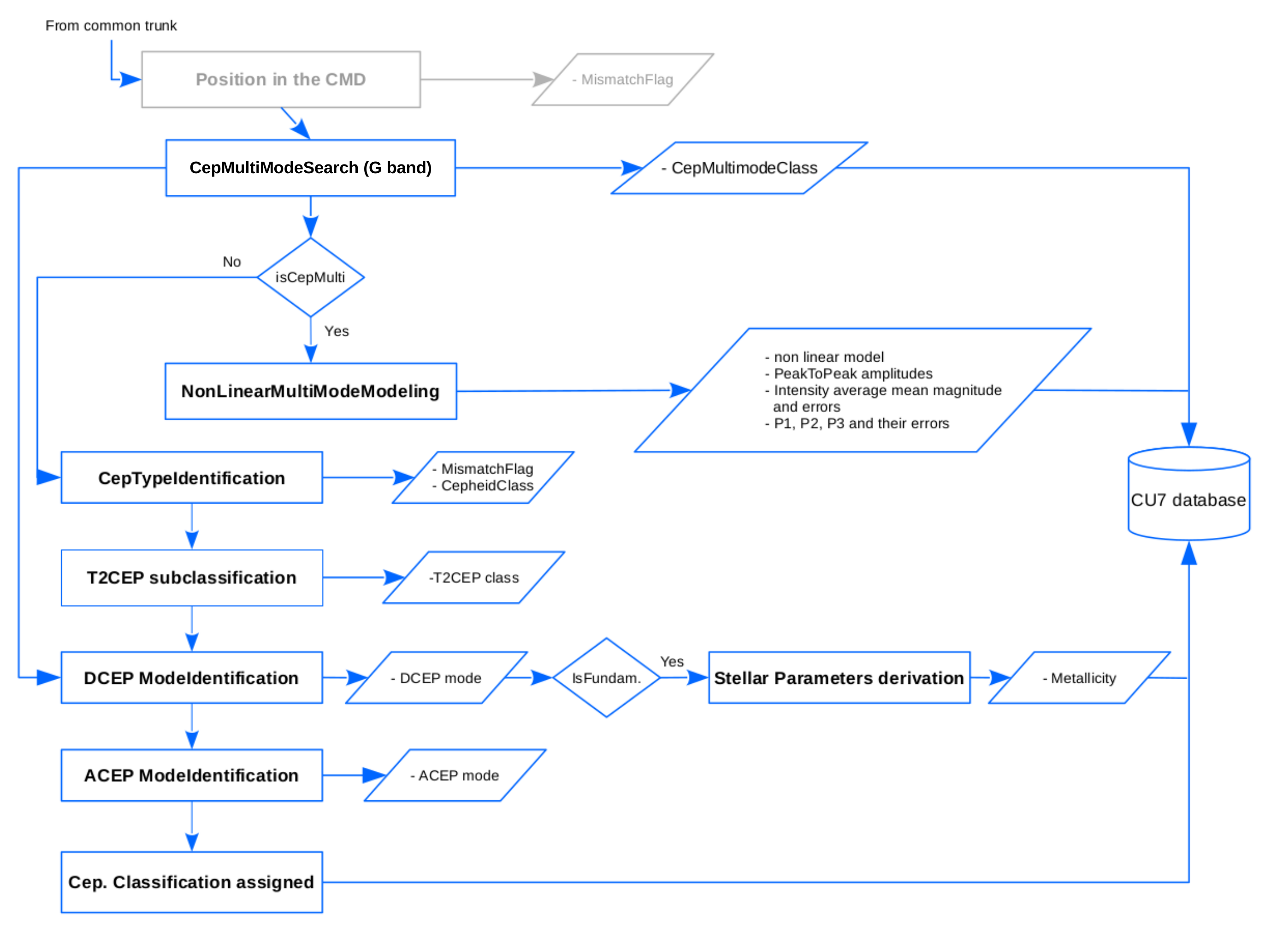}
   \caption{Flow chart of the Cepheid branch in the SOS Cep\&RRL pipeline as properly modified from fig.~3 in 
   Paper~1 to process the DR2 photometry.  Layout and colour-coding are the same as in Fig. ~\ref{decomposition2}.
   }
              \label{decomposition3}
    \end{figure*}

Main purposes, tools and steps of the SOS Cep\&RRL pipeline are extensively  described in Sect. 2 
 of  
Paper~1, to which the interested reader is referred for details.
Here we only summarise main differences with respect to the DR1 processing. 
An overview of the different parts of the SOS Cep\&RRL processing is presented in Figs.~\ref{decomposition1}, ~\ref{decomposition2} and ~\ref{decomposition3}, which were updated from  
figs.~1, 2 and 3 of 
Paper~1 to summarise changes operated in the pipeline, such as parts that were activated or tools newly developed
to allow for processing  the multi-band ($G$, $G_{\rm BP}$ and $G_{\rm RP}$)  time-series photometry of all-sky sources  
that are released in DR2.

As schematically summarised in fig. 2 of \citet{holl2018},    
 the SOS Cep\&RRL  pipeline received as input for  the DR2 processing 
the calibrated $G$ and integrated $G_\mathrm{BP}$, $G_\mathrm{RP}$ time series photometry \citep{evans2018, riello2018}  
collected by \textit{Gaia} in the 22 months between 25 July 2014 and 23 May 2016 
of  the sources  pre-classified as candidate 
Cepheids and RR Lyrae stars by  the classifiers of the  
 general variability pipeline  (\citealt{eyer2017b}, \citealt{rimoldini2018}). 
Specifically,  SOS Cep\&RRL processed candidate Cepheids and RR Lyrae stars with $\geq$ 20 (hereafter geq20) $G$-band measurements (or FoV transits)  identified by the supervised 
classification of the general variability pipeline (\citealt{eyer2017b}) and candidates with $\geq$ 12 measurements (hereafter geq12) identified by the semi-supervised classification 
described in \citet{rimoldini2018}, for a total of about 640,000 candidate RR Lyrae stars and more than 72,000 candidate Cepheids, after removing overlaps between geq20 and geq12 samples. 
A preliminary version of the DR2 astrometry was also used in order to test our classification of the  All-Sky Cepheids (Sect.~\ref{allskycep}).
However, due to the tight DR2 data-processing schedule we could not use for processing the final astrometric
solution \citep{lindegren2018} nor could we access the astrophysical parameters (e.g., reddening values, temperatures, \citealt{andrae2018}) and 
radial velocity measurements \citep{sartoretti2018} that are published in DR2. 

Main outliers and measurements of insufficient quality are removed from the $G$, $G_{\rm BP}$ and $G_{\rm RP}$ time-series by specific operators of the general variability pipeline (see sect. 4.1 in \citealt{holl2018}). However, we performed an additional cleaning of the $G_{\rm BP}$ and $G_{\rm RP}$ time-series by setting more stringent limits, specifically  tailored for Cepheids and RR Lyrae stars, to the cleaning operators before the  source time-series were entered in the main trunk of the SOS pipeline that is common to both Cepheids and RR Lyrae stars (Fig.~\ref{decomposition1}).   

The determination of the source main periodicity is the initial step of the SOS Cep\&RRL pipeline (see Fig.~\ref{decomposition1})  and 
the  subsequent classification and characterisation steps   
significantly rely on the accuracy and precision of the SOS period  which ultimately depends  upon the number of epoch-data available for the sources. A comparison between periods derived by the SOS pipeline 
and the  literature periods  
for a sample of 37,941 RR Lyrae stars in common between \textit{Gaia} and the OGLE catalogues for the LMC, Small Magellanic Cloud (SMC) and Galactic bulge,  shows that agreement can already be satisfactory if at least 20 epochs are available, and that results definitely improve 
for sources with 30 or more epoch-data. 
Unless differently stated all periods along this paper are those derived with the SOS Cep\&RRL pipeline ($P_{SOS}$).

In the SOS Cep\&RRL main trunk the source main periodicity is determined with the Lomb-Scargle algorithm (\citealt{lomb}, \citealt{scargle}, see  Section 2.1 in 
Paper~1, for details) looking for a periodicity in the range 
0.2 $\leq P <$ 1 day (frequency between 1 and 5) for RR Lyrae stars and 
0.2 $\leq P <$ 333 days (frequency between 0.003 and 5) for Cepheids\footnote{The period range for Cepheids was later reduced to 0.2 $\leq P \lesssim$ 160 days during validation in account for the actual time interval spanned by the DR2 time series data.}.  
 The light curves are modelled with truncated Fourier series and the pulsation characteristics (period, peak-to-peak amplitude, epoch of maximum light, intensity-averaged mean magnitude, in each of the three passbands) are determined from the modelled light curves along with the related errors. The Fourier parameters ($\phi_{21}$, $\phi_{31}$, $R_{21}$, $R_{31}$) of the modelled $G$-band light curve are also determined. 
Secondary periodicities, if any, are identified in the 
$G$-band data  looking for one additional period in the case of RR Lyrae stars and up to two for Cepheids (see Fig.~\ref{decomposition1}).
    
A significant change with respect to DR1 introduced in the main trunk of the SOS Cep\&RRL pipeline to process all-sky sources 
 was the selection, before entering the RR Lyrae (Fig.~\ref{decomposition2}) and Cepheid (Fig.~\ref{decomposition3}) branches, of the region of the sky where the sources are
 located, whether LMC, SMC  or All-Sky.
 The sky-region corresponding  to the LMC is defined as a box with centre at RA=82.5$^{\rm o}$, DEC=$-$68.25$^{\rm o}$ and extending  from  67.5$^{\rm o}$ to 97.5$^{\rm o}$ in right ascension and from 
 $-$73.0$^{\rm o}$ to $-63.5^{\rm o}$ in declination. The SMC is defined as a box region with centre at RA=16$^{\rm o}$, DEC=$-$73$^{\rm o}$ and extending  from 2$^{\rm o}$ to 30.0$^{\rm o}$ in RA and from 
 $-$75.0$^{\rm o}$ to $-71.0^{\rm o}$ in DEC. 
The All-Sky region is defined by what is left after subtracting the LMC and SMC  selections.
  Different reference relations are used to classify the Cepheids and RR Lyrae stars in these three distinct regions, as detailed in Sects.~\ref{allskyrrl} and ~\ref{allskycep}. 
Once the region  of the sky  
has been selected, the source enters into  the RR Lyrae (Fig.~\ref{decomposition2}) or the Cepheid (Fig.~\ref{decomposition3}) branches according to the following schema. 
If the source belongs to the All-Sky region,  it is first ingested into the RR Lyrae branch (Fig.~\ref{decomposition2}) and if it is not classified as an RR Lyrae star, it is then sent for analysis to the Cepheid branch (Fig.~\ref{decomposition3}) where the star is processed 
providing that 
the parallax value conforms to  specific quality assurance conditions defined in  the astrometric processing (see Sect.~\ref{allskycep} and \citealt{lindegren2018}, for details).
If this is not the case the source is rejected.  
If the source belongs to the LMC or SMC regions, which branch to enter first is chosen based on a number of checks on the intensity-averaged mean $G$ magnitude\footnote{For sake of clarity we recall that  the intensity-averaged  or intensity average magnitude is the mean magnitude of a variable star obtained by transforming to 
  intensity each individual value of the Fourier model best fitting the light curve, then averaging all those intensities and converting the mean intensity back into a magnitude.}, the period and  the amplitude  of the $G$-band variation. 

In the RR Lyrae branch (Fig.~\ref{decomposition2}), the SOS Cep\&RRL pipeline can use specific
 features and  diagnostic tools such as: $a)$ the (apparent and/or absolute) 
   colour-magnitude diagram (CMD; $G$ vs $G_{\rm BP} - G_{\rm RP}$, $G$ vs $G - G_{\rm BP}$, $G$ vs $G - G_{\rm RP}$),  $b)$ 
the  parameters of the $G$-band light curve Fourier decomposition ($R_{21}$, $\phi_{21}$ versus $P$ and $R_{31}$, $\phi_{31}$ versus $P$ plots),   
$c)$ the {\it Gaia} period-amplitude  ($PA$) diagram, also known as Bailey diagram (\citealt{bailey1902}) in the $G$ band and  $d)$ the amplitude ratios in the different passbands
 [specifically,  {\rm Amp}($G_{\rm BP}$)/{\rm Amp(}$G_{\rm RP}$)] to classify the sources and infer their pulsation mode(s) and type(s)\footnote{All quantities (mean magnitudes, colours, amplitudes and Fourier parameters) used in these tools are inferred from the non-linear model of the light curves computed by the SOS Cep\&RRL pipeline.}. We refer the reader to sects. 2.3.2, 2.3.3, 2.4.1 and 2.4.2 in 
 Paper~1, 
 for a general description of the procedures and provide in following Sections of the present paper new relations that replace/complement those used for DR1. 
 We note that of the various tools described above,  tool $c)$ is the one most largely applicable as it only requires knowledge of the source  period and $G$-band amplitude, but it is also the least 
constraining. On the contrary, tools $a)$ and $d)$ are increasingly less applicable because require availability of well modelled $G_{\rm BP}$, $G_{\rm RP}$  light curves from which to derive mean colours 
to build the CMDs needed for tool $a)$,  and derive reliable amplitudes to compute  amplitude ratios needed for tool  $d)$, and increasingly sampled light curves  modelled with 
at least 2 or 3 harmonics  to correctly infer Fourier parameters used by tool $b)$.  This also shapes limitations of the present analysis that will be discussed in the following sections. 
Furthermore, for the SOS processing of RR Lyrae stars we did not use  the parallax information both because we could access only a preliminary version of the 
DR2 parallax values and also  because RR Lyrae stars being  at least 3 magnitudes fainter than Cepheids can more likely have parallaxes with large errors which may even scatter the parallaxes to negative values if they are very close to zero, as in the case of distant systems like the Magellanic Clouds, especially in this still initial astrometric solution.

In the Cepheid branch (Fig.~\ref{decomposition3}) to classify the sources, identify their pulsation mode and the multi-mode pulsators we used both the Fourier parameters and 
the period-luminosity ($PL$) and period-Wesenheit ($PW$) relations.  
For the LMC and SMC the zero points of the $PL$ and $PW$ relations (in the \textit{Gaia} pass-bands) were calibrated adopting 18.5 mag and 19.0 mag for the 
distance modulus of LMC and SMC, respectively. For the classification of the Milky Way All-Sky Cepheids we primarily relied on the Fourier parameter planes and, only as a test, we also looked at the $PW$ distributions, for which we had to rely on parallaxes (see Sect.~\ref{allskycep}).

 In Figs.~\ref{decomposition2} and ~\ref{decomposition3} we have indicated in light grey a number of modules not fully operational during the  
 SOS DR2 processing. 
 Among them are the one checking the position of the sources on the CMD  and the check of amplitude ratios to reject binaries contaminating the samples. Although not yet automatically activated during the pipeline processing both tools were later largely used during the validation of the sources confirmed as Cepheids and RR Lyrae stars by the SOS Cep\&RRL pipeline (see 
Sect.~\ref{validation}).

The first working step in the  RR Lyrae (Fig.~\ref{decomposition2}) and in the Cepheid (Fig.~\ref{decomposition3}) branches is to verify whether any detected secondary periodicity is  consistent with  the RR Lyrae star being a double-mode pulsator (RRd;  Fig.~\ref{decomposition2}) and the Cepheid being a multi-mode pulsator (Fig.~\ref{decomposition3}).
In the affirmative, it then follows a non-linear multimode modelling of the light curve taking into account all 
excited periodicities. We refer the reader to sects. 2.3.1, 2.4.1 of 
Paper~1 for a general  and more detailed presentation of the algorithms used in the SOS Cep\&RRL  pipeline to identify and characterise double/multi mode sources. 

The subsequent step in the Cepheid branch (Fig.~\ref{decomposition3}) is the identification of the Cepheid type, whether DCEP, ACEP, or T2CEP (see Table~B.1 
 for the meaning of the acronyms),  and pulsation mode (only for DCEP, ACEP) for which the path was as it is described in Sect. 2.4.2 of Paper~1, with the additional use of $PW$ relations to distinguish the different types and differentiation of the $PL$ and $PW$ relations according to the location of the sources on the sky. The reference $PL$ and $PW$  relations used in DR2 are described in detail in Sect.~\ref{allskycep}.  A further minor change 
with respect to  DR1 was to use a stricter limit in period  of   0.234 d$ < P < 6$ d when $R_{21}$ $ < 0.214$ to identify  DCEPs that pulsate in the first overtone (1O) using the $R_{21}$ vs $P$ diagram.
Regarding the RR Lyrae branch  (Fig.~\ref{decomposition2}), since the SOS Cep\&RRL modules for detection of the Blazhko effect and calculation of the amplitude ratios (see Sect.2.3.2   in Paper~1)  were not activated for DR2, the next active step  
is  the RR Lyrae mode identification,  that follows the path described in sect.~2.3.3 of Paper~1 with only the following change,  a new line to separate RRc and RRab types was adopted that is described by the equation:  ${\rm Amp(}G{\rm )}$ = $-$3.5$\times  P$ +2.08.

 \subsection{Stellar Parameters Derivation} \label{stellar-param} 
A major addition in the SOS Cep\&RRL processing of DR2 sources was the activation  of a module  for  the derivation of stellar parameters for the confirmed 
Cepheids and RR Lyrae stars. 
This occurs in  the  {\it StellarParametersDerivation} module of the Cepheid and RR Lyrae branches, where stellar intrinsic parameters are derived through a variety of methods appropriate
for RR Lyrae stars or Cepheids. For DR2, we specifically implemented  the estimate of the photometric metal abundance ([Fe/H]) from the $\phi_{31}$ parameter of the Fourier 
light curve decomposition of fundamental-mode (RRab) and first-overtone (RRc) RR Lyrae stars,  and from the $R_{21}$ and $R_{31}$ parameters for fundamental-mode classical Cepheids with period shorter than 6.3 days.  For RRab stars we also activated a tool to 
estimate the absorption in the $G$ band, $A(G)$, from the light and colour curves. 

    \subsubsection{Metallicity}\label{metallicity}
\citet{jurcsik1996} were the first to devise a method for inferring a photometric metal abundance ([Fe/H])  from the $\phi_{31}$ parameter of the visual light curve Fourier decomposition 
 of fundamental-mode RR Lyrae stars. The method was later extended also to RRc stars by \citet{morgan2007}. \citet{nemec2013} provide a revision and recalibration of the $\phi_{31}$ - [Fe/H] relations based on  
 very accurate light curves of RRab and RRc stars observed by Kepler, along with metallicities derived from abundance analysis  of high resolution spectroscopy. 
 
In the {\it StellarParametersDerivation} module of the RR Lyrae branch,  photometric metal abundance of RRab and RRc stars observed by \textit{Gaia} 
are derived from the $\phi_{31}$ parameter of the Fourier 
$G$-band light curve decomposition using relations derived by \citet{nemec2013} for RRc and RRab stars, separately.
Specifically, in  \citet{nemec2013} the
values of [Fe/H] 
are estimated using the $\phi_{31}$ parameters  calculated by fitting the observed time series 
in the Kepler photometric system 
through a Fourier series of sine functions for RRab stars and cosine functions  for RRc sources.
In order to use eqs. (2) and (4) of \citet{nemec2013}, which are valid for RRab and RRc stars, respectively, we first have transformed the $\phi_{31}$ parameters from the 
\textit{Gaia} $G$ band into
the Kepler photometric system according to the following steps: i) the $\phi_{31}$ parameter in the $G$ band was first transformed
into the $V$ band using the relation $\phi_{31}$($V$)= $\phi_{31}$($G$) $- 0.104$; ii) the $\phi_{31}$ parameter  in the
Kepler system was then obtained using the following relations: $\phi_{31}^s = \phi_{31}$($V$)+ $\pi + 0.151$
and $\phi_{31}^c = \phi_{31}$($V$) + 0.151 \citep{nemec2011}  for RRab and RRc stars, respectively, where the 
superscript {\it s} stands for sine function while {\it c} indicates the use of the cosine function.
The uncertainties of the estimated [Fe/H] values were derived via  Monte Carlo simulations: 
i) the $G$-band $\phi_{31}$ parameter was varied using 100 random shifts extracted from a normal distribution with a
standard deviation equal to the error on the Fourier parameter itself; ii) the metallicity was recalculated
for each simulated $\phi_{31}$ value and the standard deviation, $\sigma_{sim}$, of these 100 values was estimated. The
final uncertainty of the derived [Fe/H] values was obtained as the sum in quadrature of the $\sigma_{sim}$ error (derived above) and of a conservative systematic
error assumed to be of 0.2 dex on account for systematics in the various calibrations and passband transformations\footnote{ 
\citet{gratton2004} and \citet{difabrizio2005} find average differences between photometric (from the $\phi_{31}$ parameter) and spectroscopic metallicities on the order of  0.30 $\pm$ 0.07 dex from a sample of RR Lyrae stars in the LMC.}.
We also recall that according to \citet{cacciari2005} photometric metallicities inferred with this method are better suited to describe the average metal abundance of a population of RR Lyrae stars rather than individual metallicities. As a check we have listed in Table~\ref{fourier-metal} the mean photometric metallicity derived  by the SOS Cep\&RRL pipeline for RR Lyrae stars in a number of Galactic globular clusters and dSphs observed by \textit{Gaia} and released in DR2 (see also Figs.~\ref{M3_isto_met},~\ref{M62_isto_met} 
and ~\ref{Sculptor_isto_met} in Sect.~\ref{gcs-dsphs}). 

The metallicity of fundamental-mode classical Cepheids   was calculated    
using the relations derived by \citet{klagyivik2013}. These authors calculate equations
to estimate the metallicity ([Fe/H]) for classical Cepheids with period  $\log P < 0.8$ ($P$= 6.3 days) using the Fourier parameters $R_{21}$ and/or $R_{31}$.
To use \citet{klagyivik2013}'s equations 
we first transformed the $G$-band $R_{21}$ and $R_{31}$ parameters into the corresponding $V$-band values through the 
equations $R_{21}$($V$) = 0.985 $\times R_{21}$($G$) and $R_{31}$($V$) = 0.982 $\times R_{31}$($G$) + 0.0098 that were obtained by inverting eqs.(9) and (10) in Paper~1. 
The errors in metallicity were estimated via Monte Carlo simulation applied to the $R_{21}$ and $R_{31}$ Fourier parameters and, as for the RR Lyrae stars, adding in quadrature 
a systematic error of 0.2 dex. 

The reliability of individual metallicities  inferred with the above methods also significantly depends on the reliability and accuracy of the $G$-band  Fourier parameters ($\phi_{31}$ for RR Lyrae stars and  $R_{21}$ and $R_{31}$ for Cepheids). We thus advise users to check errors of the Fourier parameters and inspect the light curves, before blindly trusting  
the published photometric metallicities, which are  inferred by automatically processing through the {\it StellarParametersDerivation} module of the SOS Cep\&RRL pipeline all sources for which  those Fourier parameters are available. 

\subsubsection{Absorption in the $G$ band}\label{absorption}

The absorption in the $G$ band,  $A(G)$,  of fundamental-mode RR Lyrae stars was determined from the following 
 empirical relation derived by \citet{piersimoni2002}:

\begin{equation}
\begin{split}
(V-I)_{0} = (0.65\pm0.02) -(0.07\pm0.01)\times{\rm [Amp(}V{\rm )]}\\  + (0.36\pm0.06)\times \log(P)\\ 
~~~~(\sigma = 0.02)\\
\end{split}
\end{equation}

\noindent However, to use this relation, we first had to transform amplitudes and colours from the  Johnson to the {\it Gaia} passbands. To this purpose we 
used the following transformation equation for the amplitude in  the $V$ band  (inverting Eq.~3 of Paper~1): \\

\begin{equation}
\begin{split}
{\rm Amp(}V{\rm )}= (1.081\pm0.003 )\times {\rm Amp(}G{\rm )} + (0.013\pm0.003)\\ 
~~~~(\sigma = 0.012)\\
\end{split}
\end{equation}

\noindent Then,  we calculated the $(V-I)$ colours from the $(G-G_{\rm RP})$ ones:
\begin{equation}
\begin{split}
(V-I) = (0.027\pm0.003) + (1.13\pm0.02)\times(G-G_{\rm RP})\\ + (0.55\pm0.03)\times(G-G_{\rm RP})^{2}\\   
~~~~(\sigma = 0.013)\\
\end{split}
\end{equation}

\noindent and the conversion between the $G$ absorption, $A(G)$,  and the reddening  $E(V-I)=(V-I)-(V-I)_0$
\begin{equation}
\begin{split}
A(G) = [(2.3116\pm0.006) - (0.3097\pm0.0011)\times(V-I)_0]\\ \times E(V-I)\\   
~~~~(\sigma = 0.013)\\
\end{split}
\end{equation}

\noindent which are both based on \citet{jordi2010} passband transformations\footnote{\citet{jordi2010} transformations are superseded by new transformations published in \citet{evans2018}. 
These new transformations became available when the whole variability processing for DR2 had already been completed. We are currently updating the SOS Cep\&RRL pipeline to the new transformations,  in preparation for next \textit{Gaia} releases.}. 

 In practice, for each ab-type RR Lyrae, the observed amplitude in the $G$ band, ${\rm Amp(}G{\rm )}$,  and the star period give the expected 
intrinsic value of $(V-I)_0$ through Eqs.~1 
and ~2. 
 The ``observed'' Johnson $(V-I)$ colour 
is calculated from the observed {\it Gaia} $(G-G_{\rm RP})$ colour  by means of Eq.~3.  
Finally, the calculated 
$(V-I)_0$ and $(V-I)$ colours are inserted into Eq.~4  
 to obtain the source $A(G)$ absorption.   

 \section{Application of the SOS Cep\&RRL pipeline to \textit{\textbf{Gaia}} DR2 dataset: source selections  and processing simplifications}\label{s3new}
 
The dataset  processed by the SOS Cep\&RRL pipeline to produce results published in {\it Gaia} DR2 consisted of $G$, $G_{\rm BP}$ and $G_{\rm RP}$  time series photometry\footnote{Each point in the $G$-band
time series is the mean of the nine measurements taken in the Astrometric Field (AF)
CCDs and collected during one observation of a source by {\it Gaia}, while each $G_{\rm BP}$, $G_{\rm RP}$ measurement is integrated over the
low-resolution spectra collected in the Blue and Red Photometer (BP and RP) CCDs.} 
 of candidate Cepheids and RR Lyrae stars, observed by {\it Gaia} in the  22 months between  July 2014 and  May 2016. The time series data  provided in units of flux by the photometry pipeline,  were  converted  
 into magnitudes by the variability processing prior SOS, using the magnitude zero-points defined in \citet{evans2018}. 
For DR2 the minimum number of data points in the $G$-band time series of sources to be fed  
into the SOS Cep\&RRL pipeline
was reduced to 12, as this limit was deemed to be sufficient for a reliable estimate of the period and other (pulsation) characteristics of confirmed Cepheids and RR Lyrae stars, based upon 
experience of the DR1 analysis and results (see Paper~1).  
 
 Sources  with more than 12 epoch-data in the $G$ band, pre-classified as candidate 
Cepheids and RR Lyrae stars by  the classifiers of the  
 general variability pipeline following the two separate paths described in Sect.~\ref{s2} (\citealt{eyer2017b}, \citealt{rimoldini2018}) were ingested into the SOS Cep\&RRL pipeline (see fig. 2 in \citealt{holl2018})   for a total of  639,828 univocally defined candidate RR Lyrae stars and 72,717 candidate Cepheids. 
 These rather
large numbers of candidates included even small probability levels and also candidates
flagged as class outliers 
in order to maintain a high level of  completeness and not lose potentially valid candidates. 
However, as done also in DR1 (see discussion in section 3.1 and fig. 15 of Paper~1), 
we dropped sources with ${\rm Amp(}G{\rm )}$ $\leq$ 0.1 mag.
Further cuts were applied during validation to conform to  
quality assurance limits set by \textit{Gaia} photometric and astrometric processing teams. That is we dropped candidate RR Lyrae stars and Cepheids with {\it excess flux} above the limit recommended in the photometry processing \citep{evans2018} and we also rejected All-Sky candidate Cepheids with {\it excess noise} above the limits recommended in the astrometry processing \citep{lindegren2018}.

 In order to facilitate an early release, the very tight schedule of the DR2 data-processing  did not allow a full  iteration between the 
pipelines processing  the different \textit{Gaia} data. 
In particular, since the photometry and variability pipelines  run in parallel, the SOS Cep\&RRL pipeline could be tested only on  a preliminary version of the multiband, time-series 
photometry of the candidate Cepheids and RR Lyrae stars and it was not possible to update the SOS Cep\&RRL pipeline and compute its diagnostic relations  
(see, e.g.,  Sects.~\ref{absorption}  
and ~\ref{allskycep}) using directly the \textit{Gaia}  $G$, $G_{\rm BP}$ and $G_{\rm RP}$ photometry of the sources.

In addition,  the SOS Cep\&RRL pipeline could access for test only  a preliminary version of the astrometric solution that differs 
from the final astrometry released in DR2. Due to the limited time available, the preiminary nature of the parallaxes (and related errors) available for the SOS Cep\&RRL initial processing  and because the vast majority of RR Lyrae stars released with \textit{Gaia} DR2 are typically in the Galactic halo and the Magellanic Clouds\footnote{The Galactic bulge region, where reddening is very high,  is only partially covered by the \textit{Gaia} DR2 observations and still with a limited number of epochs.}, where reddening is  a minor issue,  we did not implement the use of the $PW$ relations to process the RR Lyrae candidates. For similar reasons and because the Magellanic Cloud Cepheids have small true parallaxes, close to zero and with still rather large uncertainties in DR2,  
we did not use them and for the classification of the SMC and LMC Cepheids we preferred to use the $PL$ and $PW$ 
relations in apparent magnitude taken from the OGLE studies,  transformed to the \textit{Gaia} $G$, $G_{\rm RP}$  passbands and with zero points set by adopting literature values for the LMC and SMC distance moduli.

We also did not have access to any astrophysical parameters (reddening values in particular) or to radial velocity measurements of the candidates.  

In summary, for the processing of the DR2 sources we used the same pipeline as for DR1 (see Paper~1), whose relations were obtained by transforming to the {\it Gaia} passbands, via \citet{jordi2010} passband transformations, quantities and relations originally defined in the  Johnson-Cousins passbands, according to the procedures extensively described in sect. 2.2., Eqs. (2) to (19), sect. 2.4.2, Eqs. (20) to (24) and Appendix A of Paper~1.

\clearpage

 \subsection{SOS Cep\&RRL processing of all-sky RR Lyrae stars released in DR2}\label{allskyrrl}
  \begin{figure}
   \centering
  \includegraphics[trim=20 130 -10 80, width=9.5 cm,clip]{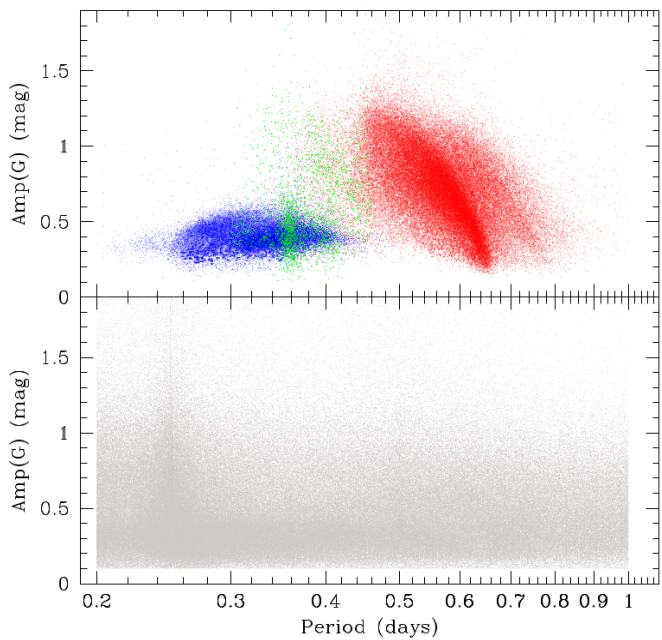}
 \caption{\textit{Upper panel}: $G$-band  $PA$ diagram of the 140,784 RR Lyrae stars confirmed by the SOS Cep\&RRL pipeline that are released in DR2 (blue: first-overtone --RRc, green: 
  double-mode --RRd, red: fundamental-mode --RRab pulsators). \textit{Lower panel}: $PA$ diagram of RR Lyrae candidates that were rejected from the SOS pipeline, 499,044 sources in 
   total (see text for details).  
}
              \label{pampRR}%
    \end{figure}

 \begin{figure}
 \centering
  \includegraphics[trim= 20 130 -10 80, width=9.5 cm,clip]{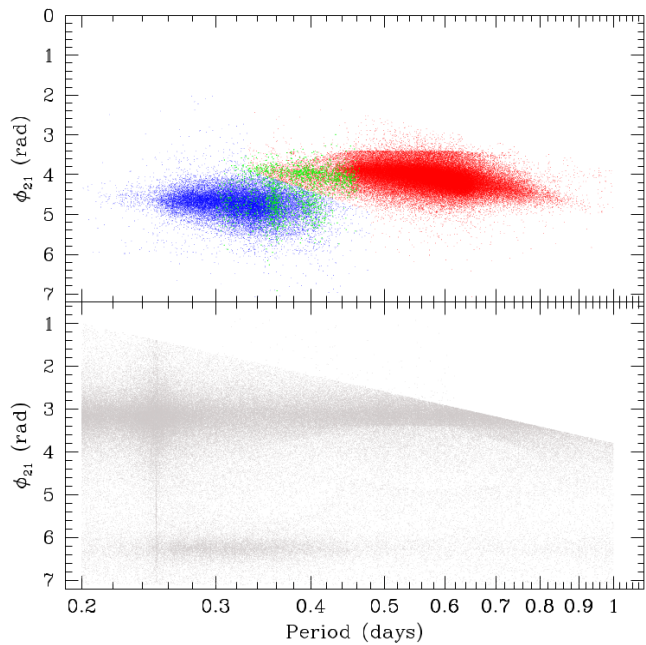}
  \caption{\textit{Upper panel}: $G$-band $\phi_{21}$ vs period   
  diagram for the RR Lyrae stars confirmed by the SOS Cep\&RRL pipeline (140,784 objects).  Colour-coding is same as in Fig.~\ref{pampRR}.  \textit{Lower panel}: $G$-band $\phi_{21}$ vs period  diagram  of RR Lyrae candidates that were rejected from the SOS pipeline, 
499,044 sources in total. The significant peak around  $P$=0.25 d  (also seen in the lower panel of Fig.~\ref{pampRR})  is an alias  
due to the  rotation period of {\it Gaia} around its axis. }
             \label{phi21RR}%
    \end{figure}
Although partially reduced by cutting in ${\rm Amp(}G{\rm )}$, {\it excess flux} and {\it excess noise}, the number of sources to process through the SOS Cep\&RRL pipeline still consisted of several hundreds of thousands. This along with the very tight schedule of \textit{Gaia} DR2, made it mandatory to run the SOS pipeline in 
 fully automatic  way limiting quality controls and verification of derived products to a detailed analysis of only  small, randomly selected, samples of specific subtypes, whose folded light curves were visually inspected during the validation of the results (see Sect.~\ref{validation}). 
The following diagnostic tools were specifically used 
and their results combined to extract from the candidates the bona fide RR Lyrae stars that are published in DR2: the $PA$,  $\phi_{21}$ vs $P$, $R21_{21}$ vs $P$,  $\phi_{31}$ vs $P$ and $R_{31}$ vs $P$ diagrams. All those tools rely on parameters derived only from the $G$-band light curves, where the time series data were  folded according to period and epoch of maximum light determined by the 
SOS Cep\&RRL pipeline and modelled by the non-linear fitting algorithm of module {\it NonLinearFourierAnalysis} (see Fig.~\ref{decomposition1}).
 Tools and diagnostics 
  of the SOS pipeline were run in fully automatic way, producing validations plots from  which the bona fide RR Lyrae stars and Cepheids were selected. 
The validation plots of the $PA$ and $\phi_{21}$ vs $P$ diagrams are shown in Figs.~\ref{pampRR} and ~\ref{phi21RR}. 
The $PA$  tool is the most applicable, but it is also the less constraining and most contaminated one  (see lower panel of Fig.~\ref{pampRR}), as it  requires knowledge of only the source period ($P$) and amplitude in the $G$ band [${\rm Amp(}G{\rm )}$]. Those quantities were available for all sources as result of the SOS Cep\&RRL processing. 
   The upper panel of Fig.~\ref{pampRR}  shows the $G$-band $PA$ diagram of the 140,784 RR Lyrae stars confirmed by the SOS Cep\&RRL pipeline that are released in DR2. The lower panel of the figure shows RR Lyrae candidates that were rejected from the SOS pipeline, 499,044 sources in total.  
The plot shows the high contamination of the RR Lyrae candidate sample
that in large part is due  to binaries and other  
type of variable sources. However, we note that a significant fraction of the sources in the lower  $PA$ diagram may be bona fide RR Lyrae stars not confirmed by the SOS  pipeline for a number of different reasons, among them  the lack of a good period determination. 
Indeed, during the validation process (Sect.~\ref{validation}), the cross-match with RR Lyrae catalogues in the literature revealed that  about 27,000 -- 30,000 sources in the lower panel of Fig.~\ref{pampRR} are known RR Lyrae stars that the SOS pipeline rejected due to a wrong period determination.  We note that, although not present in the variability tables, mean $G$, $G_{\rm BP}$, $G_{\rm RP}$ magnitudes are released in DR2 for all those rejected candidate RR Lyrae stars.

Only for $\sim$ 220,000 of the candidate RR Lyrae it was possible to model the light curve with at least two harmonics, hence measure the  $\phi_{21}$ , $R_{21}$ Fourier parameters, and construct the $\phi_{21}$ vs $P$ diagram shown in Fig.~\ref{phi21RR}{\footnote{In DR2 we did not publish the number of harmonics used to model the light curves, this parameter will be added for DR3.}. The light curves of all remaining candidates, having a sinusoidal shape, were modelled with only one harmonic and were classified only by means of the $PA$ diagram. The upper panel of Fig.~\ref{phi21RR} shows the  $\phi_{21}$ vs $P$ diagram 
for the RR Lyrae stars confirmed by the SOS Cep\&RRL pipeline processing and validation (140,784 objects) whereas the lower panel shows RR Lyrae candidates that were rejected, 
499,044 sources in total. The sharp diagonal cut in the figure is due to the limits set in the SOS Cep\&RRL pipeline to separate RR Lyrae stars from Cepheids in the 
 $\phi_{21}$ vs $P$ plane  (see Fig. 21 in Paper~1, for details). 
The significant peak around  $P$=0.25 d is an alias  due to the  rotation period of {\it Gaia} around its axis.   The two stripes running parallel to the abscissa axis, are populated by binaries (that at 
 $\phi_{21} \sim $6.28 rad) and mainly by other types of variables 
   sources (the one at $\phi_{21} \sim $3.14 rad) whose removal caused a quite sharp cut to appear in the RRab star distribution in the upper panel of the figure.

Finally, about half of the sources shown in the upper panels of Figs.~\ref{pampRR} and ~\ref{phi21RR} were modelled with 3 or more harmonics, hence,  for those sources we could also measure  the metal abundance from the $\phi_{31}$ Fourier parameter (see Sect.~\ref{results-rrl}).

In  summary, after running the  SOS Cep\&RRL on the 639,828 RR Lyrae candidates  and  final validation of the results, 140,784 were confirmed as bona fide RR Lyrae stars, and 499,044 were rejected. 
$P$ and amplitude in the $G$ band [${\rm Amp(}G{\rm )}$] are available for all 140,784 confirmed RR Lyrae stars,  
the $\phi_{21}$, $R_{21}$  Fourier parameters are available for 121,234 of them and,  the $\phi_{31}$, $R_{31}$ only for 67,681. This also bears upon quality and reliability of the SOS  Cep\&RRL classifications in type, subtype and pulsation mode(s). A complete discussion of the validation process and final results for the RR Lyrae stars is provided in Sects.~\ref{validation} and ~\ref{results-rrl}.

 \subsection{SOS Cep\&RRL processing of all-sky Cepheids released in DR2}\label{allskycep}
As anticipated at the end of Sect.~\ref{s3new}, due to the schedule constraints of the DR2 processing  
 it was not possible to update the diagnostic relations of the SOS Cep\&RRL  pipeline using 
 directly the {\it Gaia} light curves and, for the processing of the DR2 sources, we used the same version 
 of the pipeline and relations as defined for DR1 (see Paper~1).   
 In particular, for  the analysis of the DR2 candidate Cepheids and the
 classification of the confirmed ones into different types (DCEP, ACEP, and T2CEP subtypes)
 and pulsation modes, the SOS Cep\& RRL pipeline  
 relied on: i) the Fourier parameters $\phi_{21}$ and $R_{21}$ of the
 $G$-band light curve decomposition and the source position on the
 $\phi_{21}$ vs $P$, $R_{21}$ vs $P$ planes  as defined in sect.~4 and Figs. 21,
 22 of Paper~1, and  ii) the source position with respect to the
 $G$-band $PL$ and the $PW(G,G_{\rm RP})$ relations, using
 different values for the slope and zero point depending on the region of the sky 
 where the sources were located. 
 
 Specifically, for the processing of candidate Cepheids within
 the LMC region (Sect.~\ref{s2}) we used the $G$-band $PL$ relations
 described by eqs. 20, 21, 22, 23, and 24 in Paper~1 (see
 sect. 2.4.2. in that paper) and the ($G$, $G_{\rm RP}$) $PW$ relations
 defined by eqs. (6), (7), (8), (9) and (10) in the present paper.
 These relations were 
obtained by transforming to the {\it Gaia} passbands, according to \citet{jordi2010},  the relations derived   
  from  LMC Cepheids whose light curves in the Johnson-Cousins passbands 
 and pulsation characteristics have been published
 by  the OGLE team, adopting an absolute de-reddened distance modulus for the LMC of
 $DM_{LMC}=18.49$ mag from \citet{pietrz2013} and the 
 definition of the Wesenheit function in the {\it Gaia} 
 $G$, $G_{\rm RP}$ passbands provided by Eq, (5), which was derived for the present study assuming for the ratio of total-to-selective absorption the value $R_{\lambda}$=3.1.
\begin{equation}
\begin{split}
W_{(G, G_{\rm RP})} = G - 0.08193 - 2.98056 \times (G - G_{\rm RP})\\ 
~~~~~~~~~~ - ~~ 0.21906 \times (G - G_{\rm RP})^2\\
~~~~~~~~~~~~~~ - ~~ 0.6378 \times (G - G_{\rm RP})^3
\end{split}
\end{equation}
\begin{equation}
\begin{split}
{\rm DCEP_F~~}:~W_{(G, G_{\rm RP})} = 15.861 - DM_{\rm LMC} -3.317 \times \log P~~~ \\\sigma=0.069~{\rm mag}
\end{split}
\end{equation}
\begin{equation}
\begin{split}
{\rm DCEP_{1O}~~}:~W_{(G, G_{\rm RP})} = 15.365 - DM_{\rm LMC} -3.456 \times \log P~~~ \\\sigma=0.067~{\rm mag}
\end{split}
\end{equation}
\begin{equation}
\begin{split}
{\rm T2CEP~~}:~W_{(G, G_{\rm RP})} = 17.321 - DM_{\rm LMC} -2.527 \times \log P~~~ \\\sigma=0.088~{\rm mag}
\end{split}
\end{equation}

\begin{equation}
\begin{split}
{\rm ACEP_F~~}:~W_{(G, G_{\rm RP})} = 16.567 - DM_{\rm LMC} -3.19 \times \log P~~~ \\\sigma=0.15~{\rm mag}
\end{split}
\end{equation}
\begin{equation}
\begin{split}
{\rm ACEP_{1O}~~}:~W_{(G, G_{\rm RP})} = 15.995 - DM_{\rm LMC} -3.26 \times \log P~~~ \\\sigma=0.14~{\rm mag}
\end{split}
\end{equation}

\noindent {Similarly,} for candidate DCEPs within the SMC region (Sect.~\ref{s2}) we used the
$G$-band $PL$ and the $PW(G,G_{\rm RP})$ relations described by the
following set of equations:

\begin{equation}
\begin{split}
{\rm DCEP_F~~}:~M_G = 17.984 - DM_{\rm SMC} -2.898 \times \log P~~~ \\\sigma=0.266~{\rm mag}
\end{split}
\end{equation}
\begin{equation}
\begin{split}
{\rm DCEP_{1O}~~}:~M_G = 17.368 - DM_{\rm SMC} -3.155 \times \log P~~~ \\\sigma=0.271~{\rm mag}
\end{split}
\end{equation}

\noindent that were  derived transforming to the \textit{Gaia} passbands the relations of \citet{soszynski2015b} and adopting an absolute de-reddened
distance modulus of $DM_{SMC}=19.00$ mag obtained by simply adding
+0.51 mag the LMC distance modulus of \citet{pietrz2013}.

\begin{equation}
\begin{split}
{\rm T2CEP~~}:~M_G = 19.31 - DM_{\rm SMC} -1.96 \times \log P~~~ \\\sigma=0.188~{\rm mag}
\end{split}
\end{equation}

\noindent  that was derived from \citet{soszynski2015b}.\\

\begin{equation}
\begin{split}
{\rm ACEP_F~~}:~M_G = 18.33 - DM_{\rm SMC} -2.63 \times \log P~~~ \\\sigma=0.22~{\rm mag}
\end{split}
\end{equation}

\begin{equation}
\begin{split}
{\rm ACEP_{1O}~~}:~M_G = 17.81 - DM_{\rm SMC} -3.78 \times \log P~~~ \\\sigma=0.20~{\rm mag}
\end{split}
\end{equation}

\noindent  that were defined  adopting the ACEP F and ACEP FO slopes
of the $I$-band $PL$ relations from \citet{soszynski2015a} and a $(V-I)$ average colour for the ACEPs of 0.8 mag.\\

\begin{equation}
\begin{split}
{\rm DCEP_F~~}:~W_{(G, G_{\rm RP})} = 16.493 - DM_{\rm SMC} -3.46 \times \log P~~~ \\\sigma=0.155~{\rm mag}
\end{split}
\end{equation}

\begin{equation}
\begin{split}
{\rm DCEP_{1O}~~}:~W_{(G, G_{\rm RP})} = 15.961 - DM_{\rm SMC} -3.548 \times \log P~~~ \\\sigma=0.169~{\rm mag}
\end{split}
\end{equation}
\begin{equation}
\begin{split}
{\rm T2CEP~~}:~W_{(G, G_{\rm RP})} = 17.64 - DM_{\rm SMC} -2.32 \times \log P~~~ \\\sigma=0.23~{\rm mag}
\end{split}
\end{equation}
\begin{equation}
\begin{split}
{\rm ACEP_F~~}:~W_{(G, G_{\rm RP})} = 17.01 - DM_{\rm SMC} -2.85 \times \log P~~~ \\\sigma=0.15~{\rm mag}
\end{split}
\end{equation}
\begin{equation}
\begin{split}
{\rm ACEP_{1O}~~}:~W_{(G, G_{\rm RP})} = 16.64 - DM_{\rm SMC} -3.69 \times \log P~~~ \\\sigma=0.14~{\rm mag}
\end{split}
\end{equation}

\noindent  that were derived from \citet{soszynski2015b} for DCEPs and \citet{soszynski2015a}  for ACEPs and  T2CEPs. 

Candidate Cepheids  located in the LMC and SMC regions that fall within 4$\sigma$ of any  of the above $PL$, $PW$
relations were assigned the Cepheid type  and pulsation mode of the
closest $PL$ or $PW$ relation. The classification was then refined using the Fourier
decomposition parameters $R_{21}$ and $\phi_{21}$.  
Conversely, candidate Cepheids falling beyond 4 $\sigma$ were rejected as misclassified sources. 
Such large intervals were adopted in order to achieve a higher completeness introducing though a larger contamination that we have tried to mitigate 
during the validation process (see Sect.~\ref{validation}). 
As in Paper~1 the T2CEPs were further subclassified into: the BL Her, W 
Vir,  and the RV Tau type classes, depending on the pulsation 
period.  
Following \citet{soszynski2008} in the  {\it T2CEPSubclassification}
module T2CEPs with periods in the range  1$\leq P <$ 4 d were 
classified as BL Her, 
  the T2CEPs with periods in the range 4$\leq P <$ 20 d as W Vir,  and  those 
with periods equal to or longer than 20 days as RV Tau.  

Similarly, ACEPs are known to pulsate in the F and 1O modes.  The {\it
  ACEPModeIdentification} module assigns the pulsation mode to an ACEP
by combining results from the classification in types based on the
$PL$ relations ({\it CepTypeIdentification} module) and the source
period, as 1O ACEPs have periods in the range 0.35 $<$ P $\leq$ 1.20 d,
whereas F ACEPs have periods in the range 1.20 $<$ P$ \leq$ 2.5 d.
These limits were inferred from the $PL$ relations of ACEPs based on
OGLE-III data.

The selection of bona fide Cepheids from the All-Sky candidate sample (Sect.~\ref{s2}),  was made based on the star position in the $\phi_{21}$ vs $P$ and $R_{21}$ vs $P$ Fourier parameter 
 planes and retaining only candidates  located in the regions populated by Cepheids known in the literature,  as defined in sect.~4 and Figs. 21,
 22 of Paper~1.  
 As a test   we then compared the  All-Sky Cepheids selected from the Fourier parameter planes with the   $PW(G,G_{\rm RP})$ relations in Eqs. (21) - (23), which   
for  DCEPs were derived from  the TGAS
DR1 sample using the Astrometry-Based Luminosity (ABL, \citealt{Arenou1999};  see e.g. 
\citealt[][and references therein]{gaiacol-clementini}),  whereas for T2CEPs we used the 
\citet{soszynski2015a} relation, transformed to the \textit{Gaia} passbands and  with an LMC distance modulus of 18.49 mag  subtracted.\\

\begin{equation}
{\rm DCEP_F~~}:~W_{(G, G_{\rm RP})} = -3.21 -2.93 \times \log P~~~ \sigma=0.37~{\rm mag}
\end{equation}

\begin{equation}
{\rm DCEP_{1O}~~}:~W_{(G, G_{\rm RP})} = -4.31 -2.98 \times \log P~~~ \sigma=0.69~{\rm mag}
\end{equation}

\begin{equation}
{\rm T2CEP~~}:~W_{(G, G_{\rm RP})} = -1.15 -2.53 \times \log P~~~ \sigma=0.11~{\rm mag}
\end{equation}
 
 We considered only the $PW(G,G_{\rm RP})$ relation, 
because it is  
reddening free and no information on the reddening of the All-Sky Cepheids, which are highly reddened since reside mainly in the Galactic disc, was available at the time of the processing\footnote{The Wesenheit  function $W$ \citep{madore1982}  is reddening-free by construction, however, its depends on the assumed value of the ratio of total-to-selective absorption $R_{\lambda}$. In the present study we adopt $R_{\lambda}$=3.1,  which perhaps is too low for classical Cepheids in dusty regions of the Galactic disc, where a larger value might be more appropriate. For DR3 we foresee the adoption of a varying $R_{\lambda}$ value depending on the source location on the sky.}.

For the comparison with the $PW$ relations we used the parallaxes working
directly in parallax space with the ABL.  The deviation ($\Delta$) from the reference $PW$ relations was computed according to the equation:

\begin{equation}
\Delta = \left| ABL - 10^{0.2\left(a\log P + b \right)} \right|
\end{equation}

\noindent Where ABL is defined as: 
\begin{equation}
ABL = \varpi 10^{0.2 W_{(G, G_{\rm RP})} -2.0}
\end{equation}

\noindent  and $\varpi$ is the star parallax.\\

The slope of the $PW$ relation for the All-Sky fundamental-mode DCEPs [Eq. (21)] appears to be shallower than the corresponding OGLE slopes for the LMC and SMC samples and, more importantly, the scatter in the $PW$ relations for the All-Sky DCEPs [Eqs. (21) and (22)] is much larger  than the 
 LMC values,  going from $\pm$ 0.069 mag for fundamental-mode and 0.067 mag for first-overtone DCEPs in the LMC, to $\pm$ 0.37 and 0.69 mag 
 for the corresponding All-Sky samples.  
This is  due to the large uncertainty of the DR1 TGAS parallaxes used to derive the DCEP reference  $PW$ relations in Eqs. (21) and (22) and, likely, also to the adoption of  $R_\lambda$=3.1 in the definition of the $PW$ functions.\\

In total the SOS pipeline analysed 72,455 candidate Cepheids provided by the
classifiers. A total of 62,880 objects were rejected.  
 Among the remaining 9,575 objects, 3,767, 3,692 and 2,116 
are distributed in the LMC, SMC and All-Sky regions, respectively.  
The vast majority of the Cepheids of all types in the Magellanic Clouds were already known from the OGLE
survey, with the exception of 118 new objects. As for the All-Sky
sample, 998 objects are classified as Cepheids of any type in
the literature, 419 have an uncertain or other than Cepheid classification,
and  for the remaining  699 targets no cross-match with known sources was 
found (\citealt{ripepi2018-dr2}).

Sources classified as Cepheids of any type by the SOS pipeline are displayed in Figs.~\ref{plg} and ~\ref{pw}
for the $PL$ and $PW$ relations, respectively. In both figures, upper,
middle and lower panels display LMC, SMC and All-Sky sources,
respectively. Note that only objects with positive parallaxes could be 
shown in the All-Sky panels of these figures.   
An intriguing feature of Figs.~\ref{plg} and ~\ref{pw} is the presence 
of a sequence of stars running almost parallel to the abscissa axis with absolute $G$ magnitudes fainter than 2 mag.  
This sequence can also be clearly seen in  Figure~\ref{plnote}, that shows the $PL$ and $PW$ relations for the 998 All-Sky
sources that have a classification as Cepheids in the literature.  
 We have drawn dashed lines in Figs.~\ref{plg}, ~\ref{pw} and ~\ref{plnote} to better highlight the regions populated by those faint sources.
The true nature of the sources populating these faint sequences is unclear. 
 If they are bona fide Cepheids their  \textit{Gaia} DR2 parallax must be  incorrect. Conversely, if their parallax is correctly measured
they cannot be Cepheids and the SOS Cep\&RRL classification is in error as also must be some of the literature classifications.  
Indeed, we believe that the All-Sky Cepheid sample may be  significantly contaminated by spurious sources and that a large fraction  of the sources below the dashed lines in the lower panels of Figs.~\ref{plg} and ~\ref{pw} and in Fig.~\ref{plnote} might be misclassifications. In the Fourier parameters vs period planes (Figs.~\ref{CEP_21} and ~\ref{CEP_31}) these spurious sources 
share the same loci, hence are indistinguishable from bona fide Cepheids,  
 that is why they were retained in spite of them  being more than 4 sigma away from the $PW$ relations in Eqs. (21)-(23),  but they are in fact a different type of variables
 much fainter than Cepheids.  
This is indeed the case of the star with \textit{Gaia} sourceid 2077108036182676224. The SOS Cep\&RRL pipeline classifies this source as a multi-mode classical Cepheid with fundamental-mode period $P$=1.045d. However, the parallax places the star on  the lower sequences in Figs.~~\ref{plg} and ~\ref{pw}. The source was observed also by Kepler, (Kepler sourceid: KIC 6619830) and has a full Kepler light curve showing a hump in the phase interval 0.7-0.9 that indicates it is a spotted rotating star, which is consistent with its absolute magnitude 
(Tim Bedding and Dan Hey, private communication). The phase coverage of the \textit{Gaia} light curve is poor and the hump at phase 0.7-0.9 is not sampled, so it could easily be mistaken for a Cepheid. 
Perhaps some of the sources on the lower sequence of Figs.~~\ref{plg} and ~\ref{pw} are rotators with a few observations and poor coverage of the light curve that were misclassified by the SOS Cep\&RRL pipeline.
The large scatter in Figs.~~\ref{plg}, ~\ref{pw} could  also be due in part  to 
 the DR2 parallax for some of the bona fide Cepheids  being in error due to an incorrect determination  of the star mean $G$ magnitude and because the colour variation during the pulsation cycle was not taken into account in the derivation of the DR2 astrometric solution (see also https://www.cosmos.esa.int/web/gaia/dr2-known-issues and the  technical note:  GAIA-C3-TN-LU-LL-124-01).  An additional source of scatter in the $PL$ distribution in Fig.~\ref{plg} is that the reddening affecting the sources was not known, therefore we could not apply any correction for reddening.
 All these effects may have combined to 
 inflate the  dispersions observed In  the $PL$ and $PW$ distributions of the All-Sky MW samples in Figs.~\ref{plg} and ~\ref{pw},  
as it is suggested by the large scatter seen also for  the known Cepheids in Fig.~\ref{plnote}.

 We refer to \citet{ripepi2018-dr2}, 
where these issues are discussed more in detail 
and a
catalogue of bona fide DR2 new All-Sky Cepheids is presented 
after  clearing the sample from other types of  
variable sources and misclassifications.

  \begin{figure*}
   \centering
   \includegraphics[width=18.0 cm, trim= 10 140 0 80, clip]{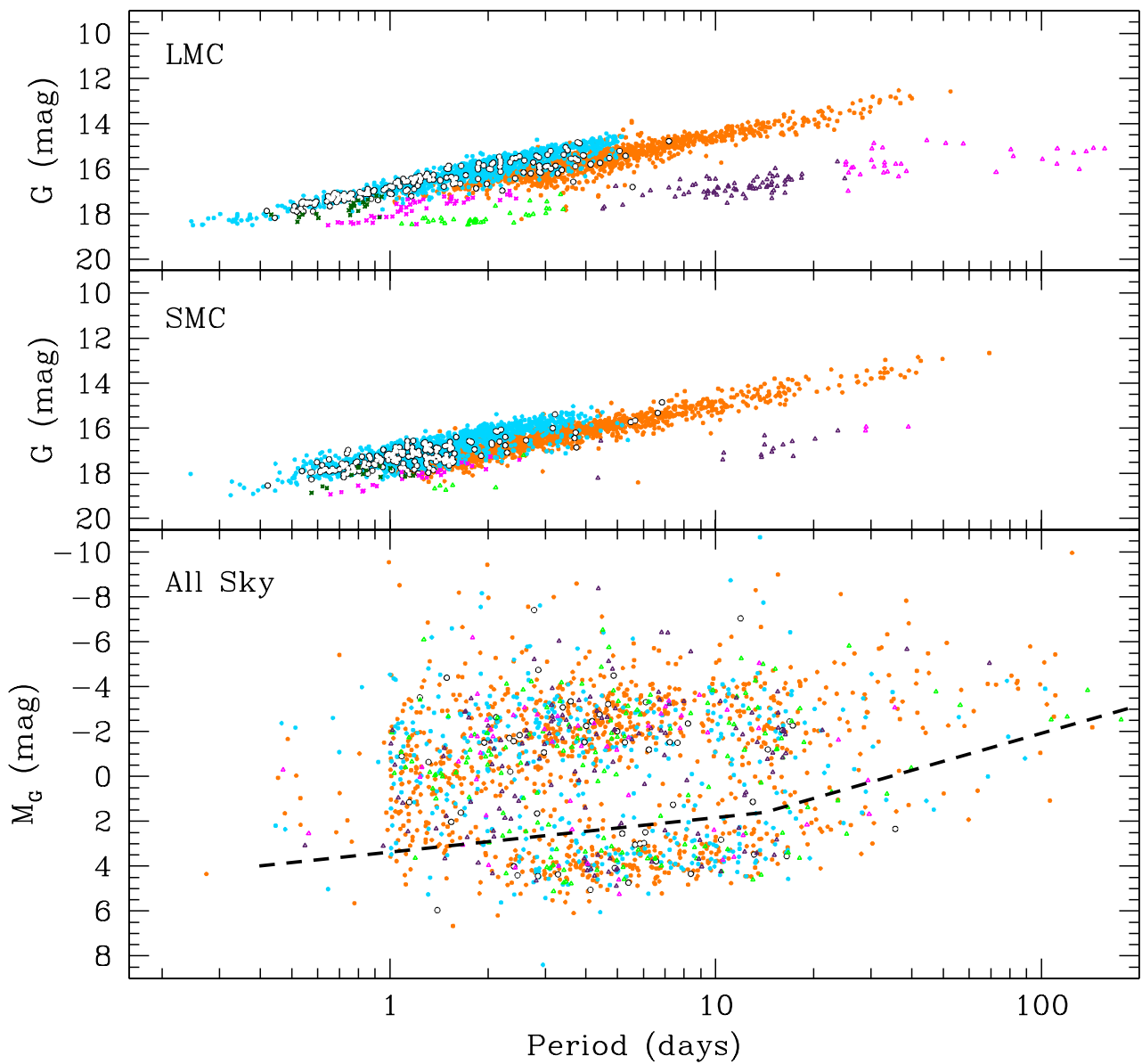}
   \caption{\textit{Upper and middle panels:} $G$-band $PL$ distribution in apparent magnitude not corrected for reddening of DCEPs,  ACEPs, and T2CEPs in the LMC and SMC, respectively. 
  \textit{Lower panel:}  $PL$ distribution in absolute $G$ magnitude (M$_G$) not corrected for extinction of All-Sky Cepheids of the different types.  
Orange filled circles: DCEPs F; cyan filled circles: DECPs 1O; blank filled circles: multimode DCEPs; magenta four-starred symbols: ACEPs F; dark green four-starred symbols: ACEPS 1O; green open triangles: BL Her; violet open triangles:
W Vir; magenta open triangles: RV Tau. The much larger scatter of the All-Sky Cepheid $PL$ distribution is clearly seen:  the  Y-axis in the lower panel of the figure spans  a magnitude range of 20.0 mag, to compare with the 11.5 mag range of the two upper panels.
Several All-Sky  sources lie below the dashed line in the lower panel 
of the figure. They are a mixing of misclassifications (spurious sources),  sources with very high reddening and Cepheids with a wrong parallax value due to  the still simplified astrometric processing applied for DR2, among which, in particular, the lack of a proper treatment of binary/multiple sources (see  Sect.~\ref{allskycep}  and \citealt{ripepi2018-dr2}, for more details).}
              \label{plg}
    \end{figure*}

  \begin{figure*}
   \centering
 \includegraphics[width=18.0 cm, trim= 10 140 0 80, clip]{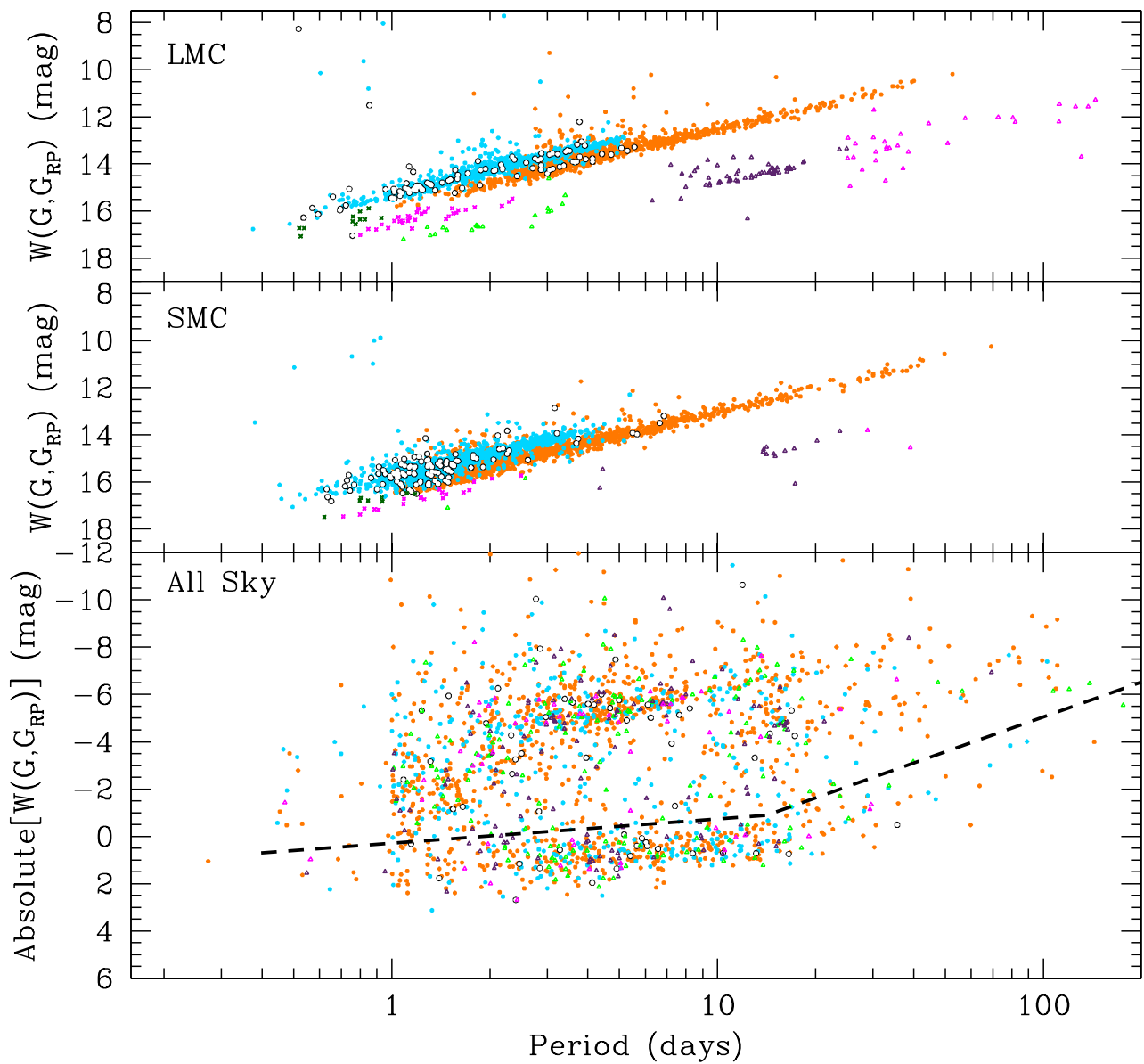}
   \caption{\textit{Upper and middle panels:}  $PW(G, G_{\rm RP})$ distribution in apparent magnitude of DCEPs, ACEPs, and T2CEPs in the LMC and SMC,
respectively.  \textit{Lower panel:} Period - Absolute[$W(G, G_{\rm RP})$]  distribution of All-Sky Cepheids of the different types. Symbols and colour-coding are the same as in Fig.~\ref{plg}. 
The much larger scatter of the All-Sky Cepheid $PW$ distribution is clearly seen:  the  Y-axis in the lower panel of the figure spans  a magnitude range of 18.0 mag, to compare with the 11.5 mag range of the two upper panels.
Several All-Sky  sources lie below the dashed line in the lower panel 
of the figure. They are a mixing of misclassifications (spurious sources) and Cepheids with a wrong parallax value due to  the still simplified astrometric processing applied for DR2, among which  in particular the lack of a proper treatment of binary/multiple sources (see  Sect.~\ref{allskycep} and  \citealt{ripepi2018-dr2}, for more details).}
              \label{pw}
    \end{figure*}

    \begin{figure*}
   \centering
   \includegraphics[width=18.0 cm, trim= 10 140 0 80, clip]{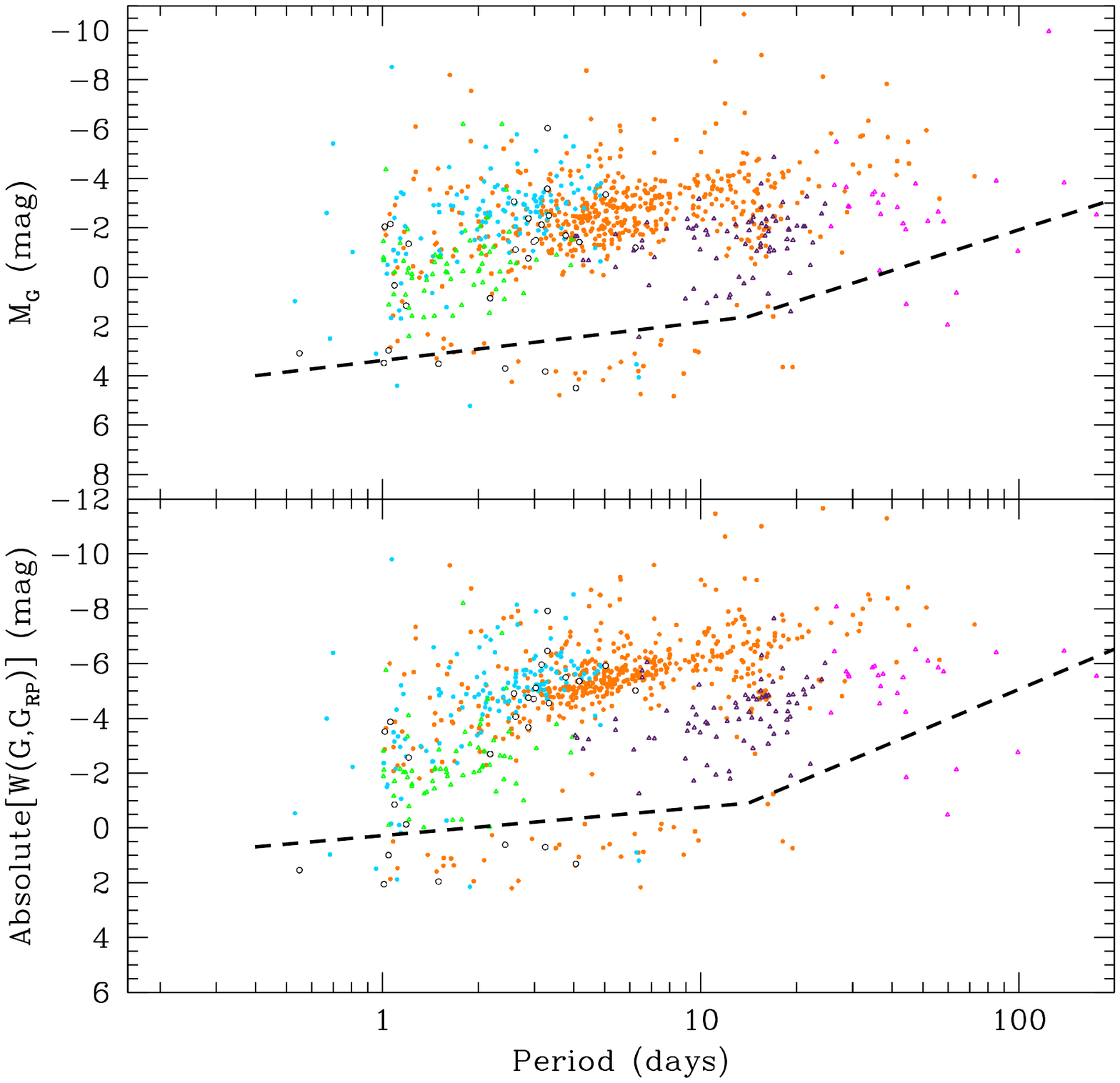}
   \caption{\textit{Upper panel:} $G$-band $PL$ distribution in absolute $G$ magnitude not corrected for extinction of All-Sky DCEPs and T2CEPs known in the literature. \textit{Lower panel:} Period - Absolute[$W(G, G_{\rm RP})$]  distribution of All-Sky DCEPs and T2CEPs known in the literature. 
  Symbols and colour-coding are the same as in Fig.~\ref{plg}.  The dashed lines in each panel  indicate the regions below which the sources may be misclassifications  (spurious sources) or  may be bona-fide Cepheids with a wrong parallax value due to  the still simplified astrometric processing applied for DR2, among which  in particular the lack of a proper treatment of binary/multiple sources (see  Sect.~\ref{allskycep} and  \citealt{ripepi2018-dr2}, for more details).
}
              \label{plnote}%
    \end{figure*}

\section{Validations and Results}\label{validation}

The SOS Cep\&RRL pipeline produced lists of confirmed Cepheids and RR Lyrae stars. 
However, as discussed in the previous sections not all  sources are correctly classified by the SOS processing. An additional  significant source of contamination, for RR Lyrae stars in particular, is due to contact binary
systems,  because  contact binaries and RRc stars populate the same regions of the $PA$ and $\phi_{21}$ vs $P$ diagrams in Figs.~\ref{pampRR} and ~\ref{phi21RR}  (specifically, the region around $\phi_{21} \sim $ 6.28 rad).  
The {\it AmplitudeRatios} module in the RR Lyrae branch of the SOS Cep\&RRL pipeline is designed to 
clear the RR Lyrae sample from contact binaries
mimicking RRc-like light curves,  based on the principle that amplitude ratios 
in different photometric bands are close to
unity for binaries. This module was not activated during the DR2 processing, nevertheless during validation we 
computed the $G_{\rm BP}$, $G_{\rm RP}$ peak-to-peak amplitude ratio 
and removed as binaries the sources with ${\rm Amp(}G_{\rm BP}{\rm )/Amp(}G_{\rm RP}$)=1$\pm$0.2.
Visual inspection of random samples of the removed sources confirmed  that they are indeed binaries.

 In order to gauge the contamination of the SOS Cep\&RRL results by other types of variables we also 
cross-matched them against catalogues of variables 
available in the literature. Main references were the OGLE catalogues of  
variable stars in the Magellanic Clouds and Galactic bulge, that we complemented with entries from  
the ASAS, LINEAR, and CTRS  
catalogues  of variables in   the Milky Way halo and with  further Galactic Cepheids  taken from SIMBAD and VSX. 

The final step of the validation process, 
was a cleaning procedure exploiting a training set of sources with well established  classification in the literature. We specifically considered the fiducial regions occupied by the training
sources in the planes of parameters used by the SOS code for classification, that is the  ${\rm Amp(}G{\rm )}$, $R_{21}$, $\phi_{21}$ and the 
$\phi_{31}$ vs period planes. 
We expect sources correctly classified to populate the same regions as  the training sources. To achieve a quantitative mathematical definition of these `fiducial'   
regions, we 
subdivided each parameter plane using a rectangular grid, assigning an occupation frequency value to every rectangular bin. The frequency of
occupation of each bin was defined as $f_i = n_i/MAX(n_i)_{\left(i=1, n_{bins}\right)}$, where $n_i$ is the number of sources that are present in the $i$-th
bin. A smoothing algorithm, given by the {\it matrixSmooth} routine \citep{RTeam}, was  applied to the resulting  frequency matrix  to avoid sharp edges for the
`fiducial' regions. The occupation frequency matrices were also calculated for the complete SOS output sample, using the same 
rectangular grids, but without the smoothing step. In order to select only sources in the defined `good' regions, we  multiplied the occupation 
matrix of the complete sample by that of the training set, performing this operation for all four parameter planes described above. Finally, we retained 
only the sources located in the bins with combined occupation frequency value larger than 0 in all parameter planes. 

The final result of the cleaning procedures described above is a validated sample comprising 140,784 RR Lyrae stars and 9,575 Cepheids, that form the final sample of SOS Cep\&RRL confirmed sources released in \textit{Gaia} DR2.

\subsection{Results for RR Lyrae stars}\label{results-rrl}

Fig.~\ref{pampRR-colour} shows the $PA$ diagrams in the $G$ (mid panel),    $G_{\rm BP}$ (upper panel) and $G_{\rm RP}$ (lower panel) bands  of the 
RR Lyrae stars confirmed by the SOS Cep\&RRL pipeline and  released in DR2.  The two panels of Fig.~\ref{RRL-F21} show the $G$-band $\phi_{21}$ vs period  (upper panel) and  $R_{21}$ vs period  (lower panel) diagrams of the subsample of 121,234 RR Lyrae stars 
whose light curves could be modelled with at least two harmonics.  Finally, the two panels of Fig.~\ref{RRL-R31} show the $G$-band $\phi_{31}$ vs period  (upper panel) and  $R_{31}$ vs period  (lower panel) diagrams for 67,681 RR Lyrae stars  
whose light curves were modelled with at least  three harmonics.
The sample of confirmed RR Lyrae stars released in {\it Gaia} DR2 includes variables in the MW (disc, bulge and halo), in the two Magellanic Clouds, in 5 dSph galaxies, 7 ultra-faint dwarfs and in 87 globular clusters (GCs). We present some tests performed on the RR Lyrae stars in GCs and one of the dSphs (Sculptor) in Sect.~\ref{gcs-dsphs}.
Examples of light curves for RR Lyrae stars in these various systems are presented in Figs.~\ref{all-sky-RRL}, 
~\ref{plotRRMC} and ~\ref{plotRRSistemi-Sculptor-no-color}. In all plots the light curves are folded according to the period and epochs of maximum light in the $G$, $G_{\rm BP}$ and  $G_{\rm RP}$ bands determined by the SOS Cep\&RRL  pipeline. 
 The  $G_{\rm BP}$ light curve of the LMC RR Lyrae star in the top-right panel of Fig.~\ref{plotRRMC} is brighter and has a lower amplitude than the $G$-band light curve, likely because the star is blended with a companion source that affects its $G_{\rm BP}$ photometry but not the other bands. This may happen in crowded fields such as the internal regions of the Magellanic Clouds,   where this specific RR Lyrae star is located, or the core of a globular cluster,  because of the longer extraction windows of the $G_{\rm BP}$, $G_{\rm RP}$ spectrophotometric data compared to the $G$-band (see 
\citealt{evans2018}).

 \begin{figure}
   \centering
   \includegraphics[trim=20 130 -10 80, width=9.5 cm,clip]{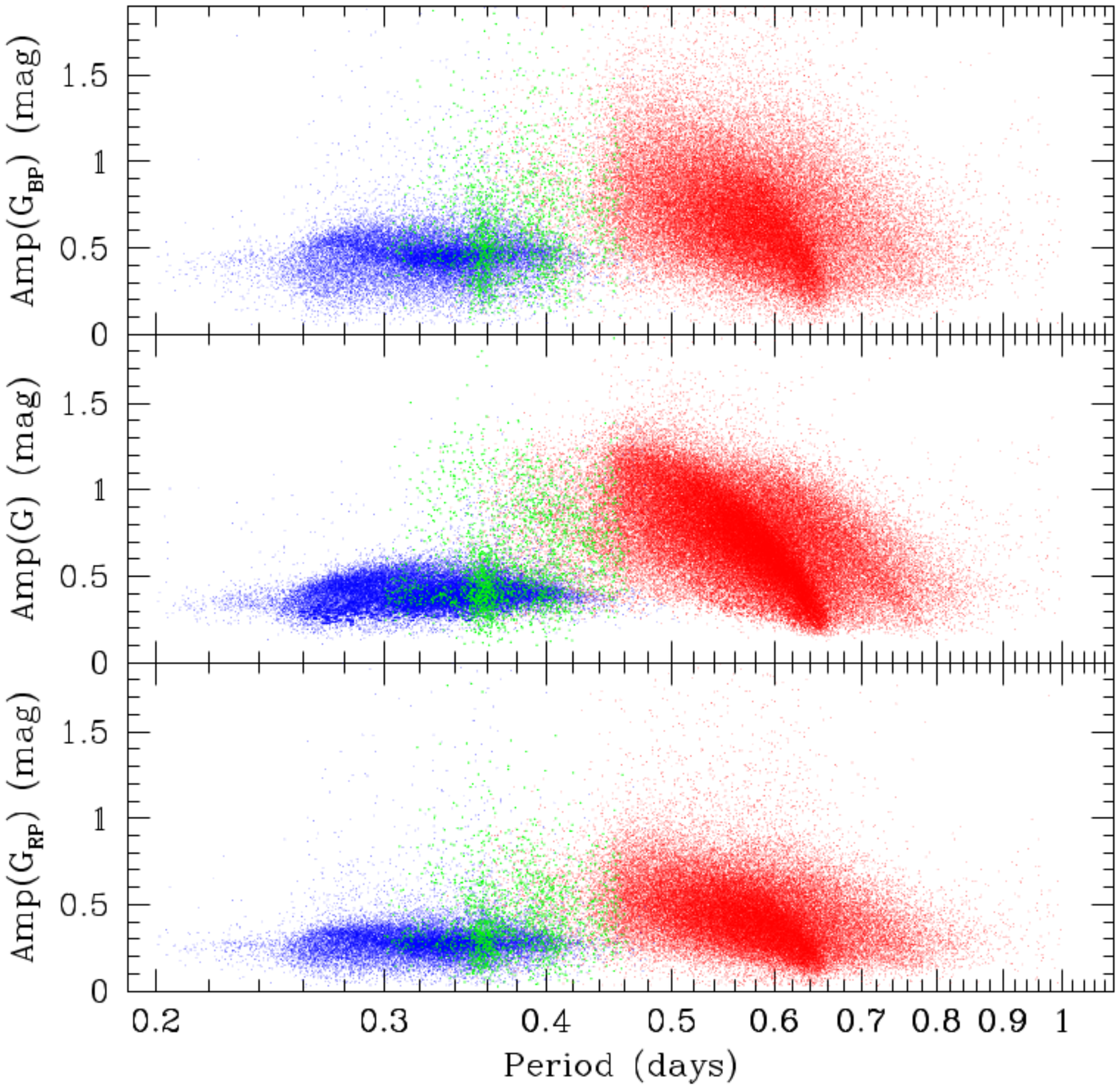}
 \caption{$PA$ diagrams in the $G$ (\textit{mid panel}),    $G_{\rm BP}$ (\textit{upper panel}) and $G_{\rm RP}$ (\textit{lower panel}) bands  of the RR Lyrae stars (140,784  in the $G$-band panel) confirmed by the SOS Cep\&RRL pipeline that are released in DR2.  Sources are colour-coded  as in Fig.~\ref{pampRR}.}
              \label{pampRR-colour}%
    \end{figure}
\begin{figure}
   \centering
   \includegraphics[trim=10 130 -10 80, width=9.5 cm,clip]{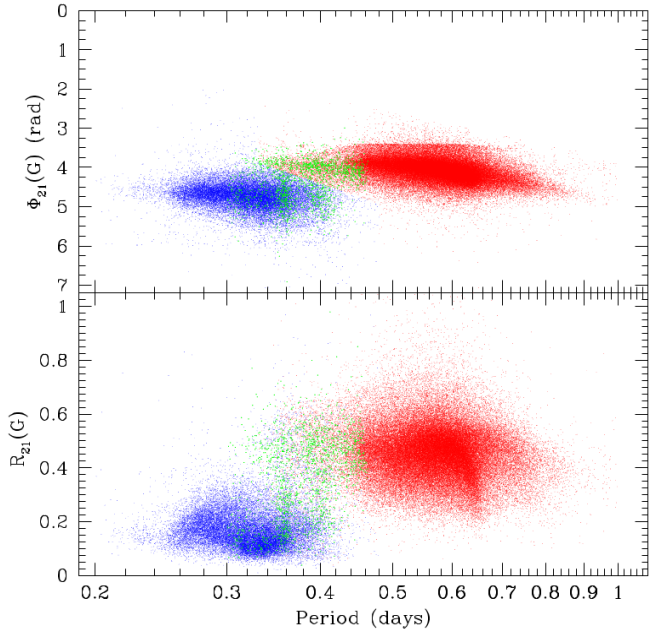}
   \caption{$G$-band $\phi_{21}$ vs period  (\textit{upper panel}) and  $R_{21}$ vs period  (\textit{lower panel}) diagrams for the RR Lyrae stars confirmed by the SOS Cep\&RRL pipeline.
Colour-coding is the same as in Fig.~\ref{pampRR}.   
   }
              \label{RRL-F21}%
    \end{figure}
\begin{figure}
   \centering
  \includegraphics[trim=10 130 -10 80, width=9.5 cm,clip]{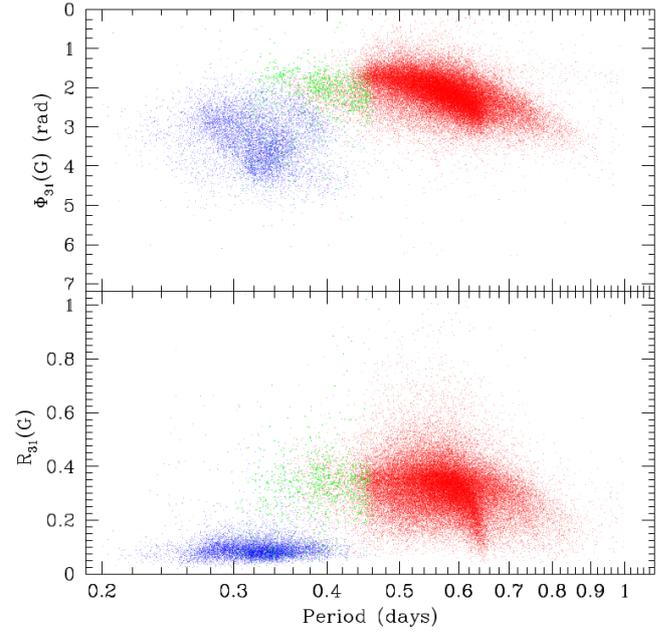}
   \caption{Same as in Fig.~\ref{RRL-F21}, for the $G$-band $\phi_{31}$ vs period  (\textit{upper panel}) and  $R_{31}$ vs period  (\textit{lower panel}) diagrams.}
              \label{RRL-R31}%
    \end{figure}
          \begin{figure*}
   \centering
    \includegraphics[width=18 cm, trim= 20 150 30 80, clip]{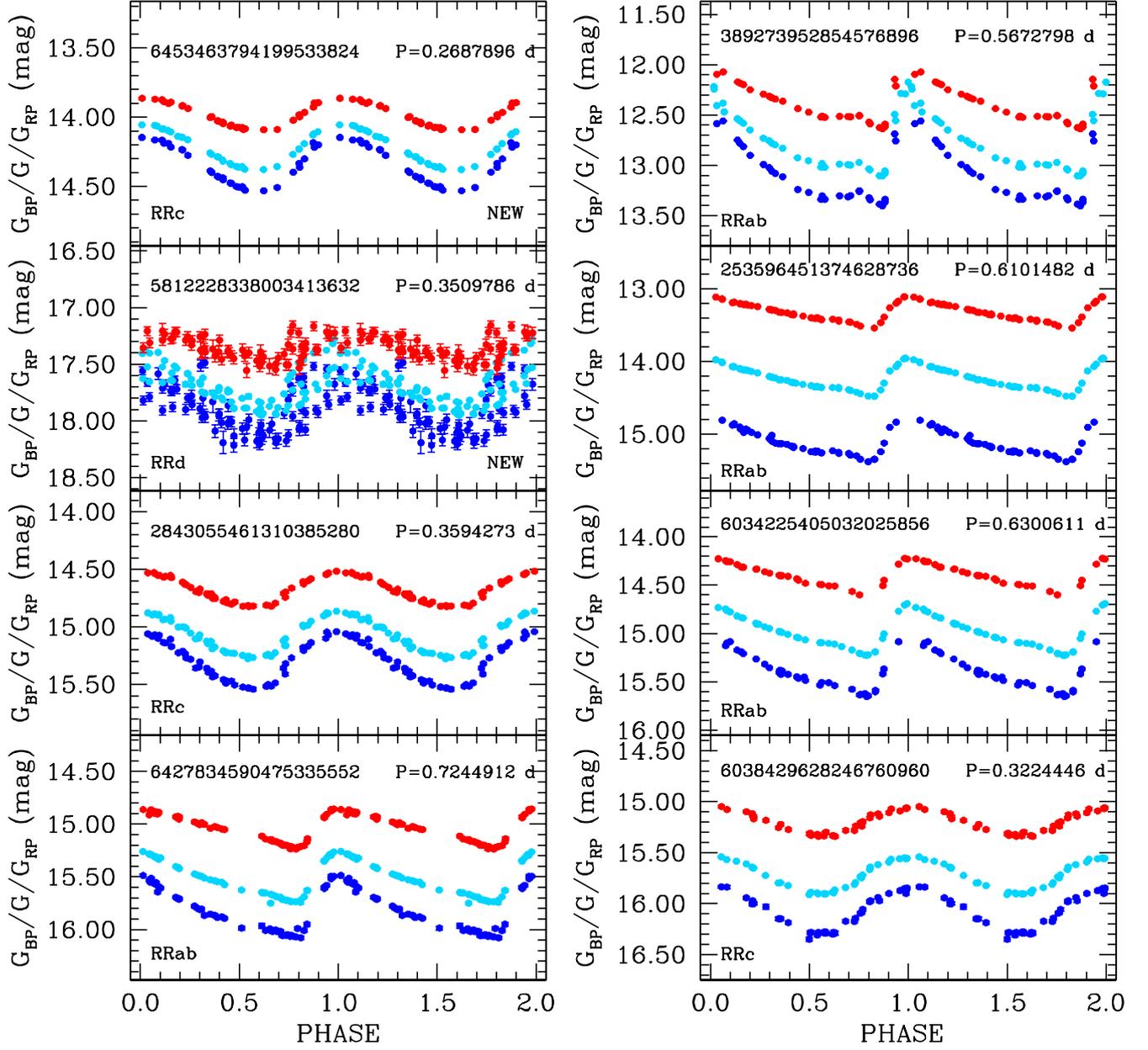}
   \caption{$G$ (cyan), $G_{\rm BP}$ (blue) and $G_{\rm RP}$ (red) light curves of All-Sky RR Lyrae stars  of different pulsation mode released in \textit{Gaia} DR2. The multi-band time series data are folded according to the period and epoch of maximum light derived by the SOS Cep\&RRL pipeline. Error bars are comparable to or smaller than symbol size. New discoveries by \textit{Gaia} are flagged.}
                \label{all-sky-RRL}%
     \end{figure*} 

          \begin{figure*}
   \centering
    \includegraphics[width=18 cm, trim= 20 150 30 80, clip]{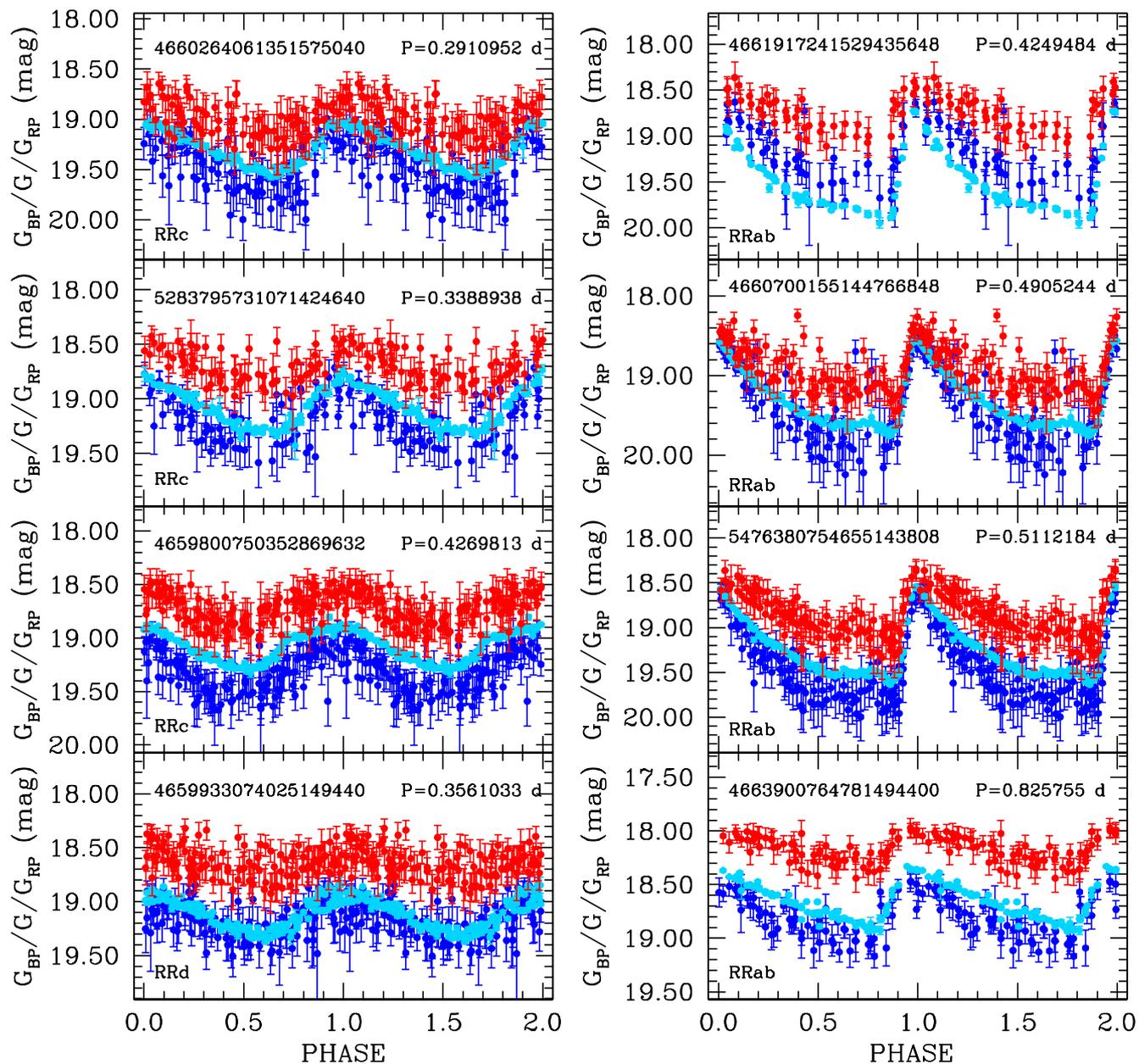}
   \caption{$G$ (cyan), $G_{\rm BP}$ (blue) and $G_{\rm RP}$ (red) light curves of RR Lyrae stars  of different pulsation mode in the LMC and SMC, released in \textit{Gaia} DR2. The multi-band time series data are folded according to the period and epoch of maximum light derived by the SOS Cep\&RRL pipeline. Notwithstanding the larger errors than in the $G$ band,  the colour light curves are well defined.  The $G_{\rm BP}$ light curve of the RR Lyrae star in the top-right panel is brighter and has a lower amplitude than the  $G$-band curve likely because the star is blended  with a companion which contaminates the $G_{\rm BP}$ photometry (see Sect.~\ref{results-rrl} for details).}
                 \label{plotRRMC}%
     \end{figure*} 

      \begin{figure*}
   \centering
  \includegraphics[width=18 cm, trim= 20 150 30 80, clip]{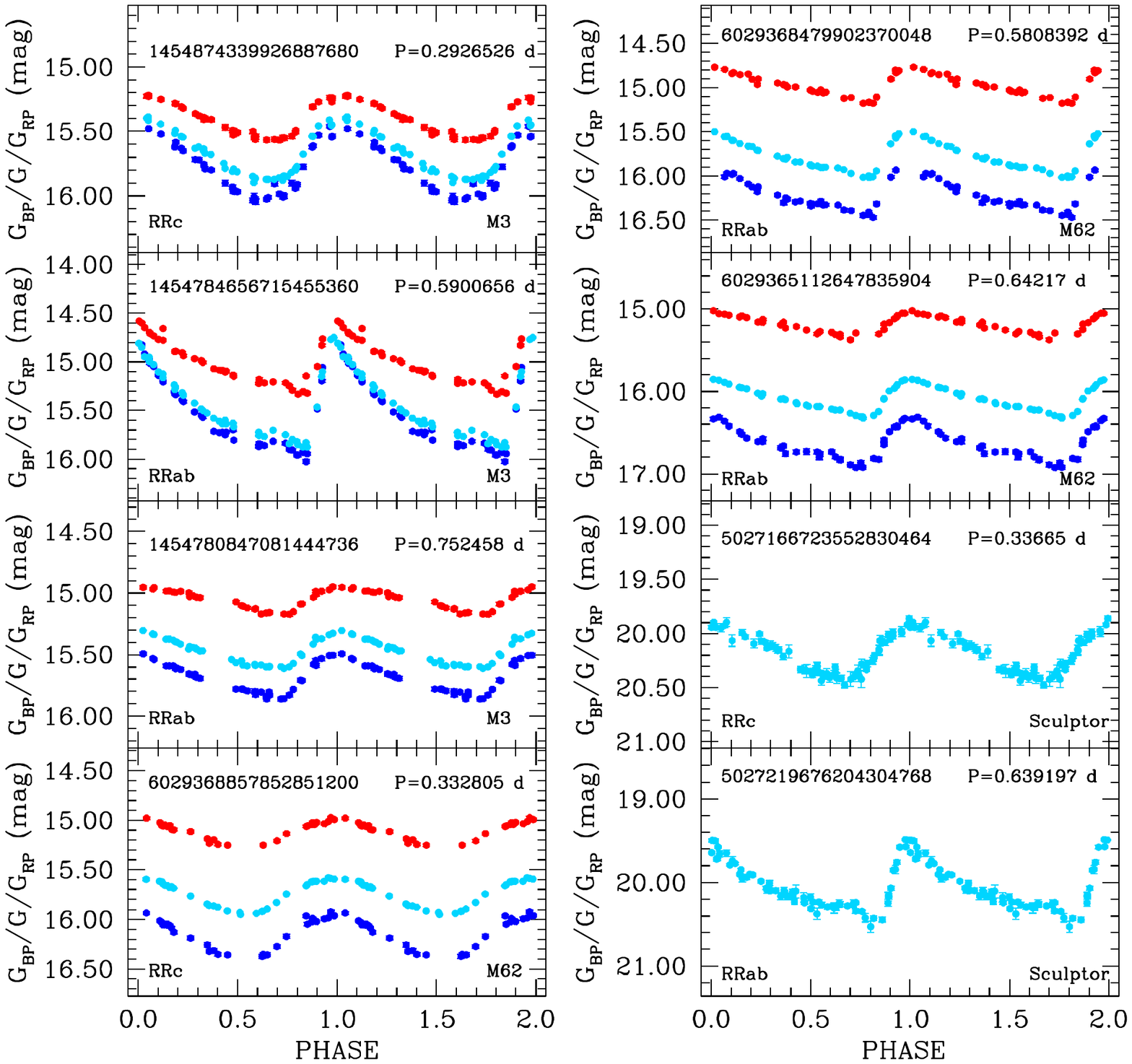}
   \caption{$G$ (cyan), $G_{\rm BP}$ (blue) and $G_{\rm RP}$ (red) light curves of RR Lyrae stars of different pulsation mode in the Galactic globular clusters M3 and M62 and in the 
   Sculptor dSph galaxy  ($G$ band only), released in DR2. \textit{Gaia} DR2 1454784656715455360 is located within the M3 half-light radius ($r_h$=2.31 arcmin, \citealt{harris1996}), star \# 6029368479902370048 is within twice the $r_h$ of M62 ($r_h$=0.92 arcmin, \citealt{harris1996}).  \textit{Gaia} DR2 5027166723552830464 in Sculptor is a new discovery by \textit{Gaia}. 
  Only the $G$-band light curves are shown for Sculptor, because of the low S/N of the $G_{\rm BP}$ and $G_{\rm RP}$ time-series  data at the faint magnitudes of the RR Lyrae stars in this  dSph galaxy.  
   The multi-band time series data are folded according to the period and epoch of maximum light derived by the SOS Cep\&RRL pipeline. Error bars are comparable to or smaller than symbol size. 
   }
              \label{plotRRSistemi-Sculptor-no-color}%
    \end{figure*} 
Figure~\ref{cmdRR} shows the CMDs defined by all the confirmed RR Lyrae stars for which $G_{\rm BP}$ and $G_{\rm RP}$ photometry is available (83,097 sources in total). The figure updates and improves fig.~3 in \citet{eyer2017a}.  A different colour-coding has been used for RRab (red), RRc (blue) and double-mode (green) pulsators.   As expected RRc and RRd pulsators are slightly bluer that RRab stars. The large concentrations of sources at $G \sim$ 19 and 19.7 mag are the LMC and SMC variables,  respectively, whereas the overdensities fainter than  $G\sim$ 20 mag are due to  RR Lyrae stars in the Draco and Sculptor dSphs. The arm extending towards redder colours is produced by reddened variables in the MW disc and bulge.  The figure confirms that sources with extreme, unphysical red colours were efficiently removed by cutting in {\it excess flux}.   CMDs in the Gaia passbands showing such a large number of All-Sky RR Lyrae stars with different pulsation type and intensity-averaged mean magnitudes and colours computed over the full pulsation cycle have never been published  before. 
Figure~\ref{cmd-gcs} shows instead 
the CMDs defined by the confirmed RR Lyrae stars in GCs (red) and dSphs (blue) for which $G_{\rm BP}$ and $G_{\rm RP}$ photometry is available (1167 sources in total). Each concentration of red points in this figure corresponds to a different GC.
Although not used for the DR2 processing, CMDs like those in Figs.~\ref{cmdRR} and ~\ref{cmd-gcs} will  be the first tool used by the SOS Cep\&RRL pipeline 
(see Fig.~\ref{decomposition2})  for the classification of RRLs during the  processing for the next  \textit{Gaia} release (DR3) which is currently foreseen for the first half of 2021.

Individual  photometric metallicities ([Fe/H])  were derived from the $\phi_{31}$ parameter  of the light curve Fourier decomposition for 64,957 of  the confirmed RR Lyrae stars. 
 The corresponding metallicity distributions are shown in Fig.~\ref{RRMet-histo} where the variables have been divided according to the three separate regions (LMC, SMC, All-Sky) defined in Sect.~\ref{s2}. 
The three distributions pick at mean values of [Fe/H]$\sim -1.15 \pm 0.6$, $\sim -1.3 \pm 0.7$ and $\sim -1.6 \pm 0.7$ dex, for  MW, LMC and SMC RR Lyrae stars, respectively.
\clearpage
   \begin{figure*}
   \centering
    \includegraphics[width=18.0 cm, trim= 10 140 30 70, clip]{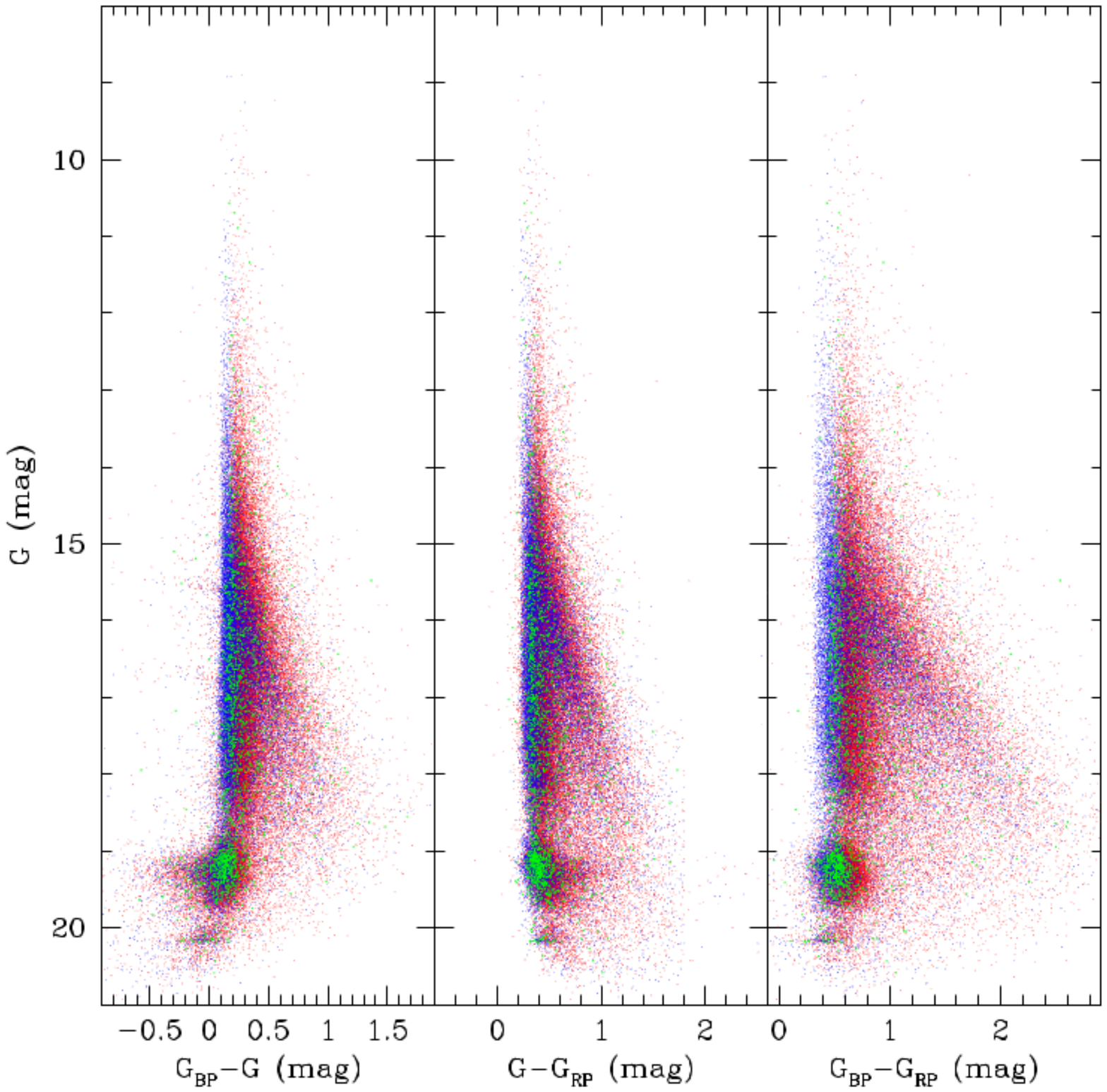}
   \caption{$G$, $G_{\rm BP}$-$G$; $G$, $G$-$G_{\rm RP}$ and $G$, $G_{\rm BP}-G_{\rm RP}$  CMDs of the RR Lyrae stars confirmed by the SOS  Cep\&RRL pipeline for which the colour information is available (83,097 sources out of 140,784).  Blue: RRc stars; green: RRd stars;  red: RRab stars. The large concentrations of sources at $G \sim$ 19 and 19.7 mag are the LMC and SMC variables,  respectively, whereas the overdensities below  $G\sim$ 20 mag are  due to  RR Lyrae stars in the Draco and Sculptor dSphs. The arm extending towards redder colours is produced by reddened variables in the MW disc and bulge.}
              \label{cmdRR}%
    \end{figure*}
   \begin{figure*}
   \centering
    \includegraphics[width=18.0 cm, trim= 10 140 30 70, clip]{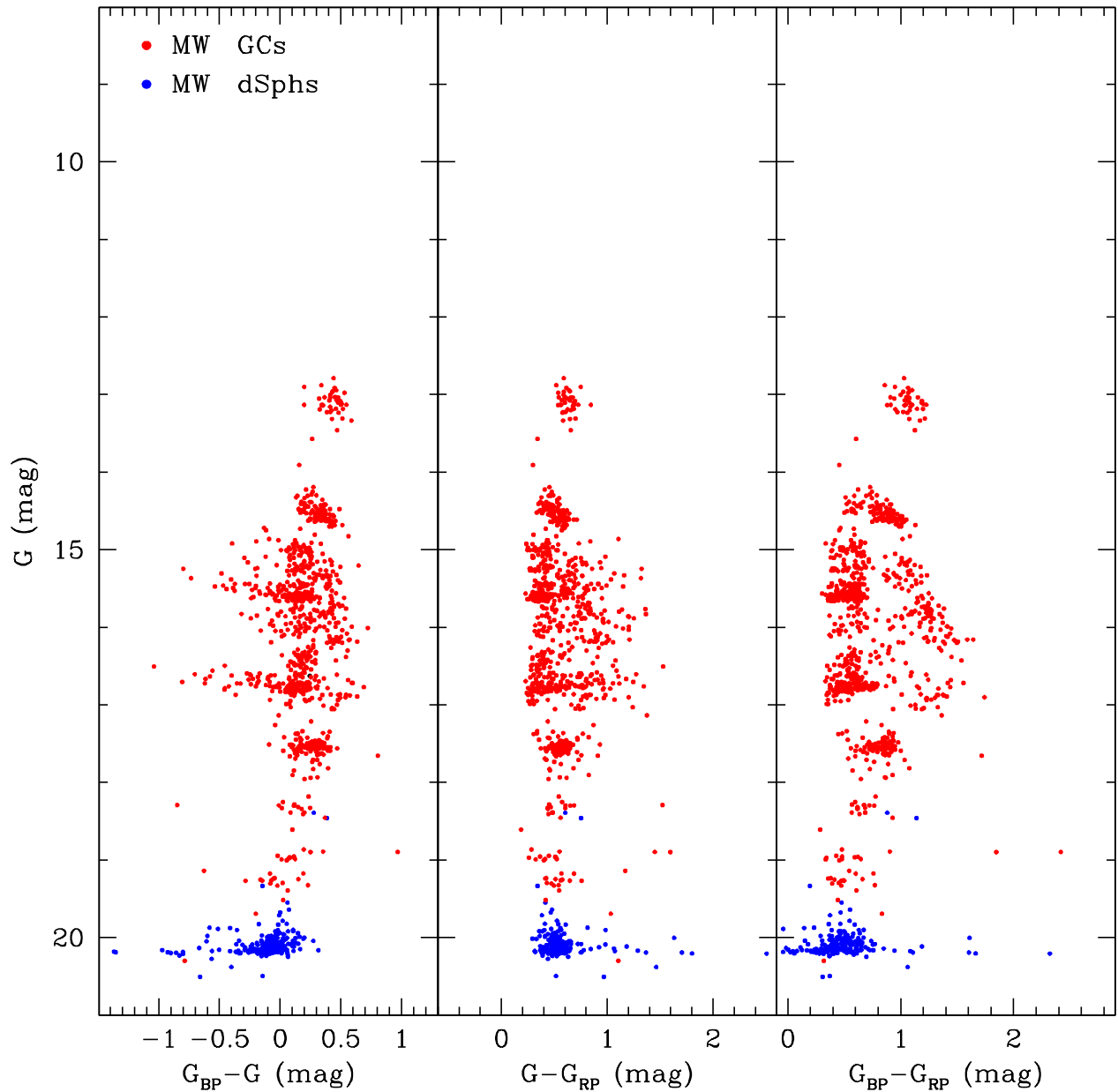}
   \caption{Same as in Fig.~\ref{cmdRR} but for  the RR Lyrae stars confirmed by the SOS  Cep\&RRL pipeline for which the colour information  is available  (1167 sources  out of 1986) that are in globular clusters (red points) and dwarf spheroidal galaxies (blue points). Each concentration of red points corresponds to a different GC, the blue overdensity below $G\sim$ 20 mag is due to RR Lyrae stars in the Draco and Sculptor dSphs.}
              \label{cmd-gcs}%
    \end{figure*}
     \begin{figure}
   \centering
  \includegraphics[width=8.5 cm, trim=0 180 0 190,clip]{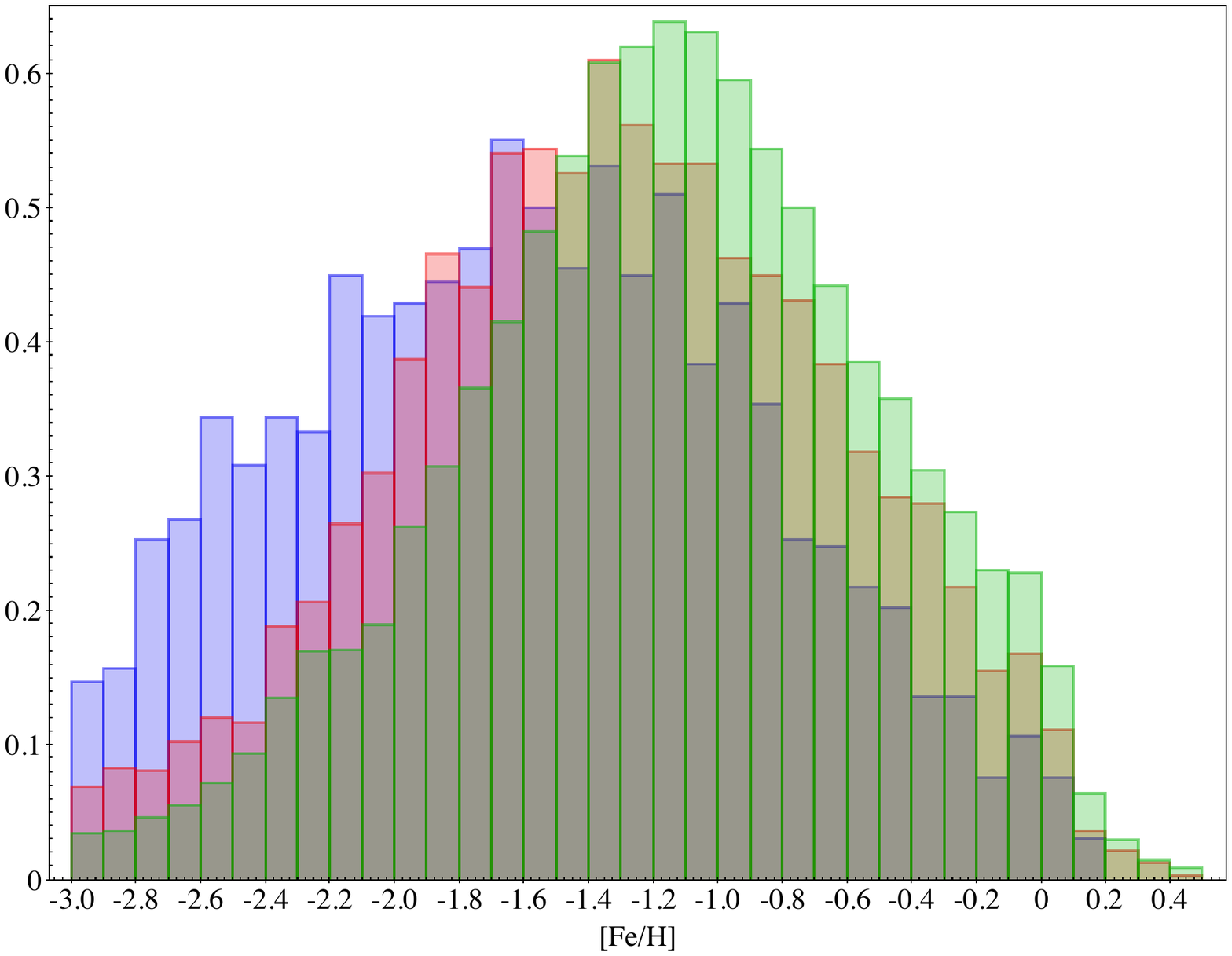}
   \caption{ Normalised metallicity distribution of  64,957  RR Lyrae stars,  in the sample of   140,784 confirmed ones, for which a photometric [Fe/H] value was inferred from the $\phi_{31}$ Fourier parameter of the $G$-band light curve. The sources were divided  according to the three separate regions (All-Sky, LMC, SMC) defined in Sect.~\ref{s2}.    
Green, pink, and blue histograms represent All-Sky, LMC and SMC variables, respectively.  The three distributions pick at mean values of: [Fe/H]$\sim -1.15 \pm 0.6$, $\sim -1.3 \pm 0.7$ and $\sim -1.6 \pm 0.7$ dex, for  MW, LMC and SMC RR Lyrae stars, respectively.
  }
              \label{RRMet-histo}%
    \end{figure}

\subsubsection{Results for double-mode RR Lyrae stars published in  DR2}

The SOS Cep\&RRL pipeline detected a secondary periodicity and classified as double-mode pulsators 2,378 among the 140,784 confirmed RR Lyrae stars. According to the comparison with the literature, this rather large number of RRd stars comprises 558  RRd pulsators already known in the literature (\textit{Gaia} light curve for  one of them  is shown in the bottom left panel of Fig.~\ref{plotRRMC}),  1,067 cross-matches with known RR Lyrae stars classified as single-mode pulsators in the literature, either of RRab or RRc types, and the  remaining 753 sources without a  counterpart in the literature. 
We specifically note that SOS Cep\&RRL classification as RRd pulsators  relies only on the detection of two periodicities in the proper period ratio in the time series data (see 2.3.1 of Paper~1), but presently does not take into account whether the source exhibits extra-scatter/noise  in the light curve folded according to the primary, dominant periodicity, as it is commonly observed among RRd stars. 
Hence, while those 2,378 sources all are found by the  SOS Cep\&RRL to have two periodicities in the proper ratio excited, a clear confirmation of their actual nature  as double-mode pulsators 
will require further analysis and additional data. A refinement of the SOS Cep\&RRL algorithm for the detection of double-mode pulsators is foreseen for \textit{Gaia} DR3.

     \subsubsection{RR Lyrae stars in globular clusters and dwarf spheroidal galaxies}\label{gcs-dsphs}
 
 Among the 140,784 SOS-confirmed RR Lyrae stars released in {\it Gaia} DR2,  1,986 reside in GCs and  dwarf spheroidal galaxies (classical and ultra-faint) within reach of \textit{Gaia} limiting magnitude ($G \lesssim$ 20.7 mag). Specifically, 1569 are distributed over 87 GCs and  417 over 12 dSphs,  with the largest numbers being in M3 (159), NGC 3201 (83), Sculptor (176) and  Draco (176).    Fig.~\ref{cmd-gcs} shows the CMDs defined by RR Lyrae stars in these systems.  Examples of light curves  
 in the Galactic GCs M3 and M62 and in the Sculptor dSph are shown in Fig.~\ref{plotRRSistemi-Sculptor-no-color}. 
 We specifically tested the  SOS Cep\&RRL pipeline on the RR Lyrae stars in GCs, in order to establish its performance in crowded  fields  and verify the reliability  of the derived pulsation characteristics 
 (e.g. periods, amplitudes) and stellar parameters (metallicity and $G$-band absorption)   on systems, like the GCs, for which metallicity and reddening are generally well known in the literature. 
 In  M3, a relatively low central concentration cluster (c=1.89, \citealt{harris1996})   \textit{Gaia} recovered 159 of the 222 known RR Lyrae stars. 
 They are plotted as large black filled circles in the map in Fig.~\ref{M3_mappa}, where different colours are used for  stars at different distance from the cluster centre.  \textit{Gaia} was capable to identify  RR Lyrae stars even within the  core radius of M3 ($r_{c}$ = 0.37 arcmin).  Centre and right panels of Fig.~\ref{M3_mappa} show the \textit{Gaia} CMD of M3 using the same colour-coding as in the left panel. The RR Lyrae stars brighter than the cluster horizontal branch (HB) level in the right panel of Fig.~\ref{M3_mappa} are located closer to the cluster centre where their $G_{\rm BP}$ photometry is more likely to be contaminated by companions  (a similar effect  was seen in the light curve shown in the upper-right panel of Fig.~\ref{plotRRMC} and discussed in Sect.~\ref{results-rrl}). 
The effect of contamination by companions in the M3 central regions is better seen in the two  panels of Fig.~\ref{M3_HB}  which show an enlargement of the HB region in the \textit{Gaia} CMD of the cluster.  
Although the \textit{Gaia} photometry might suffer by blending in crowded regions like the centre of a globular cluster, the periods measured by the SOS Cep\&RRL pipeline and  the literature periods for the 159 RR Lyrae stars observed in M3 by \textit{Gaia} are in excellent agreement, as shown by Fig.~\ref{M3_PeriodPeriod}.  
The three deviating objects in the figure are variable stars known to be affected by Blazhko effect (\citealt{blazhko1907}), a modulation of shape and amplitude of the light variation that may occur on time span ranging from a few
tens to hundreds of day. Finally,  metal abundance were measured from the $\phi_{31}$ Fourier parameter for 111 RR Lyrae stars in the cluster. The corresponding metallicity distribution is shown in Fig.~\ref{M3_isto_met}.
   

        \begin{figure*}
   \centering
   \includegraphics[width=18.5 cm, trim= 0 300 0 0, clip]{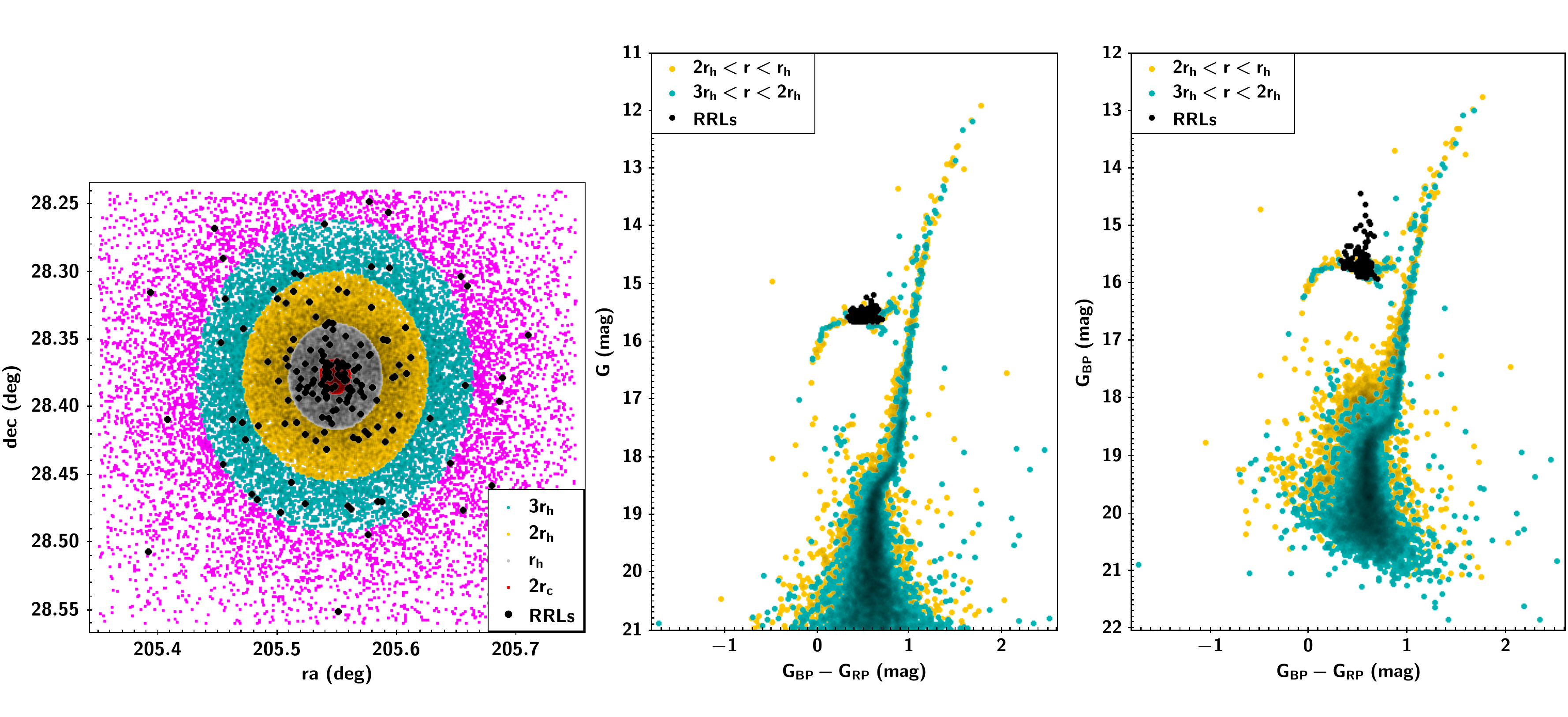}
   \caption{{\it Left panel:} Map of Gaia DR2 stars in a 0.31 $\times$ 0.30 degree$^2$ area centred on the Galactic globular cluster M3 (NGC 5272). Different colours highlight stars within twice the cluster $r_c$ (red filled circles), 
    one time (grey filled circles), twice (yellow filled cicles), three times (cyan filled circles) the cluster half-light radius ($r_h$=2.31 arcmin) and beyond 3 $\times r_h$ (magenta filled circles).
   The RR Lyrae stars observed by \textit{Gaia} in the cluster (159 over the 222 RR Lyrae stars known in M3) are marked as large black filled circles. The M3 centre coordinates and $r_c$, $r_h$ values are from \citet{harris1996}.  {\it Centre panel:} $G$ vs  $G_{\rm BP} -  G_{\rm RP}$ CMD for stars within two times the M3 $r_h$, colour-coding is as in the left panel. 
 {\it Right panel:}   Same as in the centre panel but for the cluster $G_{\rm BP}$ vs  $G_{\rm BP} -  G_{\rm RP}$ CMD.}
   \label{M3_mappa}%
    \end{figure*} 
    
  \begin{figure*}
   \centering
   \includegraphics[width=18.5 cm, trim= 0 380 0 -20, clip]{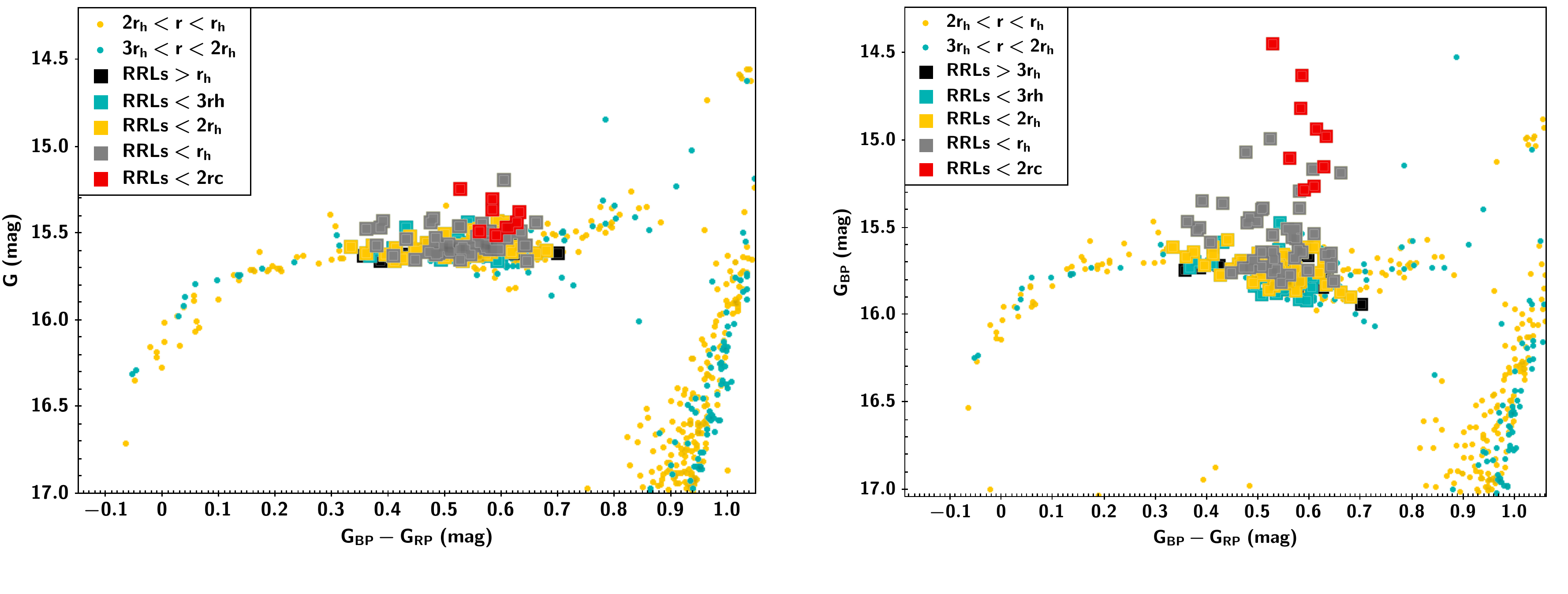}
   \caption{{\it Left panel:}  Zoom of the HB region in the $G$ vs  $G_{\rm BP} -  G_{\rm RP}$ CMD of M3. RR Lyrae stars are plotted as  filled squares and different colours according to their position with respect to the cluster centre adopting the same colour-coding as in the left panel of Fig.\ref{M3_mappa} and using black filled squares for the variables beyond 3 $\times r_h$.  
   {\it Left panel:}  Same as in the left panel but for the $G_{\rm BP}$ vs  $G_{\rm BP} -  G_{\rm RP}$ CMD.}
                   \label{M3_HB}%
    \end{figure*}

       \begin{figure}
   \centering
      \includegraphics[width=9.0 cm, trim= 0 0 0 0, clip]{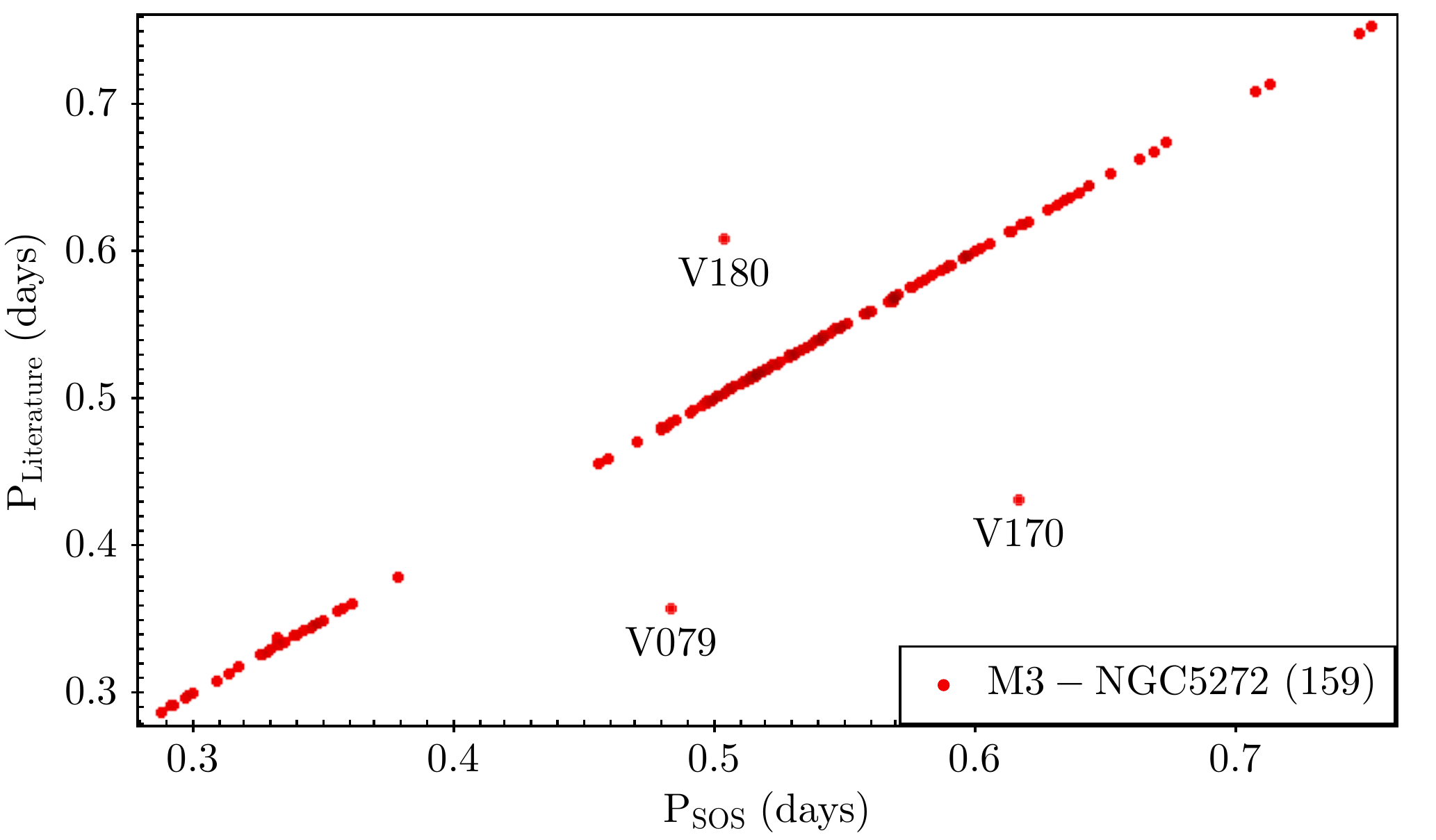}
   \caption{Comparison between periods measured by the SOS Cep\&RRL pipeline and literature periods for 159 RR Lyrae stars observed in M3 by \textit{Gaia}. 
   The three deviating objects are variable stars affected by the Blazhko effect \citep{blazhko1907}.}
               \label{M3_PeriodPeriod}%
    \end{figure} 
    
   \begin{figure}
   \centering
    \includegraphics[width=9.0 cm, trim= 0 0 0 0, clip]{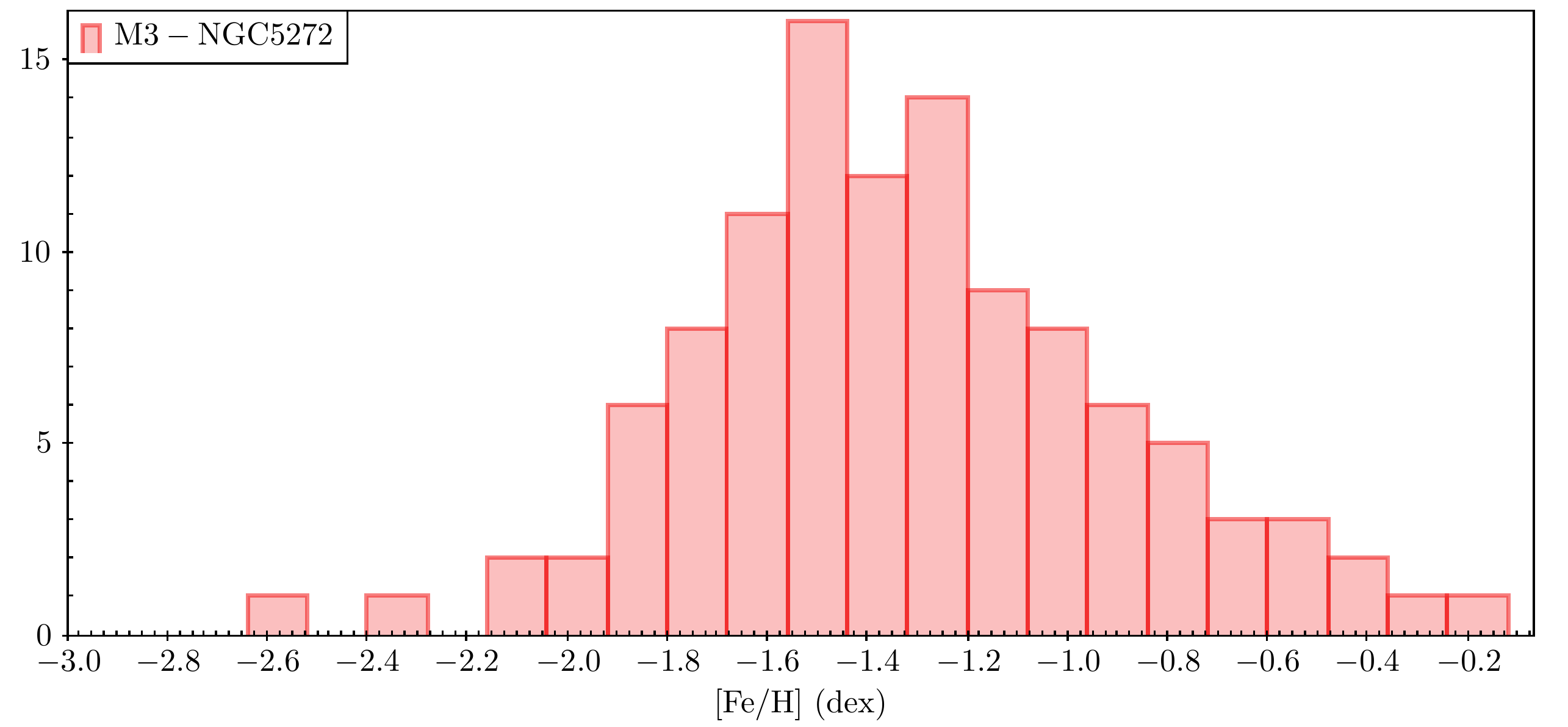}
   \caption{Metallicity distribution of  the RR Lyrae stars observed by \textit{Gaia} in M3, based on the individual measurements from the SOS Cep\&RRL pipeline.}
              \label{M3_isto_met}%
    \end{figure} 

The same 
comparison was also done for M62, a suspected core-collapsed globular cluster with c=1.71, $r_{c}$ = 0.22 arcmin and half-light radius $r_h$=0.92 arcmin (\citealt{harris1996}). 
 Due to higher concentration only 91 out of  214 ($\sim$ 42.5\%) RR Lyrae stars known in M62 were recovered, to compare with 71.6\% in M3 that hosts the same RR 
Lyrae population.  Their metallicity distribution is shown in  Fig.~\ref{M62_isto_met}.

%
    \begin{figure}
   \centering
   \includegraphics[width=9.0 cm, trim= 0 0 0 0, clip]{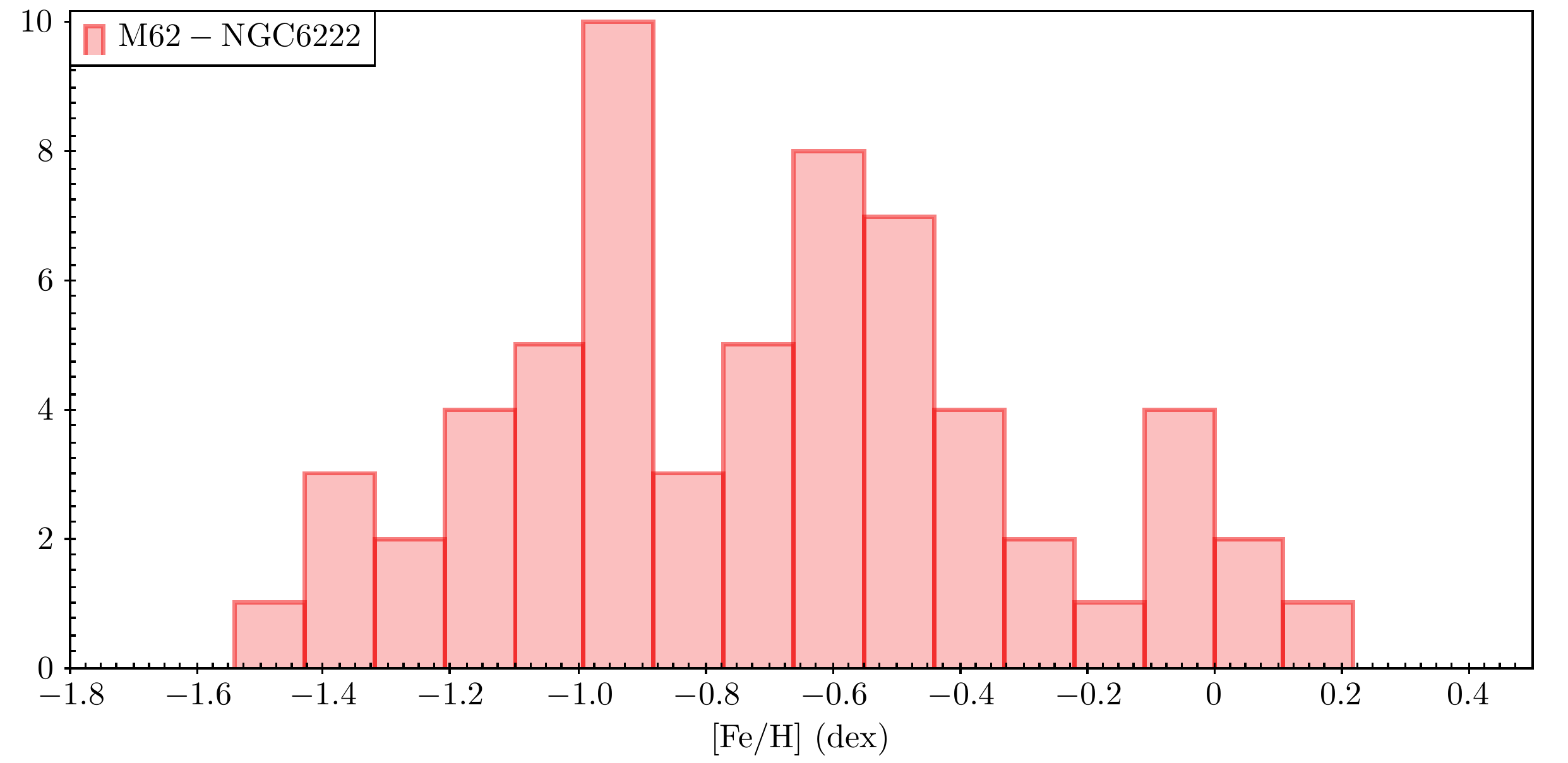}
   \caption{Same as in Fig.~\ref{M3_isto_met} but 
   for  RR Lyrae stars in M62.
 }
              \label{M62_isto_met}%
    \end{figure} 

More than 200 RR Lyrae stars 
are known in the Sculptor dSph from the study of \citet{kaluzny1995}.  We recovered 176 of them by cross-matching the SOS confirmed RR Lyrae stars against \citet{kaluzny1995}'s catalogue, 
that remains so far the only study where identification and coordinates for the Sculptor variable stars have been published. 
Fig.~\ref{sculptor_bailey} shows the $G$-band $PA$ diagram defined by the Sculptor RR Lyrae stars, where filled symbols are known variables that cross-match with  \citet{kaluzny1995}, while open symbols are new RR Lyrae stars observed by \textit{Gaia}. Fig.~\ref{Sculptor_isto_met} shows that a good agreement is found between individual metallicities measured by the SOS Cep\&RRL pipeline for RR Lyrae stars in Sculptor (blue histogram) and the corresponding spectroscopic metallicities from \citet{clementini2005} (red histogram). 
    
     \begin{figure}
   \centering
    \includegraphics[width=9.0 cm, trim= 0 0 0 0, clip]{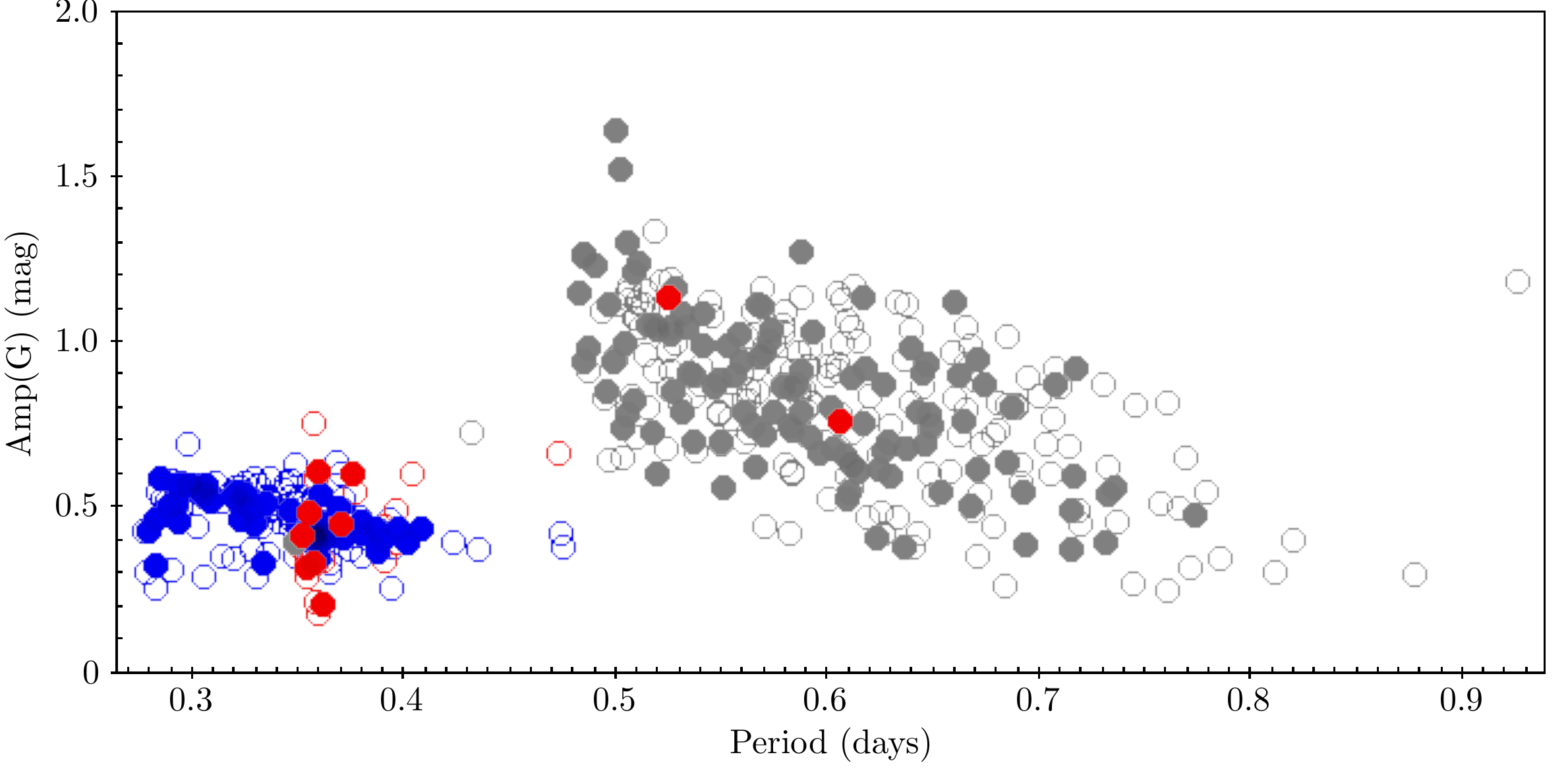}
   \caption{$PA$ diagram in the $G$ band of known (filled circles; \citealt{kaluzny1995}) and new RR Lyrae stars observed by \textit{Gaia} in the Sculptor dSph. Blue, red and grey symbols are RRc, RRd and RRab stars, respectively.}
              \label{sculptor_bailey}%
    \end{figure} 

     \begin{figure}
    \centering
    \includegraphics[width=8.8 cm, trim= 0 0 0 0, clip]{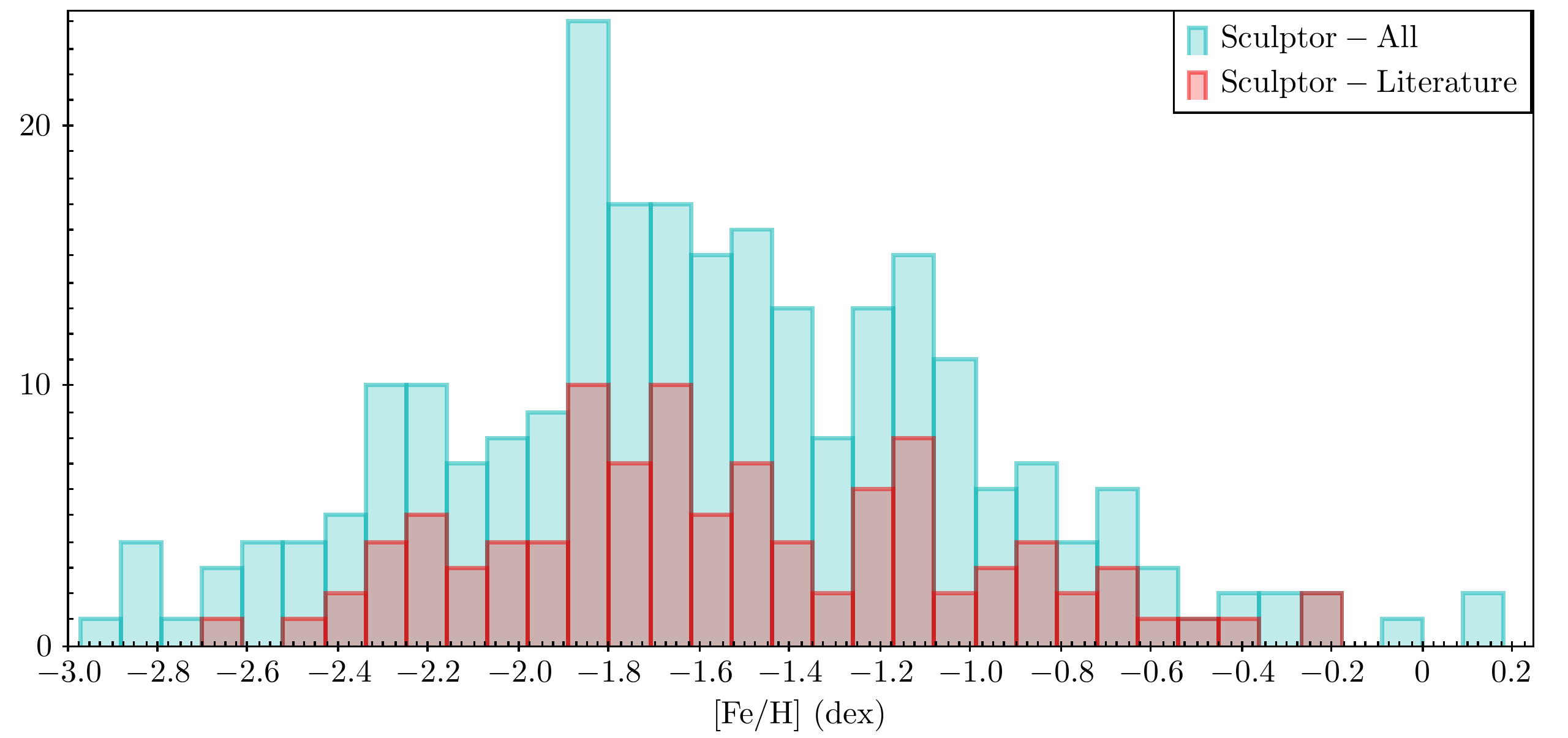}
   \caption{Metallicity distribution of RR Lyrae stars in Sculptor. The blue histogram is obtained from metallicities measured by the SOS Cep\&RRL pipeline.
   The red histogram shows the metallicity distribution of the RR Lyrae stars studied spectroscopically  by \citet{clementini2005}.}
              \label{Sculptor_isto_met}%
    \end{figure} 

For a more direct comparison in Table~\ref{metallicity-table},   we report for a few GCs and dSphs the mean metallicity and $G$-band absorption (with the related standard deviations) as derived averaging individual [Fe/H] and $A(G)$ values inferred by the SOS Cep\&RRL pipeline  
 from the RR Lyrae stars 
 (columns 3, 4 and 6, 7, respectively) and the corresponding literature reference values (\citealt{carretta2009} for the GCs metal abundances,  \citealt{clementini2005} and \citealt{kirby2011},  for the RR Lyrae stars in Sculptor,  Draco, UMa~I and ~II, respectively). 
  In the table, N$_1$ and N$_2$ are the number of variable stars on which the mean quantities were computed.   Overall agreement is quite satisfactory.   On the other hand, the rather high value of $A(G)$ derived for Sculptor is likely due to the poor S/N of the $G_{\rm BP}$, $G_{\rm RP}$ photometry at the faint magnitudes of the RR Lyrae stars in this dSph.

\begin{table*}
\caption{Metallicity and absorption in the $G$ band for  a sample of GCs and dSphs, obtained  averaging individual values derived for the RR Lyrae stars by the SOS Cep\&RRL pipeline.}\label{fourier-metal}
\label{metallicity-table}
\begin{tabular}{l c c c r c c c r r}
	\hline
    \hline
	Name & ${\rm [Fe/H]_{C09}}$& ${\rm [Fe/H]}_{Gaia}$ & stdev([Fe/H]$_{Gaia}$) & N$_1\tablefootmark{\rm (a)}$ & $A(G)$& stdev[$A(G)$] & $A(V)$ & N$_2\tablefootmark{\rm (a)}$ & Ref.\\
                   &  dex                           &   dex                           &     dex                            &             & mag     & mag                & mag     &             &        \\          
	\hline
  \noalign{\smallskip}
    NGC 1261 & $-$1.27 & $-$1.33 & $\pm$ 0.55 & 14& 1.07 & $\pm$ 1.33 & 0.03 & 10 & (1) \\ 
    NGC 1851 & $-$1.18 & $-$1.17 & $\pm$ 0.48 &  9& 0.46 & $\pm$ 0.34 & 0.06 & 8 & (2) \\
    NGC 288 & $-$1.32 & $-$1.33 & $\pm$ 0.22 &  1& 0.17 &  & 0.09 & 1& (3) \\
    NGC 3201 & $-$1.59 & $-$1.30 & $\pm$ 0.04 & 82& 0.67 & $\pm$ 0.21 & 0.74 & 73& (4) \\
    NGC 5024 & $-$2.10 & $-$1.62 & $\pm$ 0.48 &  8& 0.73 & $\pm$ 0.75 & 0.06 & 20& (5) \\ 
    NGC 5139- $\omega$ Cen & $-$1.53 & $-$1.17 & $\pm$ 0.79 &  8 & 0.32 & $\pm$ 0.14 & 0.37 & 9 & (6) \\    
    NGC 5272 (M3) &  $-$1.50 & $-$1.32 & $\pm$ 0.43 & 111 & 0.49 & $\pm$ 0.87 & 0.03 & 92 & (7) \\ 
    NGC 6266 (M62) & $-$1.18 & $-$0.71 & $\pm$ 0.39 &  81 & 2.18 & $\pm$ 0.78 & 1.46 & 52 & (8) \\ 
    IC 4499 & $-$1.53 & $-$1.47 & $\pm$ 0.53 &  56 & 0.81 & $\pm$ 0.35 & 0.71 &  54& (9) \\ 
    NGC 7078 (M15) & $-$2.37 & $-$1.26 & $\pm$ 0.87 &  5 & 0.22 & & 0.31 & 1 & (10) \\ 
    \hline
    \hline
     \noalign{\smallskip}  
    Sculptor & $-$1.68 & $-$1.52 & $\pm$ 0.53 &  102 & 0.68 & $\pm$ 0.43 &0.06 & 105& (11)  \\
    Draco&  $-$1.93& $-$1.70 & $\pm$ 0.74 &  30 & 0.54 & $\pm$ 0.32 & 0.09 & 13 & (12) \\ 
    Ursa Major I&  $-$2.18& $-$1.92 & $\pm$ 0.18 & 2 & 0.53 & $\pm$ 0.05 & 0.02 & 2& (13) \\ 
    Ursa Major II& $-$2.47& $-$2.28 & $\pm$ 0.23 & 1 & 0.33 &  & 0.30 & 1& (14) \\ 
    \hline
\end{tabular}
\tablefoot{$^{\rm (a)}$N$_1$ and N$_2$ are the number of RR Lyrae stars on which the mean [Fe/H] and $A(G)$ were computed, respectively.}\\
 \textbf{References:} (1) \citet{salinas2016}; (2) \citet{walker1998}; (3) \citet{kaluzny1997}, \citet{arellanoferro2013}; (4) \citet{layden2003}, \citet{arellanoferro2014}; (5) \citet{cuffey1996}, \citet{goranskij1976}, \citet{arellanoferro2011}; (6) \citet{braga2016}; (7) \citet{benko2006}; (8) \citet{contreras2010}; (9) \citet{walker1996}; (10) \citet{corwin2008}; (11) \citet{kaluzny1995}, \citet{clementini2005}; (12) \citet{kinemuchi2008}; (13) \citet{garofalo2013};  (14) \citet{dallora2012}.
\end{table*}

   \subsubsection{RR Lyrae stars: validation and comparison with the literature}\label{val-rrl}

In order to establish the completeness and purity of the RR Lyrae stars confirmed by the SOS Cep\&RRL pipeline and to 
estimate the number of new discoveries by \textit{Gaia} we performed a deep and careful comparison with the literature.  
As a first step the catalogue of 140,784 confirmed sources was cross-matched against all major catalogues of known 
RR Lyrae stars that are available in the literature. We primarily used the OGLE catalogues for RR Lyrae stars 
\citep[version IV of the survey,][]{soszynski2014,soszynski2016}, but we also used RR Lyrae stars 
by CTRS \citep[][]{drake2013a, drake2013b, drake2014,torrealba2015,drake2017}, 
ASAS \citep[][]{pojmanski1997,richards2012}, ASAS-SN \citep[][]{jayasinghe2018}, 
ATLAS \citep[][]{tonry2018}, IOMC \citep[][]{alfonso2012}, 
LINEAR \citep[][]{palaversa2013}, NSVS \citep[][]{kinemuchi2006}, 
Pann-Stars \citep[PS1][]{sesar2017}, as well as from the works based
 on KEPLER/K2 \citep[][]{debosscher2011,nemec2011,molnar2015a,molnar2015b,molnar2016} 
 and on the Simbad database \citep[][]{wenger2000}. 
 These cross-matches returned a list of 88,578 known RR Lyrae stars among our sample of 140,784.  
The SOS Cep\&RRL confirmed RR Lyrae stars were also cross-matched against catalogues of candidates RR Lyrae 
stars discovered by the VVV survey  (\citealt{gran2016,minniti2017}; D. Minniti, private communication)  in the  
MW disc and bulge. This returned  319 VVV cross-identified sources in the MW disk and  222 in 
the MW bulge. We thus confirm these VVV candidates. 
For known RR Lyrae stars in GCs,  the main reference was the catalogue of  \citet{clement2001} that was 
updated to the latest literature as described in Garofalo et al. (in preparation). 
For variables in dSphs we used the following references \citet[][]{kaluzny1995,clementini2005,kinemuchi2008,dallora2012,garofalo2013}. 
These latter cross-matches returned a  list of 1,986 further known RR Lyrae stars. 
At the end of the above cross-match procedure out of 140,784 RR Lyrae stars confirmed by the SOS Cep\&RRL pipeline 90,564 turned out to be already known 
in the literature and 50,220 are new discoveries by  \textit{Gaia}. 

A  detailed  confusion matrix was  derived only for the RR Lyrae stars in the Magellanic Clouds (MC) and the MW bulge because only for these variables we have available as reference  a catalogue as complete and pure as OGLE's in the LMC, SMC and MW bulge. The confusion matrix is shown in Fig.~\ref{confusionMatrixRR}.
 \begin{figure}
   \centering
   \includegraphics[width=9 cm, clip]{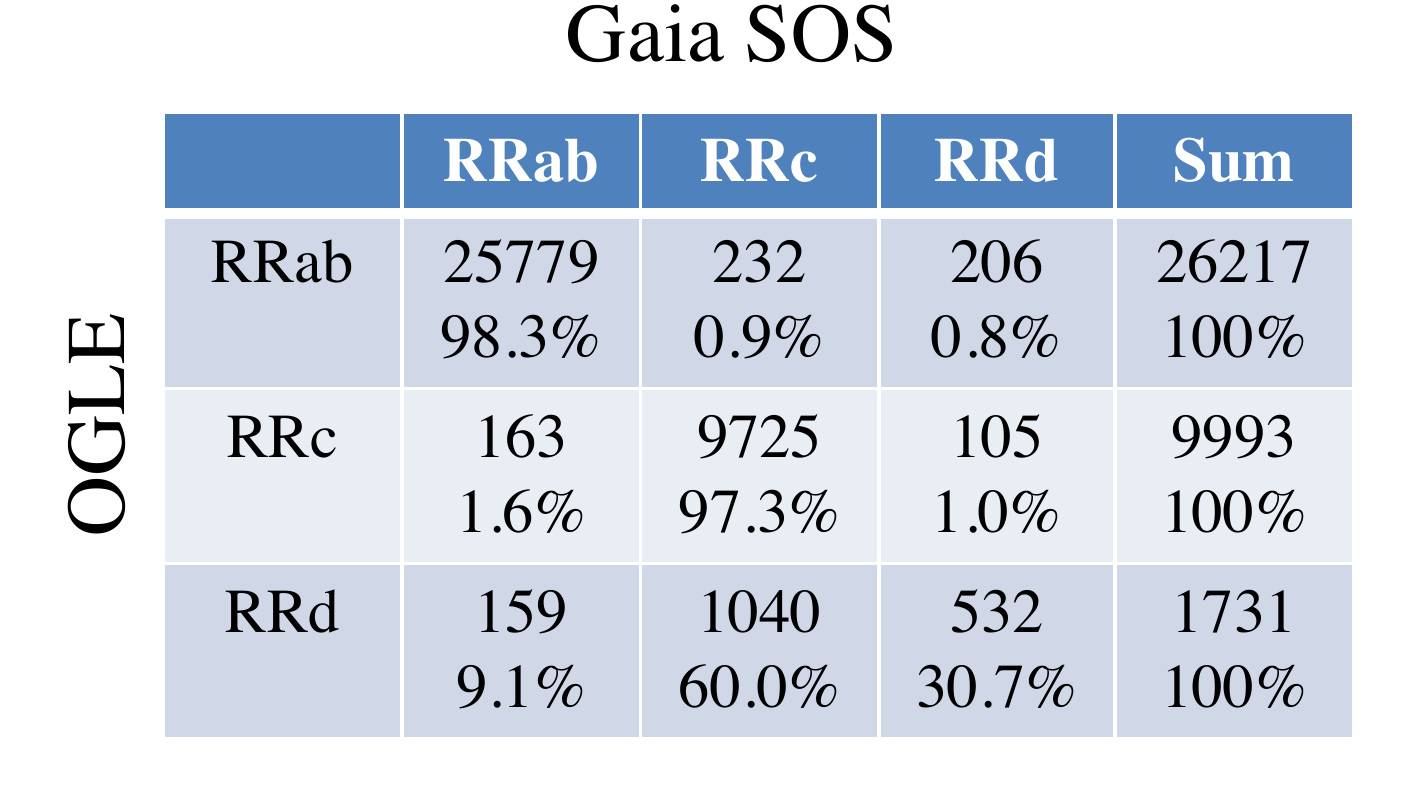}
   \caption{Confusion matrix for  the RR Lyrae stars. As control sample we used 
all the variable stars classified as RR Lyrae by the OGLE survey
in the LMC, SMC and Bulge that have a cross-match within a radius of 3 
arcsec with the SOS confirmed RR Lyrae stars published in {\it Gaia} DR2, for a total 
of 37,941 objects. Rows refer to literature results and columns to results of the SOS 
Cep\&RRL pipeline. The corresponding success percentage is shown along the diagonal cells.}
\label{confusionMatrixRR}%
\end{figure}
 For the MC RR Lyrae stars we have $<$ 0.1\% contamination. 
We achieved  such a high  purity  because binaries were removed during the validation process either using the  amplitude ratios or via cross-matching  against binaries and ellipsoidals from the OGLE catalogues. 

A more general assessment of the contamination affecting the SOS confirmed RR Lyrae sample 
was achieved 
through a procedure described in \citet{holl2018} that consisted in the visual inspection of \textit{random} and \textit{sky-uniform} samples of SOS confirmed RR Lyrae stars each composed by 500 sources without overlap between 
the two groups. Based on this  procedure we estimated a contamination of about 9\%, of which 4\% is due to faint  stars (20.0 $< G <$ 20.7 mag; see \citealt{holl2018}, for detail). 
We note that out of  140,784 RR Lyrae stars confirmed by the SOS pipeline, 8,306
are fainter than  $G \sim$20.0 mag. Colours for these sources are less reliable, hence misclassifications and contamination by other types of variable 
 sources are definitely possible.
On the other hand among these 8,306 faint sources,  16  are bona fide RR Lyrae stars in Sculptor
and 637 are OGLE confirmed RR Lyrae in the two Magellanic Clouds, thus reducing the number of possible faint misclassifications
 to 7,653.

The number of epochs available in the $G$-band light curves of the 140,784 RR  Lyrae stars confirmed by the SOS Cep\&RRL pipeline  is shown in Fig.~\ref{nepochsRRNew}. The distribution of the sources on sky is shown in Figs.~\ref{skyMapKnown+Gaia_RR} and ~\ref{skyMapKnown+Gaia_RR_1}, whereas Figs.~\ref{RRMetallicita}  and ~\ref{RRabsorption}  present their distribution on sky according to the metallicity and $G$-band absorption derived by the SOS pipeline. 

  \begin{figure*}
   \centering
  \includegraphics[width=17.0 cm, trim=0 0 0 0, clip]{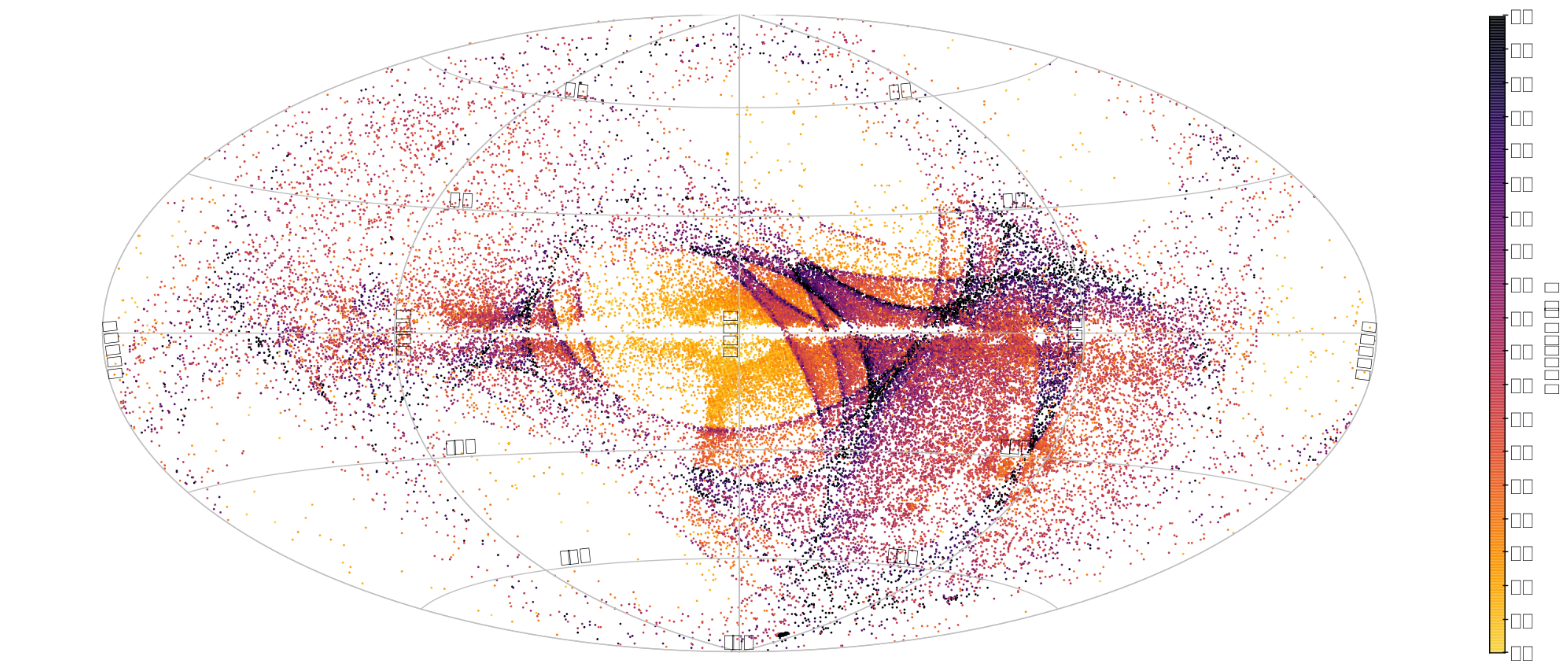}
    \caption{Map in galactic coordinates showing \textit{Gaia} newly discovered  RR Lyrae stars that were confirmed by the SOS Cep\&RRL pipeline,  colour-coded according to the number of $G$-band epochs available per source.}
              \label{nepochsRRNew}%
    \end{figure*}
    
     \begin{figure*}
   \centering
  \includegraphics[width=14.50 cm, trim=0 0 -90 0, clip]{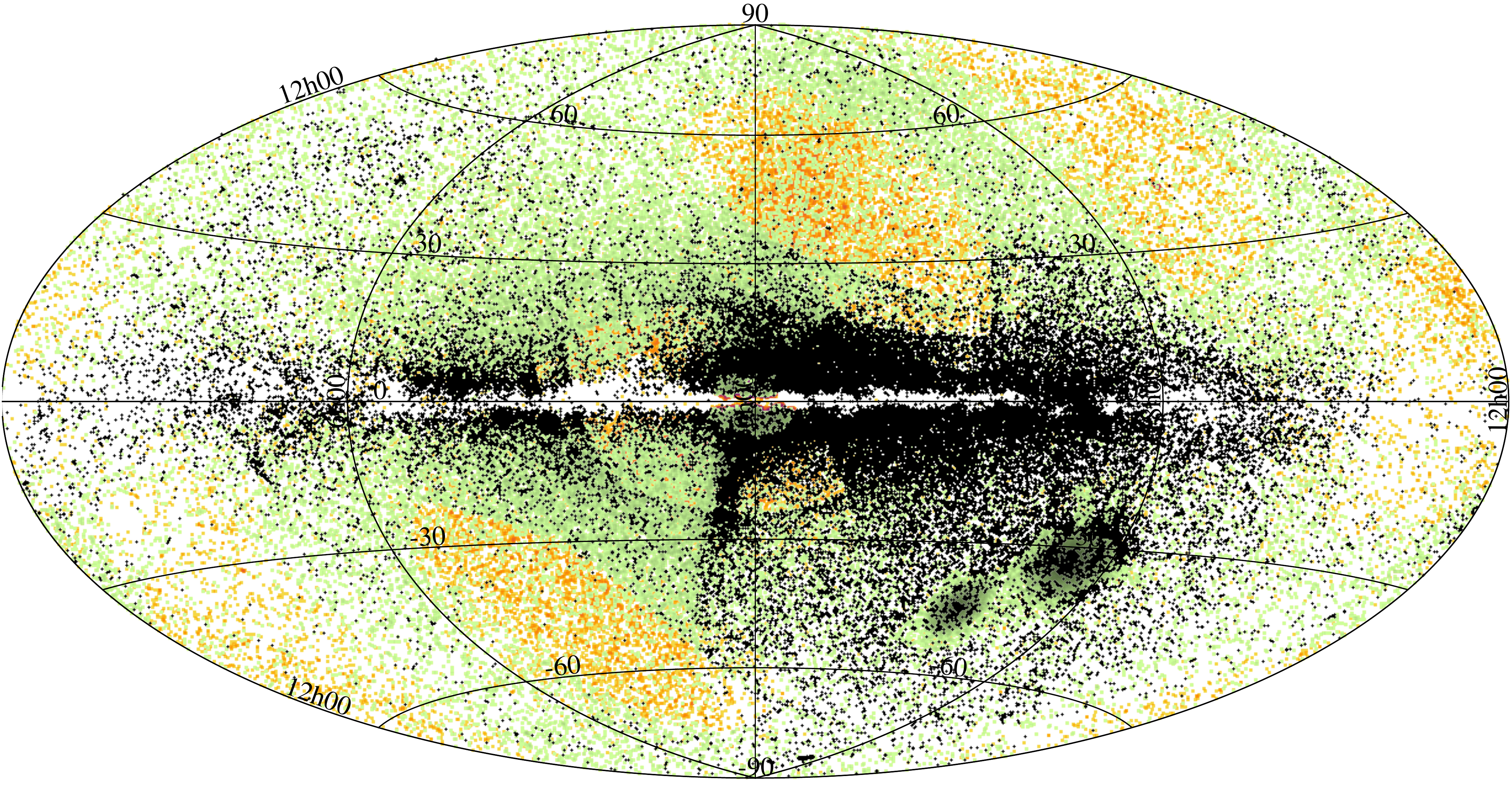}
    \caption{Sky distribution, in galactic coordinates,  of already known and \textit{Gaia} newly discovered  RR Lyrae stars confirmed by the SOS Cep\&RRL pipeline. 
    Orange points:  known RR Lyrae stars that do not have a counterpart among the SOS confirmed RR Lyrae stars published in 
     DR2.
  Green points:  known RR Lyrae stars that cross-match with SOS confirmed RR Lyrae stars. 
  Black points:  new RR Lyrae detected by \textit{Gaia} and confirmed by the SOS Cep\&RRL pipeline. Green and black points clearly reflect the pattern of  \textit{Gaia}  scanning law.
    A total number of more than 220,000 RR Lyrae stars are shown in the figure, 
     of which 46,443  
       are in the Magellanic Clouds,  2,860  are in globular clusters, 984 in classical dSphs (885) and ultra-faint dwarfs (99; Garofalo et al., in preparation) and  50,220 are  new discoveries by \textit{Gaia}. To avoid further overcrowding the figure we have not highlighted globular clusters and dSphs, but refer to Figs.~\ref{mapRR_GC_4},~\ref{mapRR_GC_3} in  Sect.~\ref{maps} for a most complete post-\textit{Gaia} DR2 view of All-Sky RR Lyrae stars down to \textit{Gaia}'s  faint magnitude limit of $G \sim$ 20.7 mag.}
              \label{skyMapKnown+Gaia_RR}%
    \end{figure*}

  \begin{figure*}
   \centering
   \includegraphics[width=14.0 cm, trim=0 0 0 0, clip]{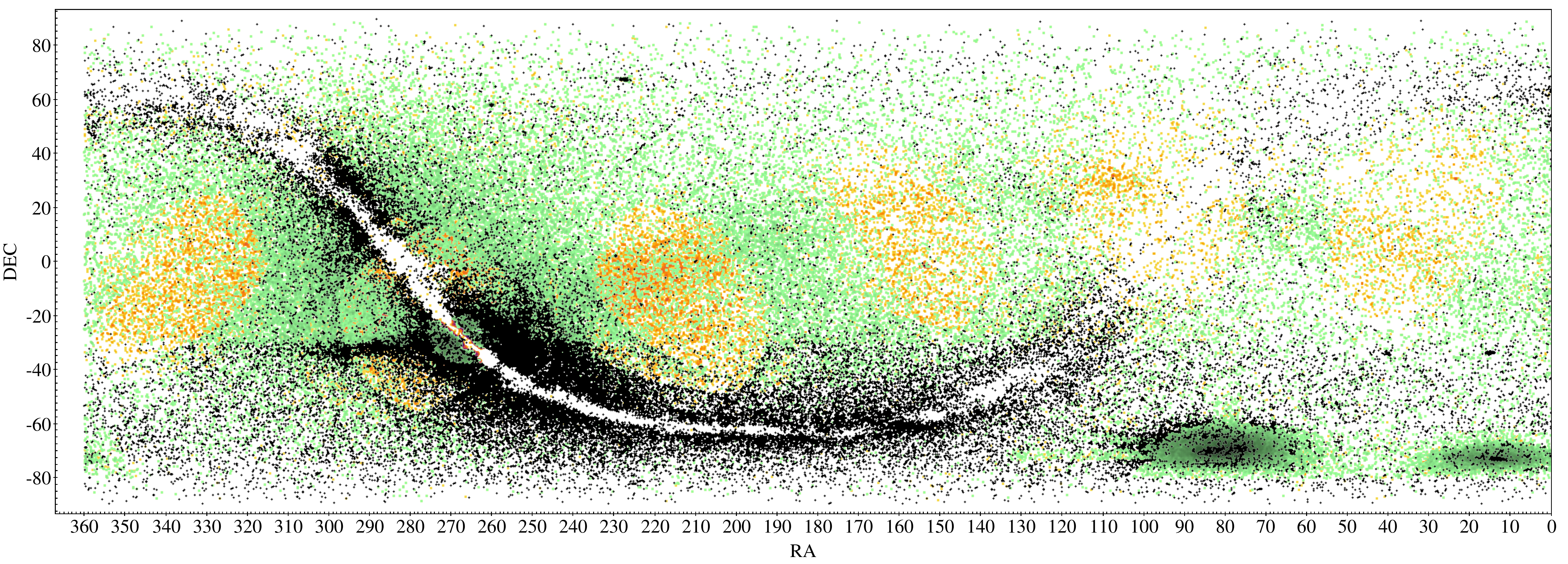}
   \caption{Same as in Fig.~\ref{skyMapKnown+Gaia_RR} but in equatorial coordinates.}
          \label{skyMapKnown+Gaia_RR_1}%
    \end{figure*}

  \begin{figure*}
   \centering
   \includegraphics[width=14.5 cm, trim=20 -30 0 -20, clip]{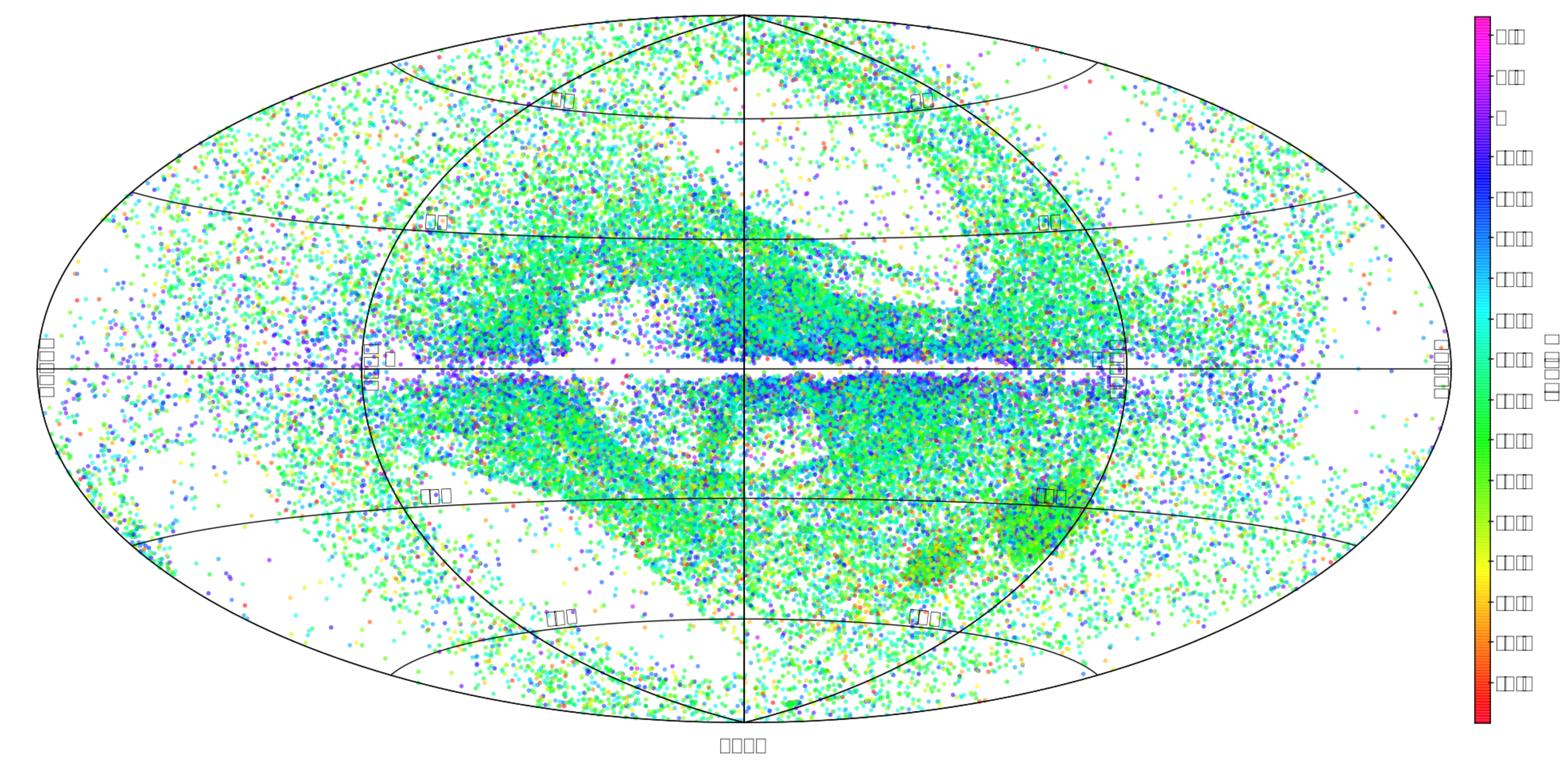}
   \caption{Map in galactic coordinates  of 64,932 RR Lyrae stars for which a photometric [Fe/H] metallicity was inferred using the $\phi_{31}$ parameter in the Fourier decomposition of the $G$-band light curve. The map is  colour coded according to the source metallicity.}
              \label{RRMetallicita}%
    \end{figure*}
  
   \begin{figure*}
   \centering
  \includegraphics[width=14.5 cm, trim=20 -30 0 -20, clip]{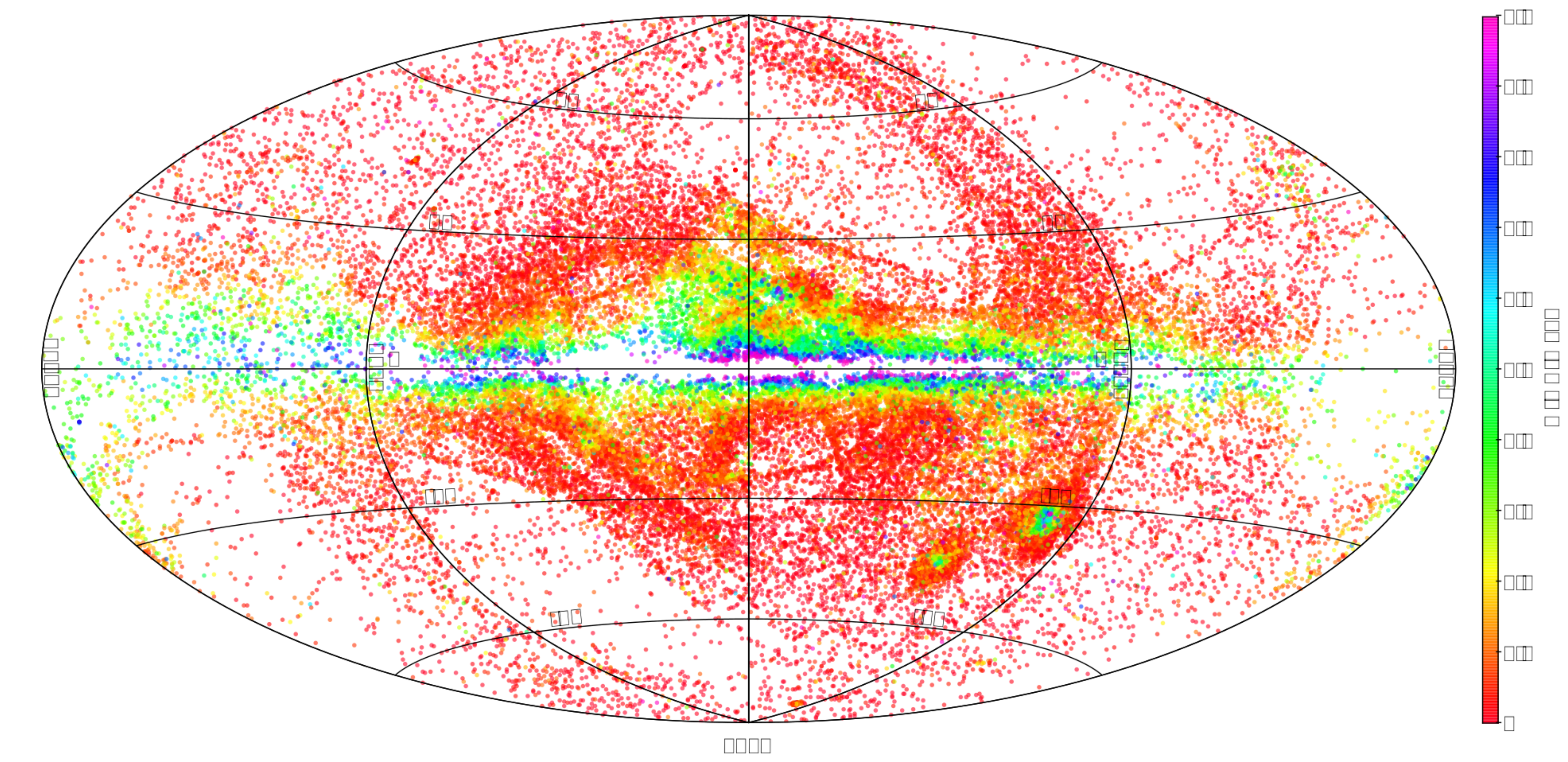}
   \caption{Map in galactic coordinates  of 54,272 fundamental-mode RR Lyrae stars with absorption in the $G$-band [$A(G)$] derived    
   from an empirical relation that 
connects  the amplitude of the light variation in the $G$-band [${\rm Amp(}G{\rm )}$]  and the star pulsation period (see text for details). 
 The map is  colour coded according to the star $A(G)$ value.
  }
              \label{RRabsorption}%
    \end{figure*}

\clearpage
 
\subsection{Results for Cepheids}\label{cep-results}

Examples of $G,G_{\rm BP}, G_{\rm RP}$ light curves for known and new Cepheids confirmed by the
SOS Cep\&RRL pipeline and released in {\it Gaia} DR2 are presented in
Figs.~\ref{plotCepNote} (known All-Sky Cepheids),
~\ref{plotCepMC-known} (known LMC and SMC Cepheids) and
~\ref{plotCepIgnote} (new All-Sky Cepheids).  Light curves are folded
according to period and epoch of maximum light determined by the SOS
Cep\&RRL pipeline.

Figs.~\ref{cmdCClmc} and ~\ref{cmdCCsmc} show the CMDs in apparent $G$
magnitude not corrected for reddening of confirmed Cepheids in the LMC
and SMC, whereas Fig.~\ref{cmdCCallsky} shows the CMD in absolute $G$
magnitude (M$_G$) not corrected for extinction of the confirmed All-Sky
Cepheids (only objects with positive parallax can be plotted). An 
inspection of Figs.~\ref{cmdCClmc} and ~\ref{cmdCCsmc} shows the
expected behaviour of DCEP F and 1O variables (redder the first,
hotter the latter), as well as of ACEPs and T2CEPs, in order of
increasing (apparent) luminosity passing from BLHER to RVTAU objects. 
This neat behaviour is less visible in the All-Sky sample displayed 
in Fig.~\ref{cmdCCallsky}, this is due to the likely contamination by other
types of variables especially for  absolute $G$ magnitudes $\gtrsim$ 2 mag.

Figs.~\ref{CEP_21} and ~\ref{CEP_31} show the distributions of the
confirmed Cepheids in the $\phi_{21}$, $R_{21}$, $\phi_{31}$, $R_{31}$
vs period planes. These figures display a very good separation between
DCEP F and 1O variables, especially in the $R_{21}$-period plane,
where the two modes have significantly different location. As with the CMDs, 
the different diagrams appear more confused in the
case of All-Sky Cepheids. 

Fig.~\ref{raDecAllCep} presents the spatial distribution
of 
the bona fide Cepheids released in \textit{Gaia} DR2 (about 8,900
sources in total) after cleaning the sample of All-Sky Cepheids from other types of variable 
objects not following the $PL$ and $PW$ relations (see
Sect.~\ref{allskycep}). Finally, Fig.~\ref{CepMet-histo}
presents the metallicity distributions of 3,738 fundamental-mode
classical Cepheids with period shorter than 6.3 days published in DR2.  Individual metallicities for these stars were estimated from the $R_{21}$ and $R_{31}$ Fourier 
parameters of the light curves according to the procedure described in Sect.~\ref{metallicity}. The sources are divided among LMC (red histogram in the top panel), SMC 
(green histogram in the mid panel)  and All-Sky (blue histogram in the bottom panel).
Their distributions have median values of [Fe/H]$\sim -$0.2,  [Fe/H]$\sim -$0.1 and [Fe/H]$\sim$0.0 dex for the SMC, LMC and All-Sky samples, respectively. 
The yellow histogram in the bottom panel highlights 235  fundamental-mode classical Cepheids in the All-Sky sample which are already known in the literature. They are   
indistinguishable  from the total sample (blue histogram). 
 We caution potential users of these
metallicity values that while the median metallicity of the All-Sky sample is consistent with the literature values, 
 there seems to be a shift of $\sim$+0.2 dex in
the LMC and SMC distributions, that might be a hint
of calibration issues in the  \citet{klagyivik2013}'s equations which are  mainly based on Galactic Cepheids that do not cover the metal poorer regime typical of the Magellanic Cloud 
Cepheids.
 
 \begin{figure*}[h!]
   \centering
\includegraphics[width=18 cm, trim= 20 140 30 80, clip]{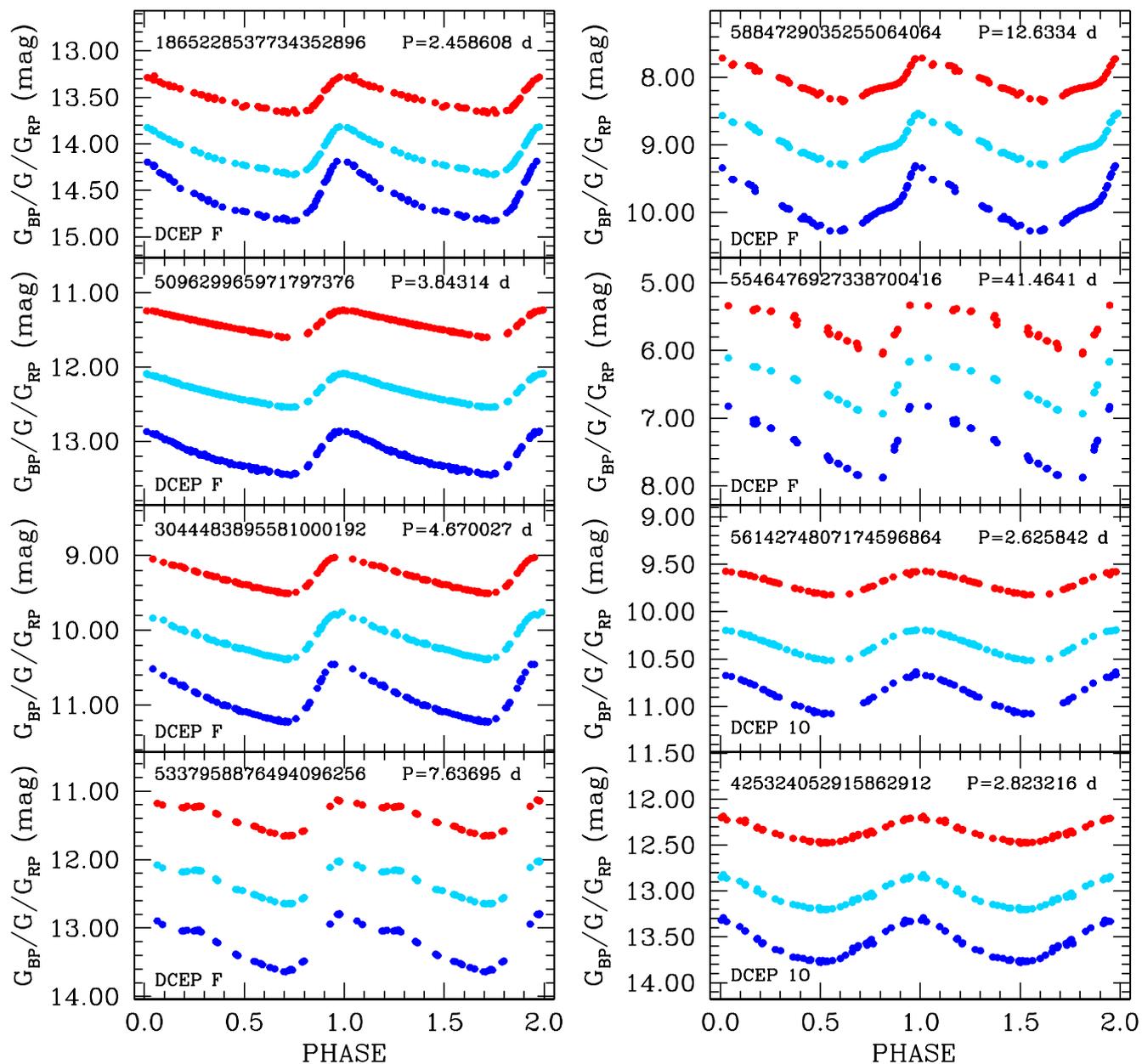}
   \caption{$G$ (cyan), $G_{\rm BP}$ (blue) and $G_{\rm RP}$ (red) light
     curves of All-Sky known 
classical Cepheids of different pulsation mode released in {\it Gaia} DR2. 
     The multi-band time series data are folded according to the
     period and epoch of maximum light derived by the SOS Cep\&RRL
     pipeline. Error bars are comparable to or smaller than symbol
     size.}
              \label{plotCepNote}%
    \end{figure*}

\begin{figure*}
 \centering
  \includegraphics[width=18 cm, trim= 20 140 30 80, clip]{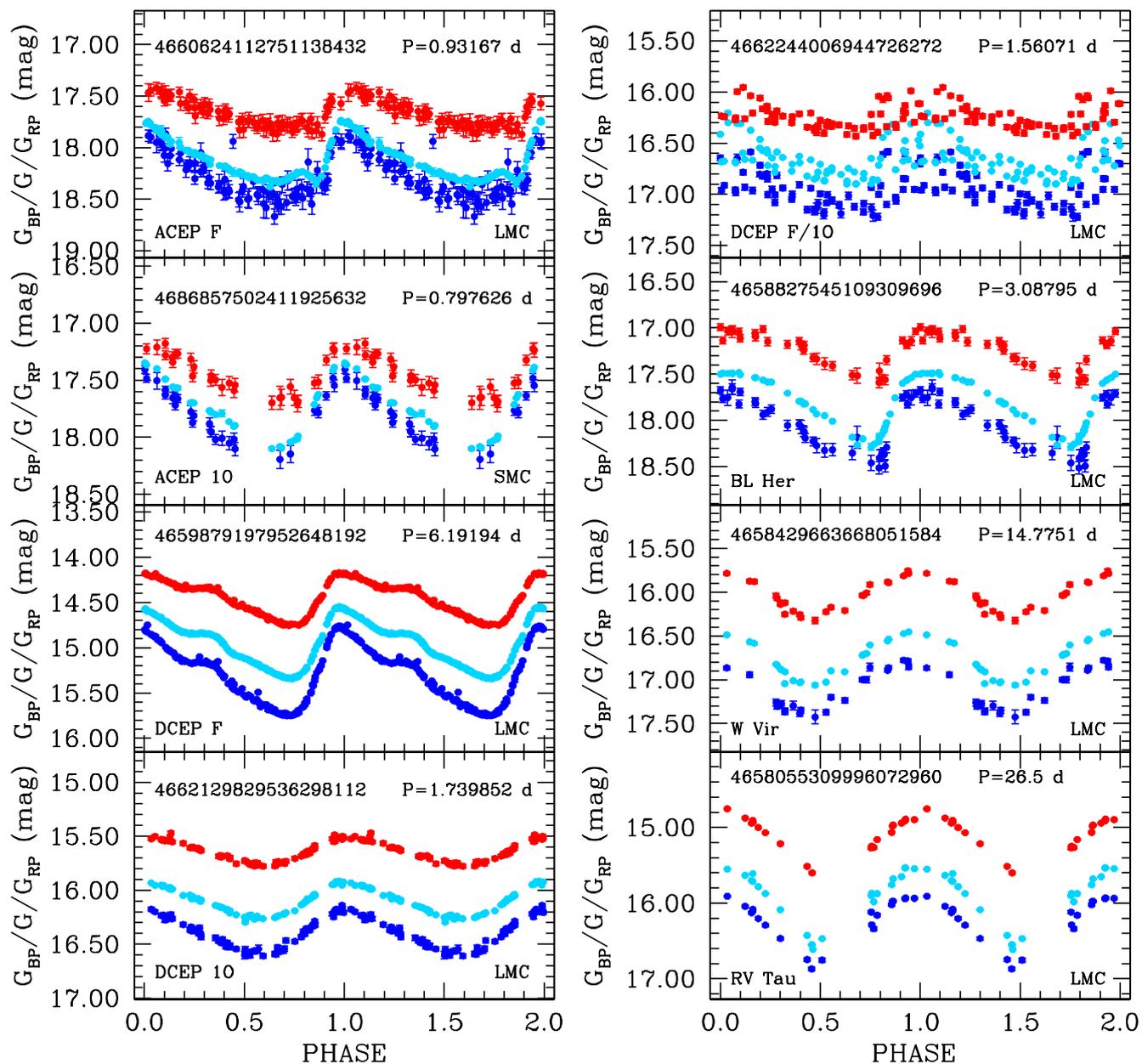}
\caption{
$G$ (cyan), $G_{\rm BP}$ (blue) and $G_{\rm RP}$ (red) light
curves of known Cepheids of different type and pulsation mode released in {\it Gaia} DR2. All of
them are in the LMC except the ACEP 1O which is in the SMC, as labelled.   The multi-band time
series data are folded according to the period and epoch of maximum light derived by the SOS
Cep\&RRL pipeline. Error bars are comparable to or smaller than symbol size except for
 the $G_{\rm BP}$ and  $G_{\rm RP}$ light curves of the two ACEPs in the upper-left panels, as
 expected, due to  the faintness of these two Magellanic Cloud variable stars.
}
              \label{plotCepMC-known}%
    \end{figure*}

 \begin{figure*}[h!]
   \centering
\includegraphics[width=18 cm, trim= 20 140 30 80, clip]{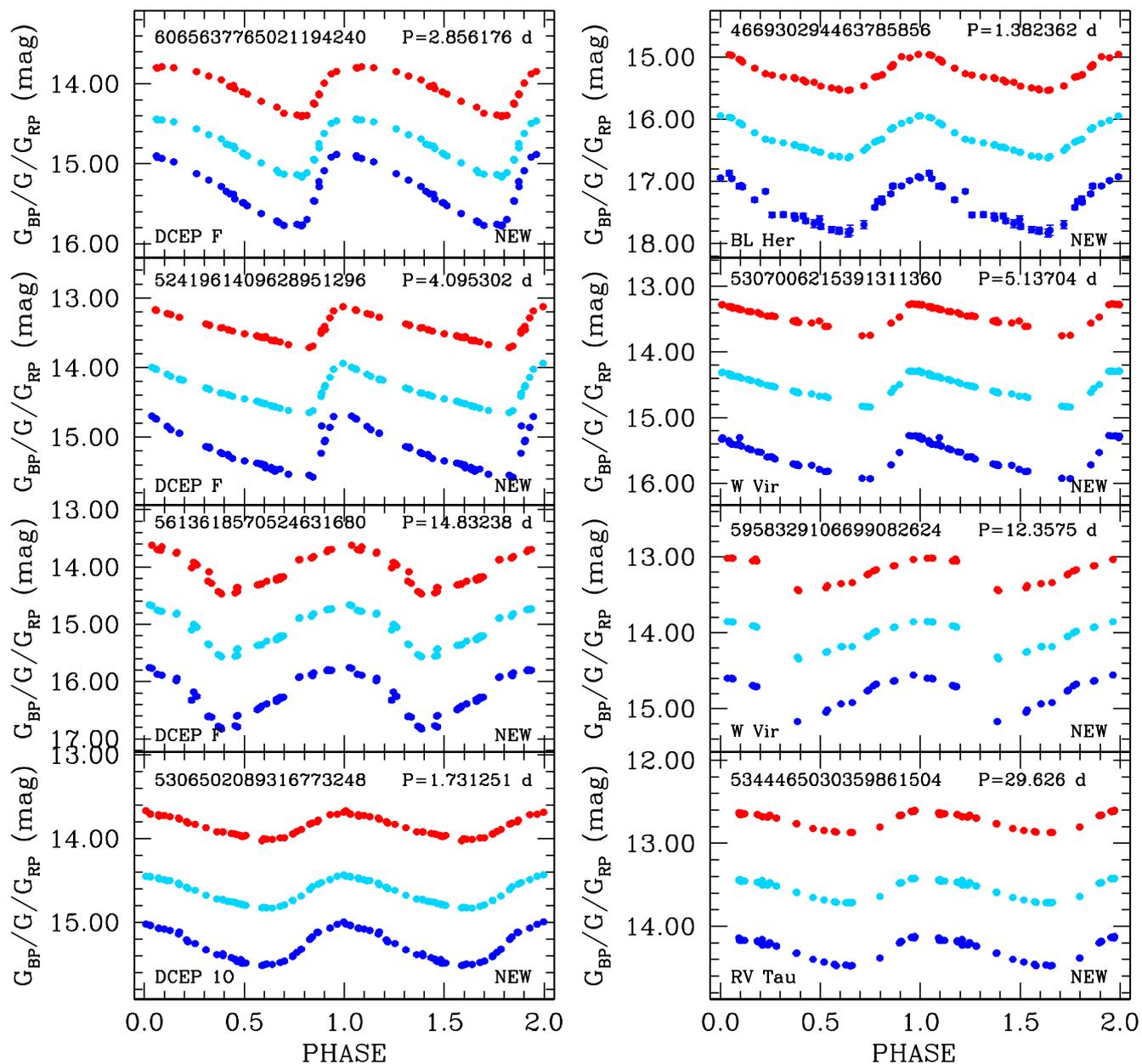}
   \caption{\textit{Left panels:} $G$ (cyan), $G_{\rm BP}$ (blue) and
     $G_{\rm RP}$ (red) light curves of All-Sky new classical  Cepheids of
     different pulsation mode released in {\it Gaia} DR2. 
   \textit{Right panels:} same as in the left panels but for  new Type~II Cepheids of different type.  
   The multi-band time series data are folded according to the period
   and epoch of maximum light derived by the SOS Cep\&RRL
   pipeline. Error bars are comparable to or smaller than symbol size.
 }
              \label{plotCepIgnote}%
    \end{figure*}

\begin{figure}
 \centering
 \includegraphics[width=9.0 cm, trim= 10 140 0 70, clip]{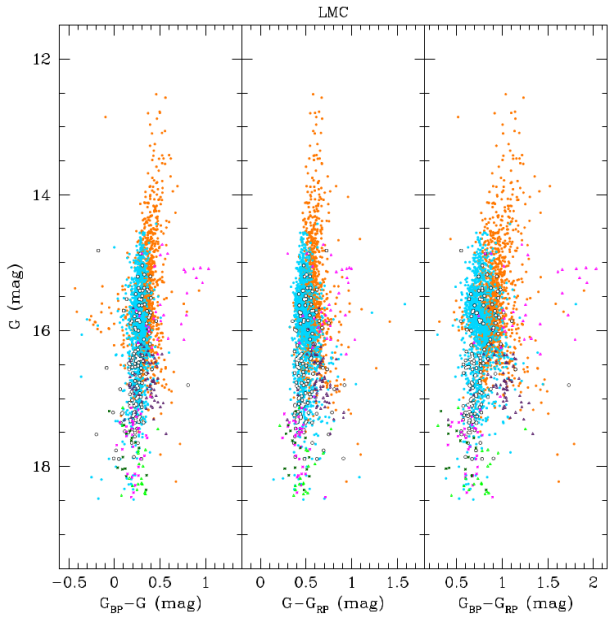}
   \caption{$G$, $G_{\rm BP}-G$; $G$, $G-G_{\rm RP}$ and $G$, $G_{\rm BP}-G_{\rm RP}$  CMDs in apparent magnitude of Cepheids in the LMC published in {\it Gaia} DR2. Symbols and colour-coding are as in Fig.~\ref{plg}.
         }
              \label{cmdCClmc}%
    \end{figure}
    
\begin{figure}
 \centering
 \includegraphics[width=9.0 cm, trim= 10 140 0 70, clip]{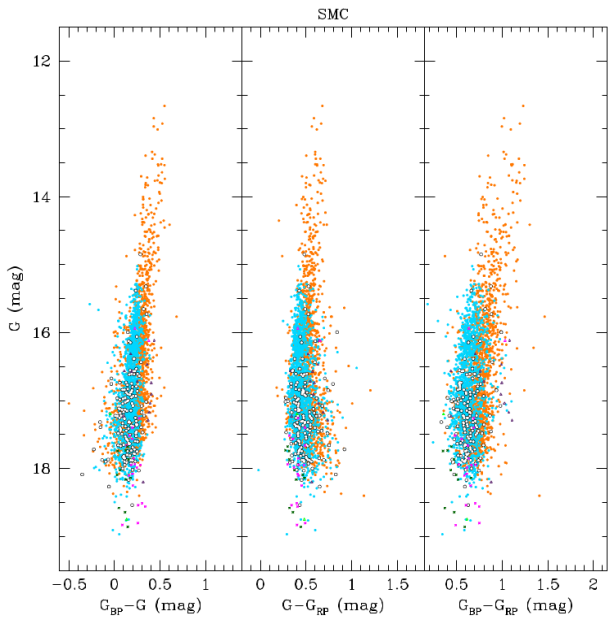}
   \caption{Same as in Fig.~\ref{cmdCClmc}  but for Cepheids in the SMC.}
              \label{cmdCCsmc}%
    \end{figure}

\begin{figure}
 \centering
   \includegraphics[width=9.5 cm, trim= 10 140 0 70, clip]{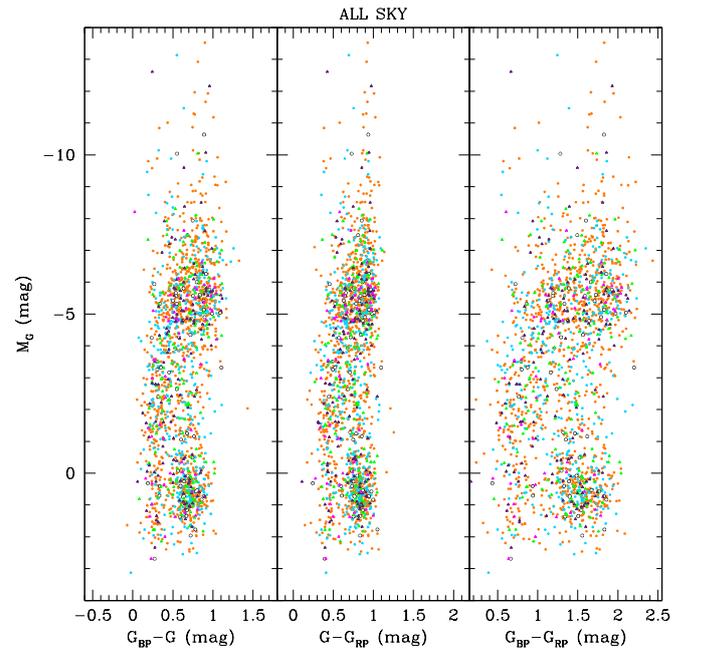}
   \caption{Same as in  Fig.~\ref{cmdCClmc}  but  in absolute $G$ magnitude (M$_G$) for All-Sky  Cepheids.}
              \label{cmdCCallsky}%
    \end{figure}

\begin{figure}
 \centering
   \includegraphics[width=9.5 cm, trim= 10 140 0 50, clip]{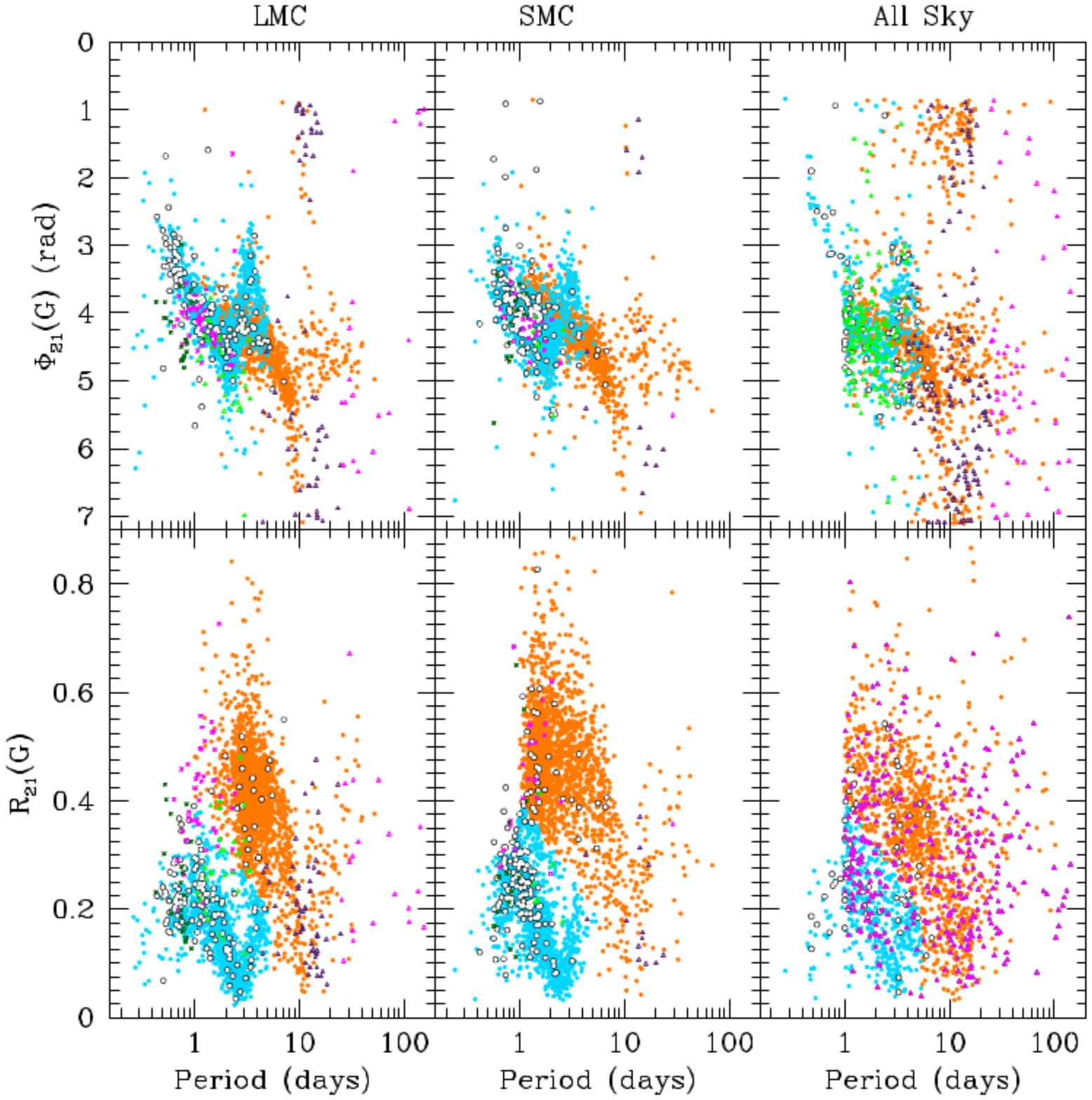}
   \caption{\textit{Upper panels:} $G$-band  $\phi_{21}$ vs period diagram for Cepheids in the LMC (\textit{upper-left}), SMC (\textit{middle}) and All-Sky (\textit{upper-right}).
   \textit{Lower panels:} same as in the upper panels but for the the $G$-band  $R_{21}$ vs period diagram. Symbols and colour-coding are as in Fig.~\ref{plg}.
         }
              \label{CEP_21}%
    \end{figure}
    
 \begin{figure}
 \centering
    \includegraphics[width=9.5 cm, trim= 10 140 0 50, clip]{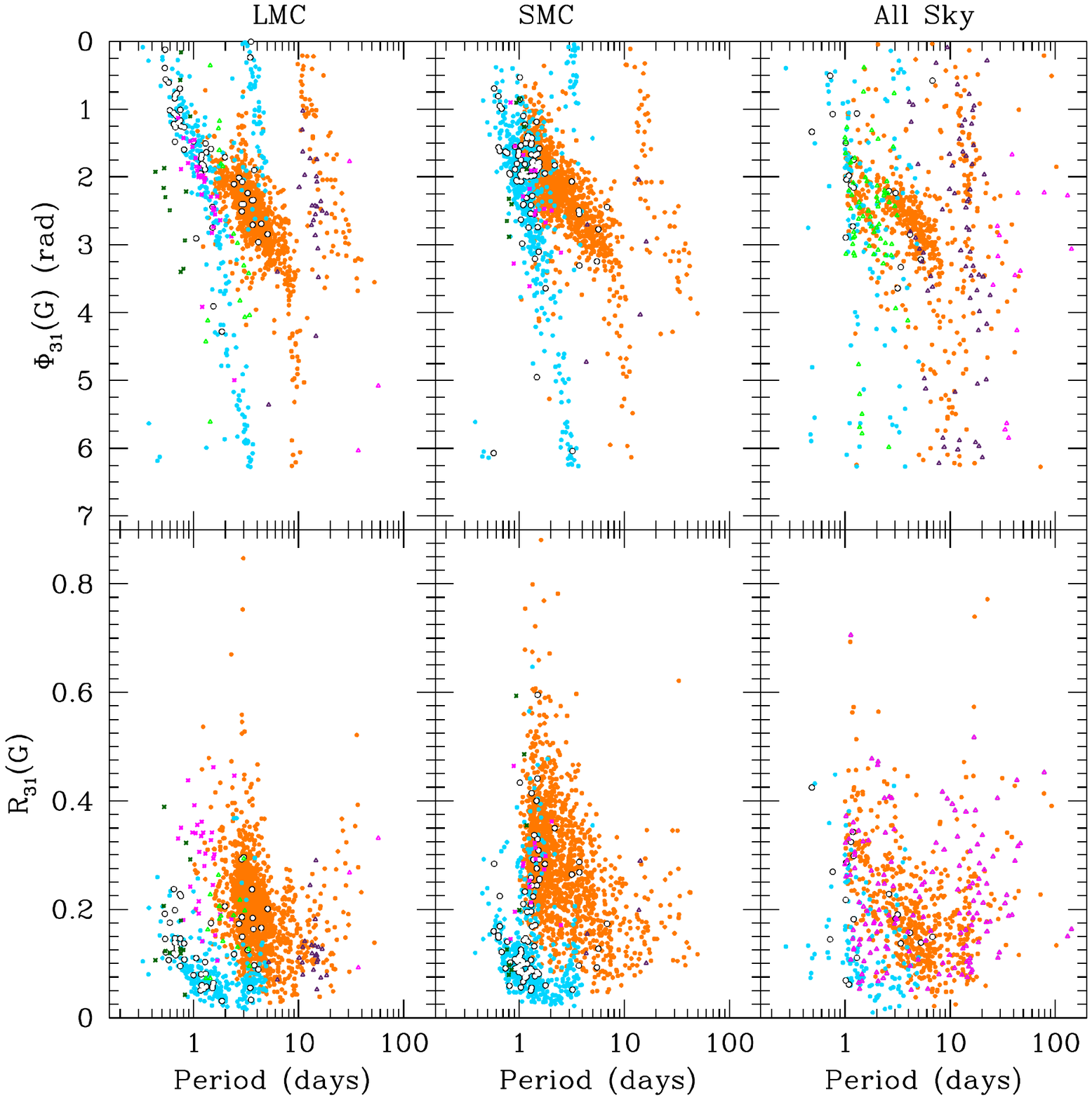}
   \caption{\textit{Upper panels:} $G$-band  $\phi_{31}$ vs period diagram for Cepheids in the LMC (\textit{upper-left}), SMC (\textit{middle}) and All-Sky (\textit{upper-right}).
   \textit{Lower panels:} same as in the upper panels but for the $G$-band  $R_{31}$ vs period diagram. Symbols and colour-coding are as in Fig.~\ref{plg}.
         }
              \label{CEP_31}%
    \end{figure}
   
\begin{figure*}
   \centering
   \includegraphics[width=16 cm,clip]{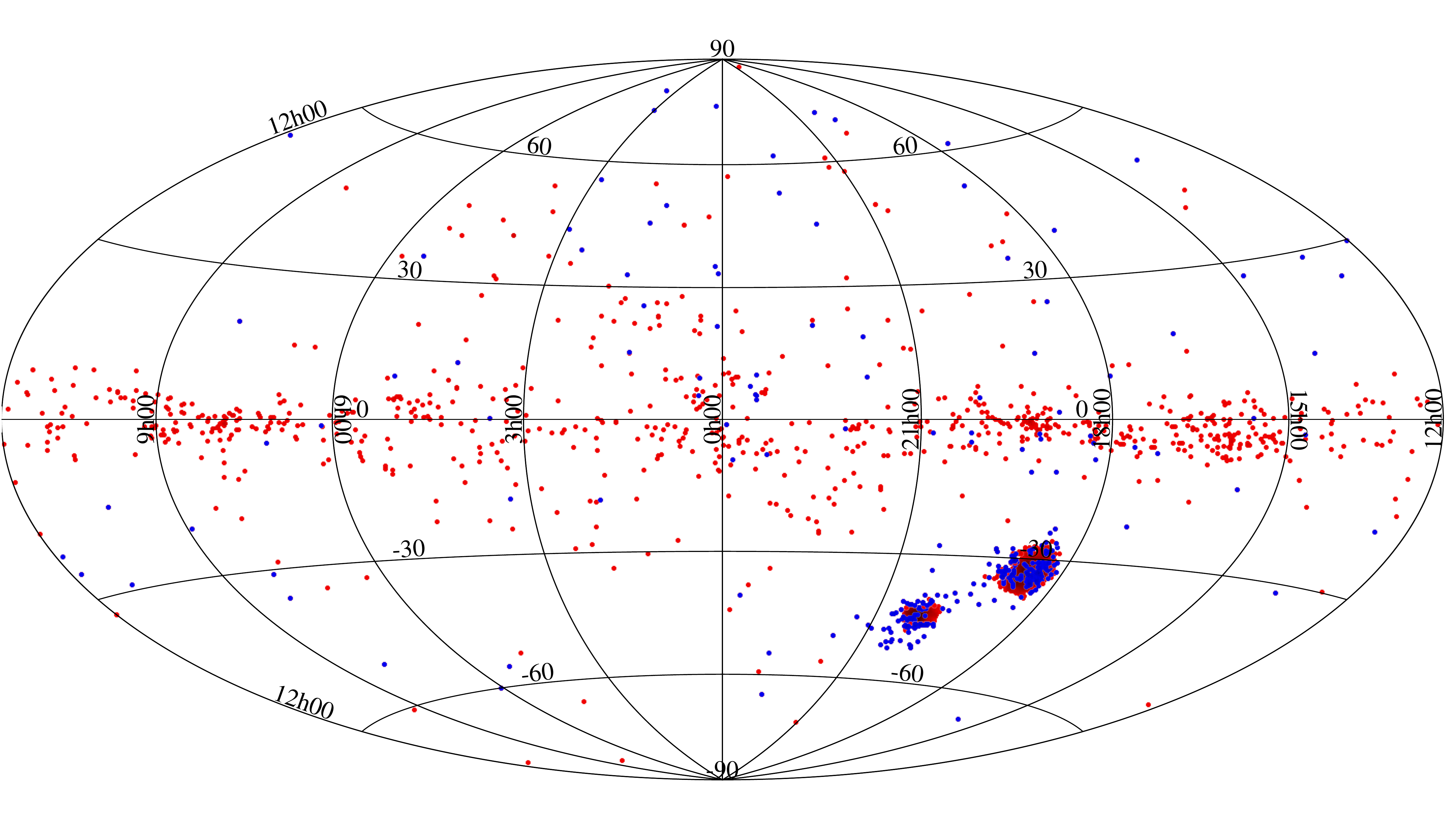}
  \caption{Spatial distribution of  the 
  bona fide Cepheids released in {\it Gaia} DR2 (about 8,900 sources in total). Red filled circles are known Cepheids in the literature (OGLE and other surveys), blue filled circles are new Cepheids detected by {\it Gaia} in the LMC, SMC (118 in total) and All-Sky (about 240 in total).
  The latter sample has been cleaned from other types of variable 
   objects not following the $PL$ and $PW$ relations, see discussion at the end of  Sect.~\ref{allskycep}.  
}
              \label{raDecAllCep}%
    \end{figure*}

 \begin{figure}
   \centering
   \includegraphics[width=9.5 cm, trim=30 140 0 60, clip]{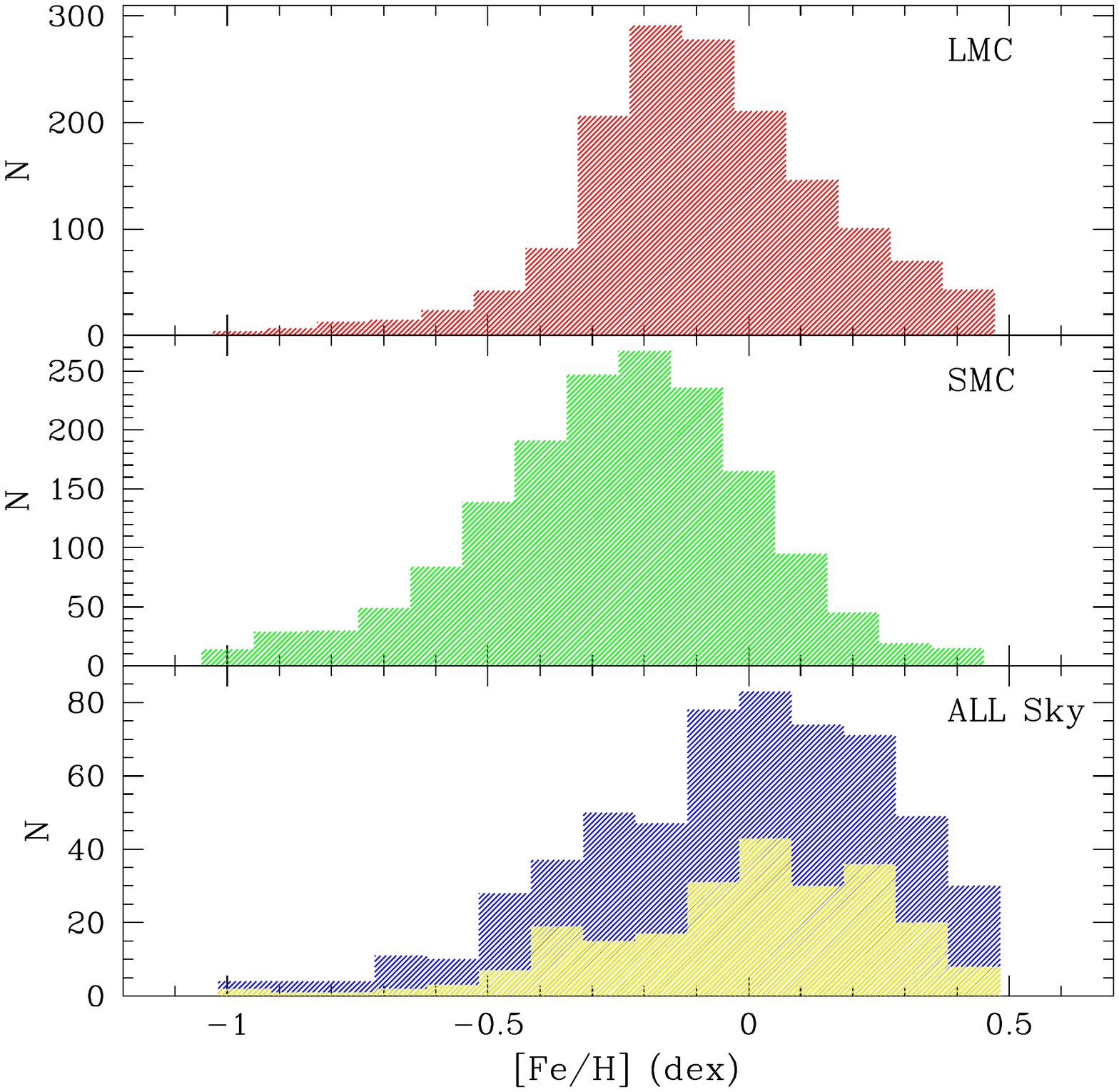}
   \caption{Metallicity distributions of  3,738 fundamental-mode classical Cepheids with period shorter than 6.3 days published in DR2. The three panels show from top to bottom the LMC, SMC and All-Sky distributions (blue histogram), respectively.  
They have median values of [Fe/H]$\sim -$0.2,  $\sim -$0.1 and $\sim$0.0 dex for the SMC, LMC and All-Sky samples, respectively. 
The yellow histogram in the lower panel highlights the metallicity distribution of 235  fundamental-mode classical Cepheids in the All-Sky sample which are already known in the literature. They are   
indistinguishable  from the total sample (blue histogram). Among them are 10 sources located on the faint sequences below the dashed lines in Fig.\ref{plnote}.}
              \label{CepMet-histo}%
    \end{figure}
     
  \subsubsection{Cepheids: validation and comparison with the literature}\label{val-cep}
  
As with the RR Lyrae stars a confusion matrix was derived only for the
Cepheids in the MCs (Fig.~\ref{confusionMatrixCep}). This is because only in the MCs we have 
complete and homogeneous reference catalogues for ACEPs, DCEPs, and
T2CEPs published by the OGLE group \citep[][]{soszynski2008,soszynski2010,soszynski2015a,soszynski2015b}. 

An 
inspection of Fig.~\ref{confusionMatrixCep} reveals a very low (almost 0\%) 
contamination level for all Cepheid types/modes, with the
exception of the multimode Cepheids, for which we have more than 50\%
false positives mostly among ordinary DCEP 1O. This is a common
problem also for double-mode RR Lyrae stars that will be mitigated in next \textit{Gaia} releases by
both improvements of the SOS pipeline and a natural increase of 
number of epochs in the light curves, a fundamental ingredient for a
successful recovery of multimode pulsators. 

Applying the same procedure to Cepheids as for the RR Lyrae stars (see Sect.~\ref{val-rrl} and \citealt{holl2018}) we estimate a contamination of about 5\%
in regions of the sky extending partially beyond OGLE-IV footprint.

\begin{figure*}
   \centering
  \includegraphics[width=16 cm, clip]{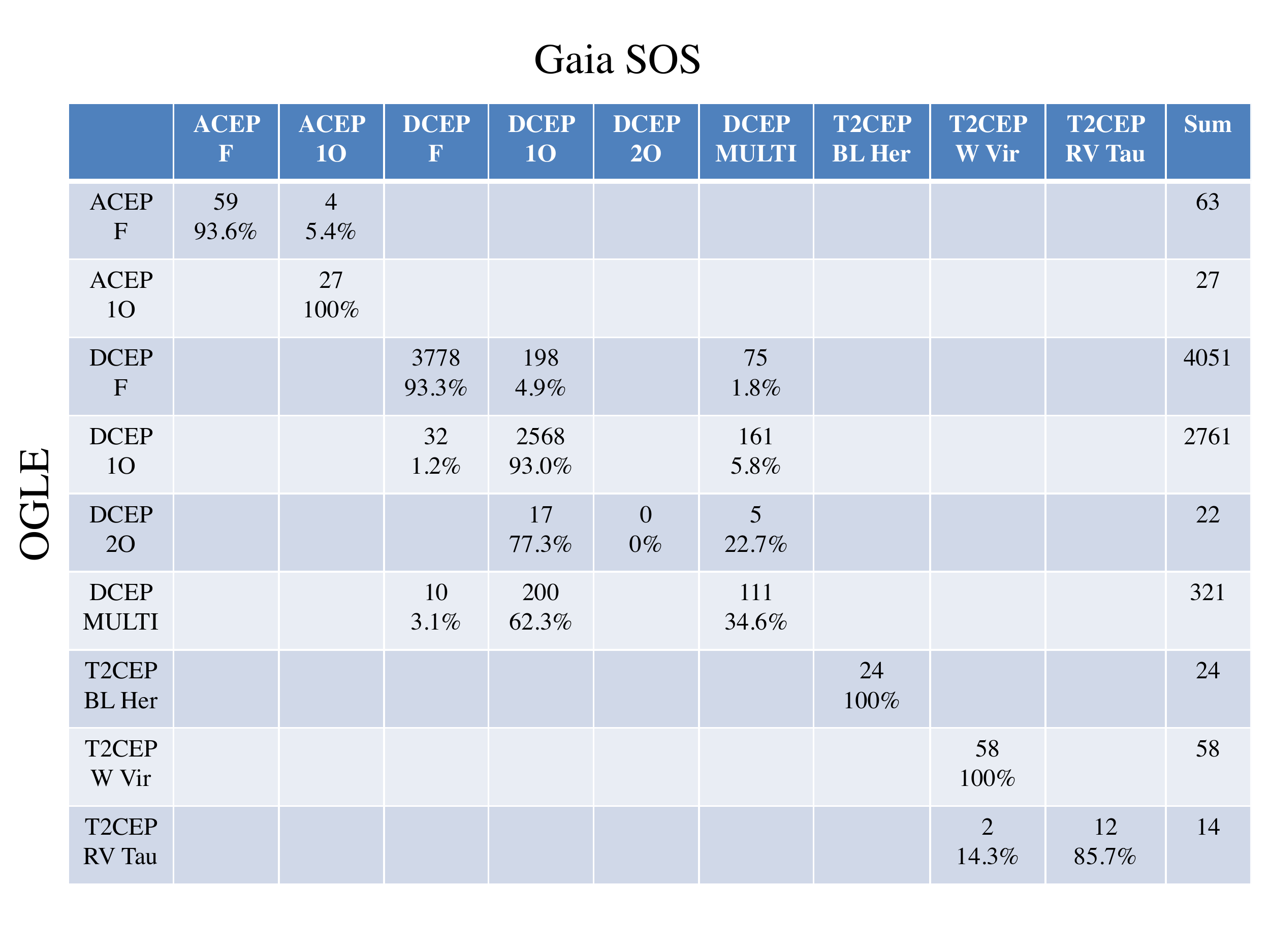}
    \caption{Confusion matrix for Cepheids. As control sample we used 
all the variable stars classified as ACEP, DCEP, and T2CEP by the OGLE 
survey in LMC and  SMC that have a cross-match within a radius of 3 arcsec with 
the Cepheids of the aforementioned types  published in {\it Gaia} DR2, 
for a total of  7,341 objects. Rows refer to literature results and 
columns to results of the SOS Cep\&RRL pipeline. The corresponding success percentage 
is shown along the diagonal cells.}
\label{confusionMatrixCep}%
\end{figure*} 

Finally, we note that a small number of new Cepheids, 118 in total,
and 1,640 new RR Lyrae stars 
were identified in the LMC and SMC areas extensively monitored by the
OGLE survey.  While we are quite confident that they are bona fide
new Cepheids and RR Lyrae stars as they did pass all diagnostics and also a very
careful visual inspection, we would be indebted to our colleagues of
the OGLE team if they could verify on their OGLE photometry, the
reliability of these
sources. 
 
\section{Results and final accounting}\label{results2}

{\rm The run of the SOS Cep\&RRL pipeline on the candidate RR Lyrae stars and Cepheids,  
combined with estensive validation procedures and  random visual inspection of some of the resulting light curves,   
produced final samples of 140,784 confirmed RR Lyrae stars and 9,575 Cepheids for a total of 150,359 sources.
Position, $G$, $G_{\rm BP}$, $G_{\rm RP}$ time series photometry, and final results of the  
SOS Cep\&RRL processing are published in {\it Gaia} DR2   for all these 150,359 sources. 

The subdivision of the 150,359  sources according to type, subtype, and pulsation mode is  summarised in Table~\ref{table:2} while Table~\ref{table:3} provides the subdivision between known and new discoveries by \textit{Gaia}.
\begin{table}[h!]
\caption{Number and type/mode classification of RR Lyrae stars and Cepheids confirmed by the SOS Cep\&RRL pipeline that are published in {\it Gaia} DR2, broken down in 
LMC, SMC and All-Sky regions of the sky.}             
\label{table:2}     
\begin{tabular}{l r r r r }       
\hline\hline                 
\noalign{\smallskip}
Type & LMC  & SMC  & All-Sky& Total \\ 
\hline                          
\noalign{\smallskip}
RRab &  20,264  & 4,629   & 73,133   &98,026 \\  
RRc &      8,395  & 1,213   &  30,772  & 40,380 \\
RRd &         693  &   164   &     1,521  & 2,378\\  
\hline                        
\noalign{\smallskip}
RR Lyrae Total & 29,352     & 6,006    & 105,426    &  140,784\\
\hline                        
\noalign{\smallskip}
DCEP F          &1,925  & 1,915    &1,158    & 4,998\\ 
DCEP 1O       & 1,461 &  1,540   & 471   & 3,472\\ 
DCEP MULTI &     175 &    177    & 68     &    420\\ 
\hline                          
\noalign{\smallskip}
DCEP Total   & 3,561    &  3,632  &  1697  & 8,890\\ 
\hline                          
\noalign{\smallskip}
ACEP F       & 43   & 23   &  --  & 66\\  
ACEP 1O       &  21  &  13  &  --  & 34\\  
\hline                          
\noalign{\smallskip}
ACEP Total   &  64   &  36  &    & 100\\ 
\hline                       
\noalign{\smallskip}
T2CEP BLHER       &  35  & 6   & 182    &223\\  
T2CEP WVIR      &  73  &  15  &  165  & 253\\   
T2CEP RVTAU      &  34  &  3  &   72  &109\\   
\hline                          
\noalign{\smallskip}
T2CEP Total   &  142   & 24   &  419  & 585\\ 
\hline
\noalign{\smallskip}
Cepheid Total & 3,767   & 3,692   & 2,116   &9,575\\
\hline  
\noalign{\smallskip}
\end{tabular}

Counts for Cepheids correspond to the  number of sources in the LMC, SMC and All-Sky regions defined in Sect.~\ref{s2}.\\
Counts for the RR Lyrae correspond to the  number of sources in regions defined by  taking into account where the number density of RR Lyrae stars in the LMC and SMC drops and becomes comparable
to  the counts in the field.\\
\end{table}

\begin{table}[h!]
\caption{Total numbers of SOS confirmed RR Lyrae stars and Cepheids published in {\it Gaia} DR2 subdivided in new and known sources.}             
\label{table:3}     
\begin{tabular}{l r r r }       
\hline\hline                 
\noalign{\smallskip}
Type & Grand total & New  & Known \\ 
\hline                          
\noalign{\smallskip}
RR Lyrae stars& 140,784  &  50,220  & 90,564\\ 
Cepheids         &     9,575  &  350      & 9,225 \\
\hline
\smallskip   
\end{tabular}

``New'' sources  mean ``new to the best of our knowledge''. \\                        
\end{table}

\noindent For these 140,784 RRLs and 9,575 Cepheids  the  following parameters, computed by the SOS Cep\&RRL  pipeline,  have been released in {\it Gaia} DR2  along with the related errors:\\ 

\noindent 
- source pulsation periods (main and secondary periodicities, if any);\\
- intensity-averaged mean $G$, $G_{\rm BP}$, $G_{\rm RP}$ magnitudes;\\
- epochs of maximum light in the 3 pass-bands;\\
- $\phi_{21}$ and $R_{21}$ Fourier parameters;\\
- $\phi_{31}$ and $R_{31}$ Fourier parameters;\\
- peak-to-peak $G$, $G_{\rm BP}$, $G_{\rm RP}$ amplitudes [${\rm Amp(}G{\rm )}$], [{\rm Amp}($G_{\rm BP}$)], [{\rm Amp}($G_{\rm RP}$)];\\
- RR Lyrae star subclassification into RRab, RRc and RRd types;\\
- Cepheid classification into DCEP, ACEP, T2CEP types;\\
- DCEPs and ACEPs pulsation mode  (F, 1O, DCEP MULTI);\\
- T2CEPs subclassification into BLHER, WVIR, RVTAU types;\\
- Absorption in the $G$ band, $A(G)$, of RRab stars;\\
- Photometric metallicity, [Fe/H],  for RRab and RRc stars;\\
- Photometric metallicity, [Fe/H],  for fundamental-mode DCEPs with $P <$ 6.3 days.\\

Gaia's sourceids, coordinates, values of the above quantities and associated statistics, along with the $G$, $G_{\rm BP}$ and $G_{\rm RP}$  time series photometry for each of the 140,784 RR Lyrae stars and 9,575 Cepheids confirmed and characterised by the SOS Cep\&RRL pipeline can be retrieved from the  {\it Gaia}  data release archive\footnote{\texttt{http://archives.esac.esa.int/gaia/}} 
  and other distribution nodes. The archives also provide tools for queries and to cross-match Gaia data with other catalogues available in the literature.
 
 We provide in Tables~\ref{table:3aRRL} and ~\ref{table:3aCep} specific links to the archive tables, and  summarise the names of the parameters computed by
SOS Cep\&RRL that can be retrieved from the archive tables. In Appendix~\ref{app:queries} 
 we provide examples of queries to retrieve some of the quantities and parameters listed in Tables~\ref{table:3aRRL} and \ref{table:3aCep}.

\clearpage

\begin{table*}
\tiny
\setlength\tabcolsep{5pt}
\caption{Links to {\it Gaia} archive tables to retrieve the 
pulsation characteristics: period(s), epochs of maximum light (E), peak-to-peak amplitudes, intensity-averaged mean magnitudes, $\phi_{21}$, $R_{21}$, $\phi_{31}$, $R_{31}$ Fourier parameters with related uncertainties,  metallicity and absorption in the $G$ band computed by the SOS Cep\&RRL pipeline for the 140,784  RR Lyrae stars confirmed by SOS and released in {\textit{Gaia}} DR2. 
To ease table access, we also provide the correspondence between parameter (period(s), E, etc.) and the name of the parameter in the \textit{Gaia}  archive table.}           
\label{table:3aRRL}     
\centering                         
\begin{tabular}{ll}       
\hline\hline                 
\noalign{\smallskip}
Table URL & \texttt{http://archives.esac.esa.int/gaia/}\\
\noalign{\smallskip}
\hline
\noalign{\smallskip}
\multicolumn{2}{c}{RR Lyrae star parameters computed by the SOS Cep\&RRL pipeline}\\
\hline
\noalign{\smallskip}
Table Name&  \texttt{\textbf{gaiadr2.vari\_rrlyrae}}\\
Source ID & \texttt{source\_id}\\
Type& \texttt{best\_classification} (one of \texttt{RRC}, \texttt{RRAB} or \texttt{RRD})\\
$Pf, P1O, P2O, P3O$ &\texttt{p\_f, p1\_o, p2\_o, p3\_o}: NB \texttt{p3\_o} is empty for RRL\\
$\sigma (Pf, P1O, P2O, P3O)$&\texttt{pf\_error, p1\_o\_error, p2\_o\_error, p3\_o\_error}: NB \texttt{p3\_o\_error}  is empty for RRL \\
E\tablefootmark{\rm (a)}($G$, $G_{\rm BP}$, $G_{\rm RP}$)&\texttt{epoch\_g, epoch\_bp, epoch\_rp} \\
$\sigma {\rm E}$($G$, $G_{\rm BP}$, $G_{\rm RP}$)  &\texttt{epoch\_g\_error, epoch\_bp\_error, epoch\_rp\_error}\\
$\langle G \rangle, \langle G_{\rm BP} \rangle, \langle G_{\rm RP} \rangle$&\texttt{int\_average\_g, int\_average\_bp, int\_average\_rp}\\
$\sigma \langle G \rangle$ , $\sigma \langle G_{\rm BP} \rangle$ , $\sigma \langle G_{\rm RP} \rangle$ &\texttt{int\_average\_g\_error, int\_average\_bp\_error, int\_average\_rp\_error} \\
{\rm Amp}($G, G_{\rm BP}, G_{\rm RP}$)&\texttt{peak\_to\_peak\_g, peak\_to\_peak\_bp, peak\_to\_peak\_rp}\\
$\sigma {\rm [Amp(}G{\rm )]}$, $\sigma {\rm [Amp(}G_{\rm BP}{\rm )]}$, $\sigma {\rm [Amp(}G_{\rm RP}{\rm )]}$&\texttt{peak\_to\_peak\_\linebreak g\_error, peak\_to\_peak\_\linebreak bp\_error, peak\_to\_peak\_\linebreak rp\_error}\\
$\phi_{21}$ &\texttt{phi21\_g}\\
$\sigma (\phi_{21})$ &\texttt{phi21\_\linebreak g\_error} \\
$R_{21}$ &\texttt{r21\_g}\\
$\sigma (R_{21})$&\texttt{r21\_g\_error}\\ 
$\phi_{31}$ &\texttt{phi31\_g}\\
$\sigma (\phi_{31})$ &\texttt{phi31\_\linebreak g\_error} \\
$R_{31}$ &\texttt{r31\_g}\\
$\sigma (R_{31})$&\texttt{r31\_g\_error} \\ 
${\rm [Fe/H]}\tablefootmark{\rm (b)}$&\texttt{metallicity}\\
$\sigma$ ([Fe/H]) &\texttt{metallicity\_error}\\
A($G$)\tablefootmark{\rm (c)} & \texttt{g\_absorption}\\
$\sigma$ A($G$) &\texttt{g\_absorption\_error}\\
$N_{\rm obs}$($G$ band) & \texttt{num\_clean\_epochs\_g}\\
$N_{\rm obs}$($G_{\rm BP}$ band) & \texttt{num\_clean\_epochs\_bp}\\
$N_{\rm obs}$($G_{\rm RP}$ band) & \texttt{num\_clean\_epochs\_rp}\\
\noalign{\smallskip}
\hline                                  
\end{tabular}
\tablefoot{$^{\rm (a)}$The BJD of the epoch of maximum light is offset by JD 2455197.5 d (= J2010.0).  
$^{\rm (b)}$ Photometric metal abundance derived from the $\phi_{31}$ Fourier parameter of the light curve for 54,272 fundamental-mode RR Lyrae stars (see Sects.~\ref{metallicity} and ~\ref{results-rrl}).  $^{\rm (c)}$ Absorption in the $G$ band computed from a relation that links the star intrinsic colour to the period and the amplitude of the light variation (see Sects.~\ref{absorption} and ~\ref{results-rrl}).}\\
\end{table*}

\begin{table*}
\tiny
\setlength\tabcolsep{5pt}
\caption{Links to {\it Gaia} archive tables to retrieve 
the pulsation characteristics: period(s), epochs of maximum light (E), peak-to-peak amplitudes, intensity-averaged mean magnitudes, $\phi_{21}$, $R_{21}$, $\phi_{31}$, $R_{31}$ Fourier parameters with related uncertainties and metallicity 
computed by the SOS Cep\&RRL pipeline for the 9,575 Cepheids confirmed by SOS and released in \textit{Gaia} DR2. 
To ease table access, we also provide the correspondence between parameter (period(s), E, etc.) and the name of the parameter in the \textit{Gaia}  archive table.}
\label{table:3aCep}     
\centering                         
\begin{tabular}{ll}       
\hline\hline                 
\noalign{\smallskip}
Table URL & \texttt{http://archives.esac.esa.int/gaia/}\\
\noalign{\smallskip}
\hline
\multicolumn{2}{c}{Cepheid parameters computed by the SOS Cep\&RRL pipeline}\\
\hline
\noalign{\smallskip}
Table Name&  \texttt{\textbf{gaiadr2.vari\_cepheid}}\\
Source ID & \texttt{source\_id}\\
Type& \texttt{type\_best\_classification} (one of \texttt{T2CEP}, \texttt{DCEP} or \texttt{ACEP})\\
Type2& \texttt{type2\_best\_classification} (for type-II Cepheids, one of \texttt{BL\_HER}, \texttt{W\_WVIR} or \texttt{RV\_TAU})\\
Mode& \texttt{mode\_best\_classification} (one of \texttt{FUNDAMENTAL}, \texttt{FIRST\_OVERTONE}, \texttt{SECOND\_OVERTONE}\\
{}	& \texttt{MULTI}, \texttt{UNDEFINED}, or \texttt{NOT\_APPLICABLE})\\
Multi-mode& \texttt{multi\_mode\_best\_classification} (for multi-mode $\delta$ Cepheids, one of \texttt{F/1O}, \texttt{F/2O}, \texttt{1O/2O}, \\
{}	& \texttt{1O/3O}, \texttt{2O/3O}, \texttt{F/1O/2O}, or \texttt{1O/2O/3O})\\
$Pf, P1O, P2O, P3O$ &\texttt{p\_f, p1\_o, p2\_o, p3\_o} \\
$\sigma (Pf, P1O, P2O, P3O)$&\texttt{pf\_error, p1\_o\_error, p2\_o\_error, p3\_o\_error} \\
E\tablefootmark{\rm (a)}($G$, $G_{\rm BP}$, $G_{\rm RP}$)&\texttt{epoch\_g, epoch\_bp, epoch\_rp} \\
$\sigma {\rm E}$($G$, $G_{\rm BP}$, $G_{\rm RP}$)  &\texttt{epoch\_g\_error, epoch\_bp\_error, epoch\_rp\_error}\\
$\langle G \rangle, \langle G_{\rm BP} \rangle, \langle G_{\rm RP} \rangle$&\texttt{int\_average\_g, int\_average\_bp, int\_average\_rp}\\
$\sigma \langle G \rangle$ , $\sigma \langle G_{\rm BP} \rangle$ , $\sigma \langle G_{\rm RP} \rangle$ &\texttt{int\_average\_g\_error, int\_average\_bp\_error, int\_average\_rp\_error} \\
Amp($G, G_{\rm BP}, G_{\rm RP}$)&\texttt{peak\_to\_peak\_g, peak\_to\_peak\_bp, peak\_to\_peak\_rp}\\
$\sigma {\rm [Amp(}G{\rm )]}$, $\sigma {\rm [Amp(}G_{\rm BP}{\rm )]}$, $\sigma {\rm [Amp(}G_{\rm RP}{\rm )]}$&\texttt{peak\_to\_peak\_\linebreak g\_error, peak\_to\_peak\_\linebreak bp\_error, peak\_to\_peak\_\linebreak rp\_error}\\
$\phi_{21}$ &\texttt{phi21\_g}\\
$\sigma (\phi_{21})$ &\texttt{phi21\_\linebreak g\_error} \\
$R_{21}$ &\texttt{r21\_g}\\
$\sigma (R_{21})$&\texttt{r21\_g\_error}\\ 
$\phi_{31}$ &\texttt{phi31\_g}\\
$\sigma (\phi_{31})$ &\texttt{phi31\_\linebreak g\_error} \\
$R_{31}$ &\texttt{r31\_g}\\
$\sigma (R_{31})$&\texttt{r31\_g\_error} \\ 
${\rm [Fe/H]}\tablefootmark{\rm (b)}$&\texttt{metallicity}\\
$\sigma$ ([Fe/H]) &\texttt{metallicity\_error}\\
A($G$)\tablefootmark{\rm (c)} & \texttt{g\_absorption}: empty for Cepheids\\
$\sigma$ A($G$) &\texttt{g\_absorption\_error}: empty for Cepheids\\
$N_{\rm obs}$($G$ band) & \texttt{num\_clean\_epochs\_g}\\
$N_{\rm obs}$($G_{\rm BP}$ band) & \texttt{num\_clean\_epochs\_bp}\\
$N_{\rm obs}$($G_{\rm RP}$ band) & \texttt{num\_clean\_epochs\_rp}\\
\noalign{\smallskip}
\hline                                  
\end{tabular}
\tablefoot{$^{\rm (a)}$The BJD of the epoch of maximum light is offset by JD 2455197.5 d (= J2010.0). $^{\rm (b)}$ Photometric metal abundance derived from the Fourier parameters of the light curve for  
3,738 fundamental-mode DCEPs with period shorter than 6.3 days  (see Sect.~\ref{metallicity} and ~\ref{cep-results}).}\\ 
\end{table*}
    
\subsection{Sky maps}\label{maps}

Figs.~\ref{raDecMC_RR} and ~\ref{raDecMC_RR_WithMag} show sky maps  of the SOS confirmed  RR Lyrae stars released in DR2 in the  region of the  Magellanic Clouds. In the latter map
variables are colour-coded according to their apparent magnitude. Fig.~\ref{raDecMC} shows the same region of the sky as mapped by the SOS confirmed Cepheids. Remarkable is the rather smooth distribution of RR Lyrae stars around the two Clouds, very nicely shaping   
    the  far-extended halo  surrounding the LMC.

\begin{figure}
   \centering
   \includegraphics[width=9.0 cm,clip]{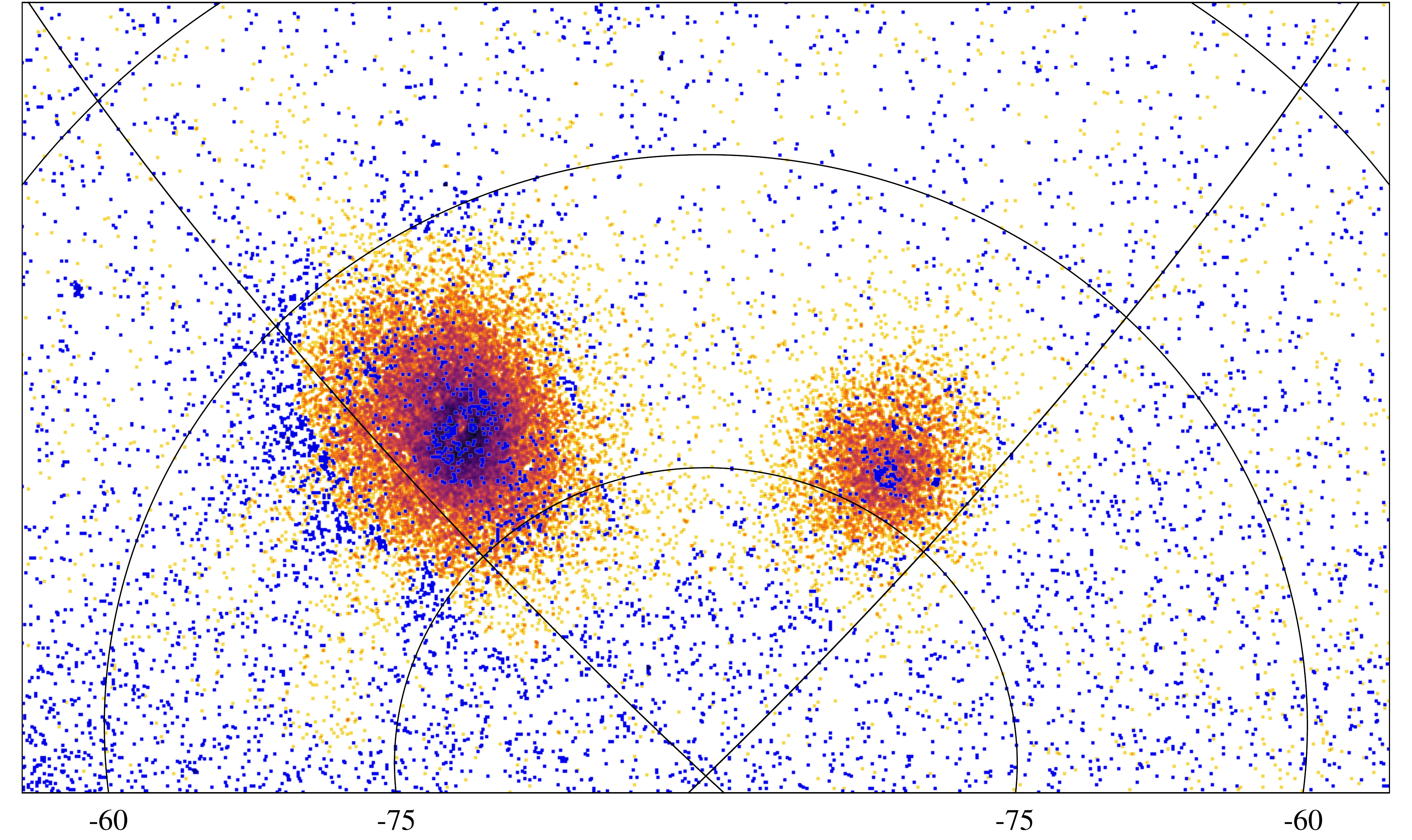}
   \caption{Spatial distribution of  RR Lyrae stars in the Magellanic Cloud region released in {\it Gaia} DR2. Orange dots: known RR Lyrae stars from the OGLE surveys; blue dots:
   RR Lyrae stars identified by  \textit{Gaia} and confirmed  by the SOS Cep\&Cep pipeline.
 }
              \label{raDecMC_RR}%
    \end{figure}
      
\begin{figure}
   \centering
   \includegraphics[width=9.0 cm,clip]{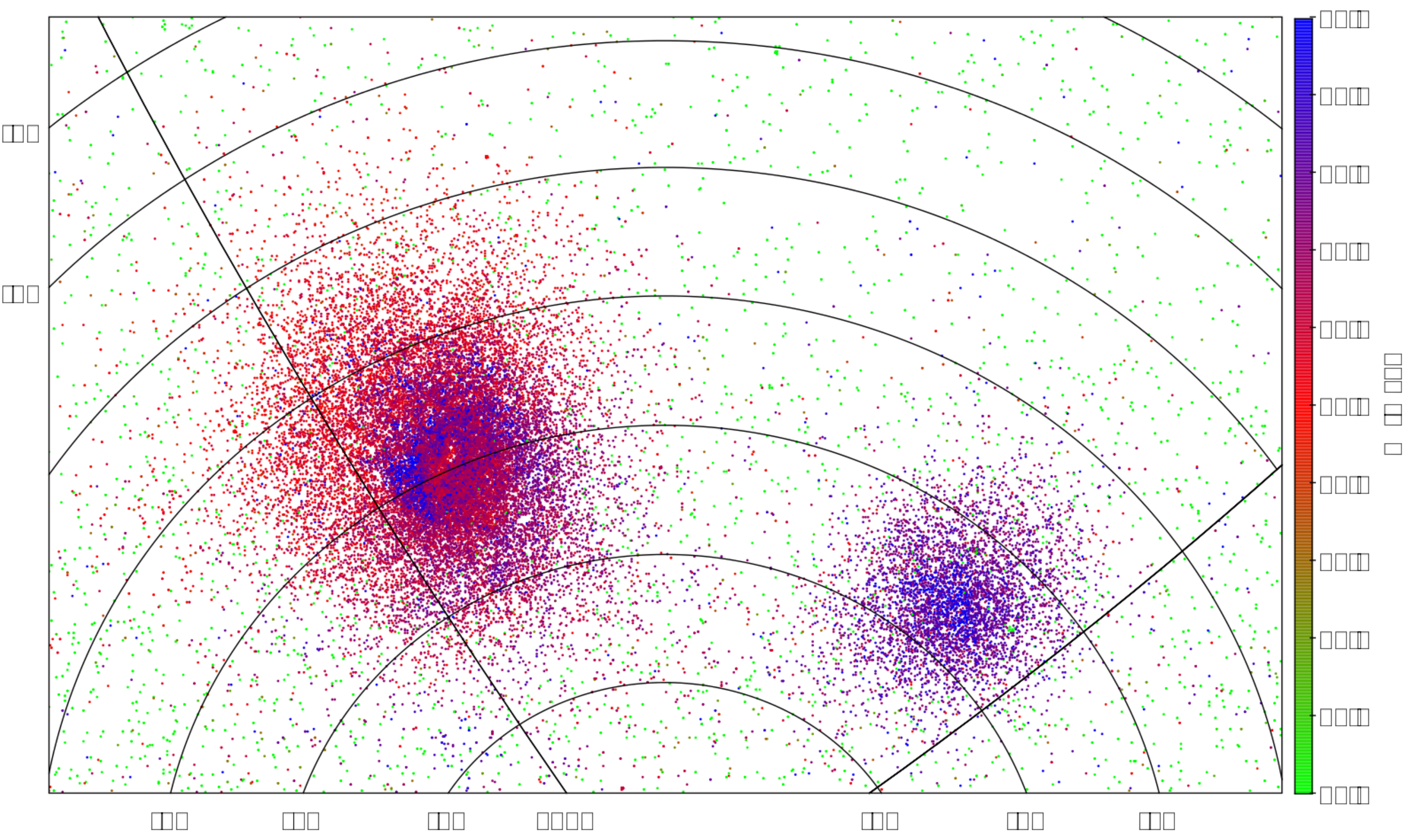}
   \caption{Same as in Fig.~\ref{raDecMC_RR} but with the RR Lyrae stars colour  coded according to their apparent magnitude.  Notable is the rather smooth distribution of RR Lyrae stars around the two Clouds, very nicely shaping the  far-extended halo  surrounding the LMC.} 
              \label{raDecMC_RR_WithMag}%
    \end{figure}

  \begin{figure}
   \centering
   \includegraphics[width=9.0 cm,clip]{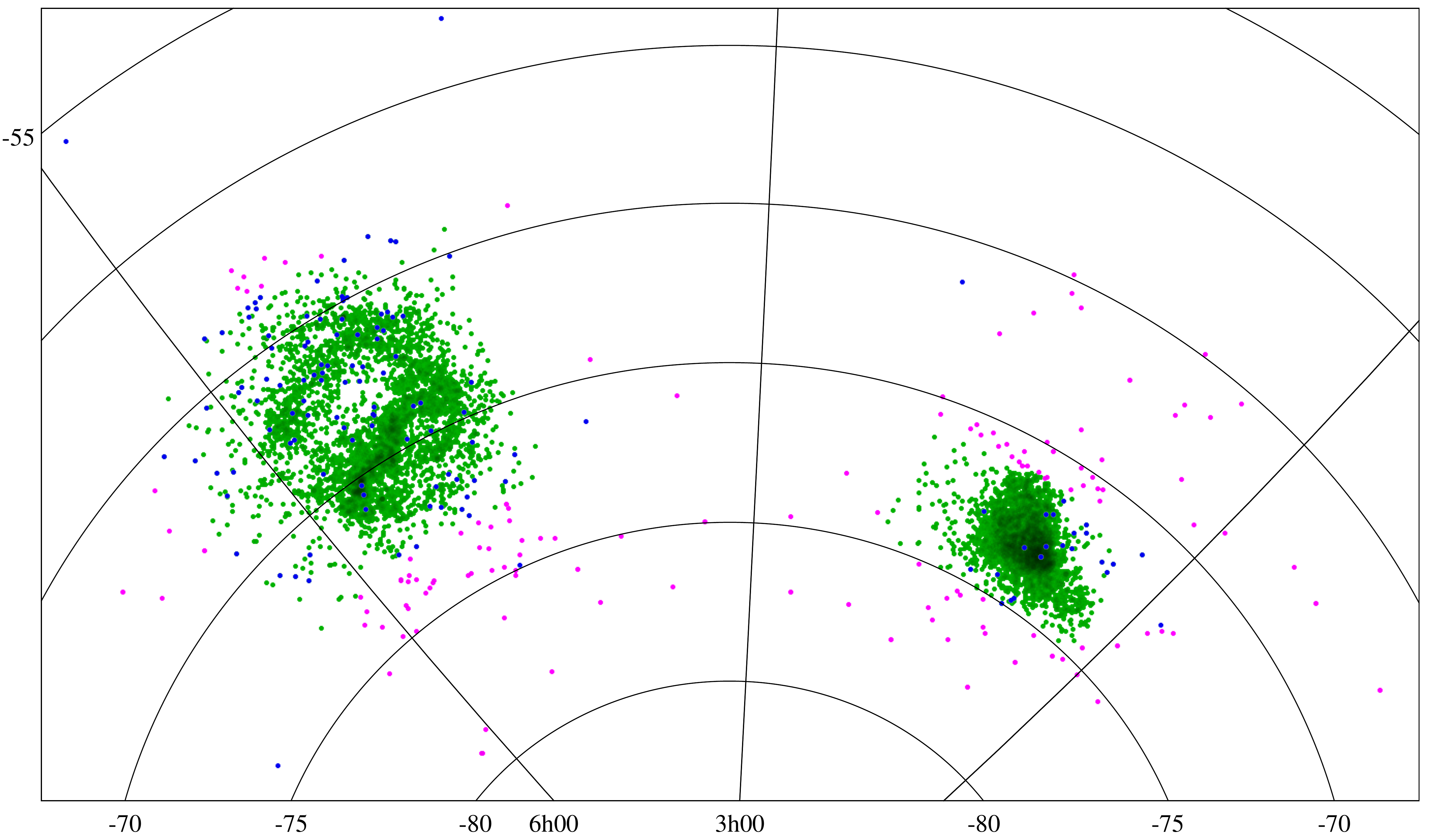}
   \caption{Spatial distribution of {\it Gaia} DR2 Cepheids located in the region of the two Magellanic Clouds. Green filled circles: known Cepheids observed in the LMC and SMC regions defined in Sect.~\ref{s2} by the OGLE survey;  magenta filled circles:  All-Sky known (from OGLE or other surveys) Cepheids;  
blue filled circles:  new Cepheids identified  by {\it Gaia} in the LMC  and SMC (118 in total).}
               \label{raDecMC}%
   \end{figure}

Figs.~\ref{mapRR_GC_4} and ~\ref{mapRR_GC_3}  show the distribution on sky, in galactic and equatorial coordinates, respectively, of RR Lyrae stars within the limiting magnitude of \textit{Gaia} (orange points). The map combines known literature (both with and without a \textit{Gaia} counterpart) and new RR Lyrae stars discovered by \textit{Gaia} and confirmed by the SOS Cep\&RRL pipeline for more than 223,000 
RR Lyrae stars in total. 
   This number favourably compares with estimates of the total number RR Lyrae stars in \citet{holl2018} and \citet{rimoldini2018}.      Blue filled dots and magenta filled squares indicate 87 GCs  and 12 dSphs  (classical and ultra faint) in which \textit{Gaia} observed SOS confirmed RR Lyrae stars   that are published in DR2. 
   This figure presents a post-DR2 update of  fig.~4  in \citet{clementini2018}, displaying the largest ever census of  RR Lyrae stars in our Galaxy and its close companions.

  \begin{figure*}
   \centering
  \includegraphics[width=16 cm,clip]{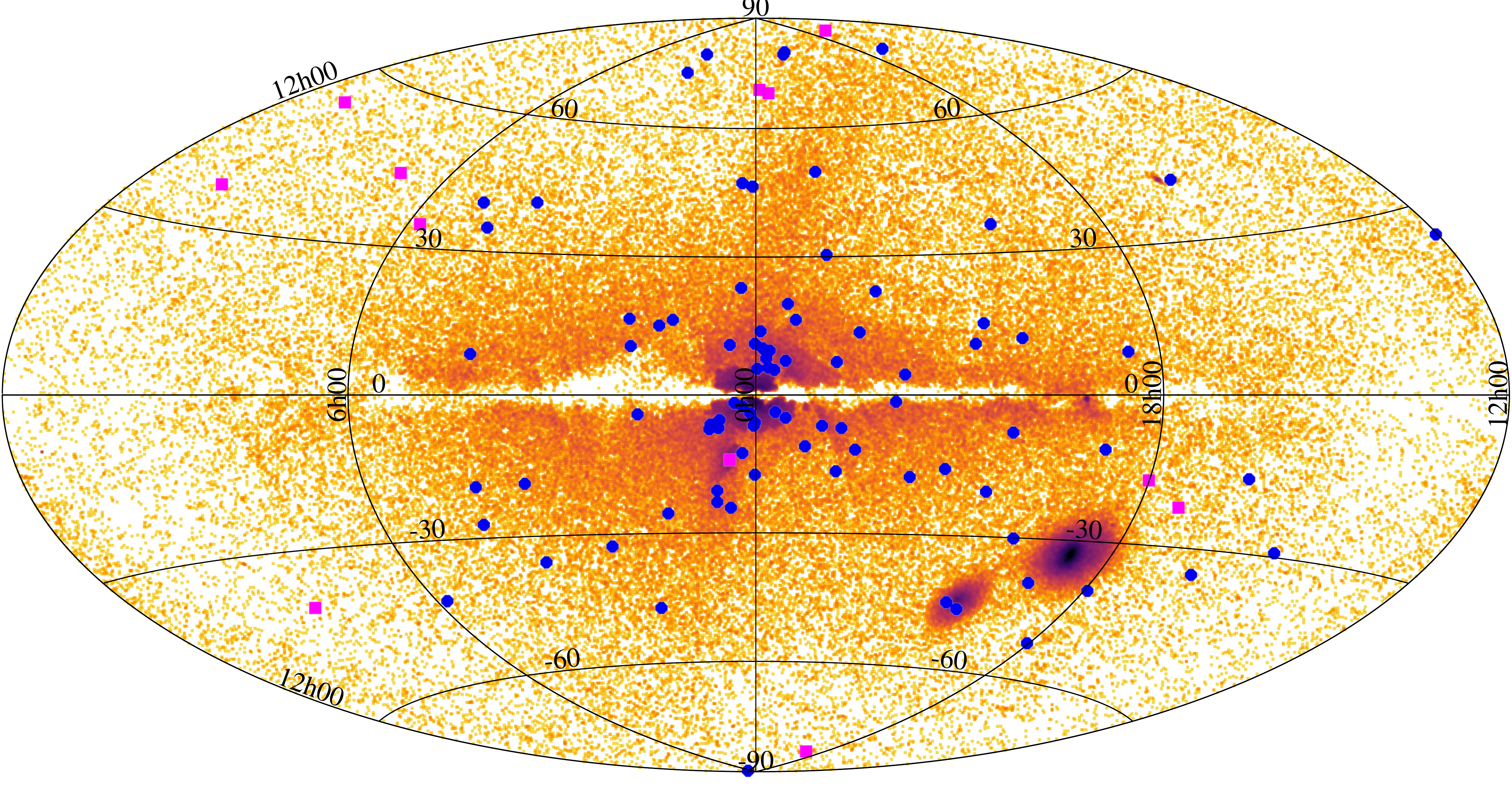}
   \caption{Distribution on sky, in galactic coordinates, of RR Lyrae stars within the limiting magnitude of \textit{Gaia} (orange points). The map combines known literature both with and without a \textit{Gaia} counterpart and new RR Lyrae stars discovered by \textit{Gaia} and confirmed by the SOS Cep\&RRL pipeline, for more than 223,000  
RR Lyrae stars in total. 
   Blue filled dots and magenta filled squares indicate 87 globular clusters and 12 dwarf spheroidal galaxies (classical and ultra faint) in which \textit{Gaia} has observed RR Lyrae stars   that are confirmed by the SOS Cep\&RRL pipeline. 
 }
              \label{mapRR_GC_4}%
    \end{figure*}
         
\begin{figure*}
   \centering
   \includegraphics[width=16 cm,clip]{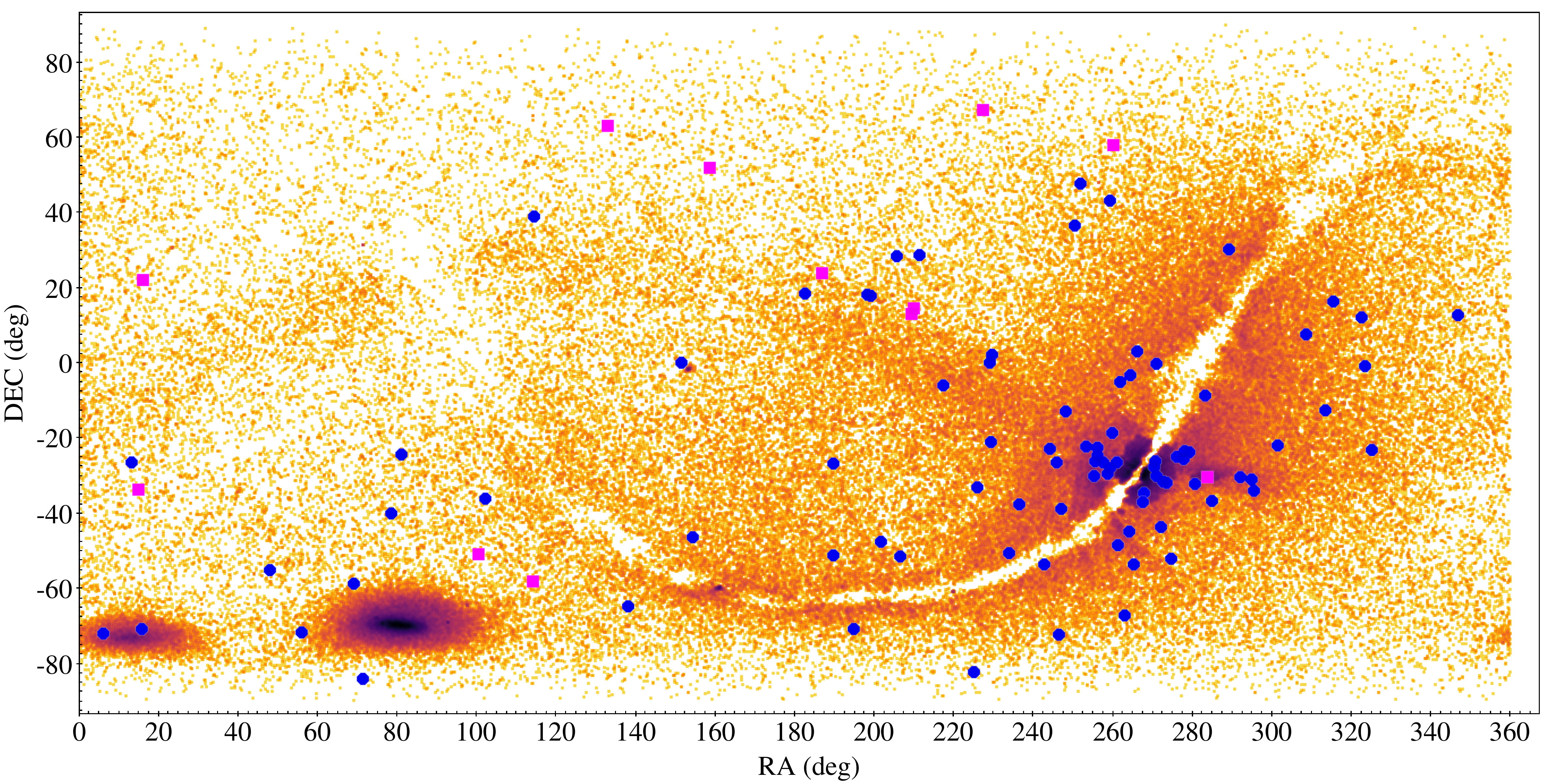}
     \caption{Same as in Fig.~\ref{mapRR_GC_4}, but in equatorial coordinates.}
              \label{mapRR_GC_3}%
    \end{figure*}

\subsection{Limitations of the SOS Cep\&RRL results for DR2}\label{DR2limitation}
 
The catalogues  of SOS Cep\&RRL confirmed Cepheids and RR Lyrae stars released in 
\textit{Gaia} DR2 have some limitations. The major issue  is  incompleteness, mainly due to the still small number of
epochs available for a significant fraction of the sky (see
e.g. Fig.~\ref{nepochsRRNew}) as per  {\it Gaia} scanning
law. Furthermore, we recall that: i)  objects with less  than 12 epoch data were not
processed by the SOS Cep\&RRL pipeline and ii) a sizeable sample of true
pulsators with less than $\sim$ 20 epochs were rejected because both
the small number of observations and the uneven sampling of the time series data caused by the scanning law
 resulted in an incorrect period determination. Hence, completeness is a strong function of the source 
position on the  sky. This can be easily checked comparing the DR2 SOS Cep\&RRL 
RR Lyrae catalogue to OGLE data for the LMC, SMC and Bulge. In these systems our
recovery percentages are 67\%, 82\% and 15\%, respectively. Taking into
account that the OGLE data is complete more or less at the same
magnitude of {\it Gaia} data, that the  typical magnitudes of the RR 
Lyrae variables in the
MCs are generally fainter than in the Bulge and that the level of
contamination is similar, it is clear that the very low recovery rate
in  the Bulge is almost entirely due to the lack of enough epoch data for the Bulge region due to {\it Gaia} scanning
law. Similar considerations can be drawn for  the rest of the sky
(mainly the MW halo), even though for the Galactic halo 
sky zones rich and poor in number of epochs are more or less the same, hence giving in the end an  indication of the average completeness of the DR2
catalogue. For example,  the comparison with the ATLAS and ASAS-SN surveys
returned a recovery rate of 66\% and 64\%, 
respectively.  A general comparison with all our literature
collection,  but excluding OGLE,  returns a completeness on the
order of 60\%. This is of course only an indicative number, as the
literature data is a mixture of different surveys with very different
characteristics, but we are confident that this estimate should not be too  far from 
reality.

As for Cepheids, the same line of reasoning returns  74\%, 73\% and
3\% percentages of recovery in the LMC, SMC and Bulge,
respectively. While percentages in the MCs are similar to those of
 RR Lyrae variables, that in the Bulge is very low. 
  This is mainly due to the small number of epochs in this region.

Estimating the percentage of recovery over the whole MW is rather difficult,
since field Cepheids are not easy to detect  and there are no
systematic and complete surveys over the whole sky to compare with, {\it Gaia}'s indeed
will be the first one. To give some rough numbers at least for DCEPs, 
we  compared the SOS Cep\&RRL Cepheids published in DR2 with a sample of 417 DCEPs with metallicity
measured by \citet{genovali2013}  and found  that 68\% of the sources are in 
common, a percentage in line with what found for field RR Lyrae variables.

We also  note that at the bright extreme of the distribution, we missed some famous
RR Lyrae stars and Cepheids. For example, RR Lyr itself is not present in the SOS Cep\&RRL RR Lyrae tables.
 This is because during validation of the SOS Cep\&RRL results the star was dropped  as the value of the $\phi_{21}$ Fourier
parameter  put the star at the limit of  a  region of heavy contamination
($\phi_{21} \sim$ 3.14 rad) where sources are  automatically rejected.
Although not present in the DR2 variability tables, RR Lyr is present in  the DR2 photometry and astrometry 
catalogues but with values for magnitudes and astrometry to be taken with caution (\citealt{Arenou2018} and 
\citealt{gaiacol-brown18}).

Finally, 
comparison of the DR2 SOS Cep\&RRL confirmed RR Lyrae stars and Cepheids with the sources 
released in \textit{Gaia} DR1  shows that out of the 2,595 RR Lyrae stars published in DR1, 2,517 are also 
confirmed in DR2, and 78 are missing. Out of the 599 DR1 Cepheids, 533 are also confirmed in DR2 and 66 are missing.
This can be due to several
reasons. For instance, the increased  number of epoch data available  for  DR2 along with 
modifications of both the  general variability pipeline prior the SOS processing and updates 
of the SOS Cep\&RRL  pipeline (see Sect.~\ref{s2}), may  have resulted for instance in 
a different period estimate, and in turn,  in a different position in the
diagrams we use for classification. 
Specifically, out of the 78 missing DR1 RR Lyrae stars, in DR2 twenty-three were no longer classified as RR Lyrae by the classifiers
hence never made it to the SOS Cep\&RRL  pipeline (of them 11 are bona fide RR Lyrae stars known from OGLE)   and 55 were rejected during the SOS processing.
Among those 55, 49 are bona fide RR Lyrae stars known from OGLE, and 6 are DR1 RR Lyrae that hence are not confirmed. 
Out of the 66 missing DR1 Cepheids, 31 never made it to the SOS pipeline (among them 3 are Cepheids confirmed by OGLE), and of the remaining 35,
twenty-eight are Cepheids confirmed by OGLE. 
Hence, working out the above  numbers, the total contamination of the new RR Lyrae stars released in DR1 turns out to be less than 0.7\% and that of the DR1 Cepheids is 
around 5.8\%.

 After the opening of the \text{Gaia} DR2 archive on 25 April 2018 we received feedback from the users about RR Lyrae stars and Cepheids which may have  been misclassified by the 
SOS Cep\&RRL pipeline. We provide in Appendix C the sourceids of a number of these possibly misclassified sources.

\section{Conclusions and future developments}\label{conclusions}
\textit{Gaia} DR2 represents a significant step forward in our  knowledge of the RR Lyrae and Cepheid census in the Milky Way and its close neighbours.
The number of  RR Lyrae stars confirmed by the SOS Cep\&RRL pipeline that have been released in DR2 along with those already known in the literature provides already 
 the largest ever census of  RR Lyrae stars in our Galaxy and its close companions. This number will increase with further  \textit{Gaia} releases as it is also expected 
 to increase the number and accuracy of the parameters derived for these sources by the SOS Cep\&RRL pipeline.
 In particular, for \textit{Gaia}  Data Release 3 (DR3) time series radial velocities will be processed by the SOS pipeline besides  the multiband photometry. This will allow us  
  to improve the source characterisation and will also open the path to the derivation 
 of additional stellar parameters  such as gravities, temperatures and absolute magnitudes independent of parallaxes.
 
A number of  improvements of the SOS Cep\&RRL pipeline are already being implemented in view of \textit{Gaia}  DR3 forthcoming release. They include the following:

\begin{enumerate}

\item
The relations used for the classification and characterisation of the sources in the SOS Cep\&RRL pipeline will be computed directly from the \textit{Gaia} light curves of
confirmed Cepheids and RR Lyrae stars.

\item
Parts of the SOS Cep\&RRL pipeline dedicated to the processing of the radial velocity time series  will be activated.

\item
Identification and characterisation of double-mode RR Lyrae stars  and multi-mode classical Cepheids (F/1O, 
1O/2O, etc.) will be activated only for sources with sufficient number of epochs 
 and 
improving the detection algorithm by properly taking   
into account the scatter in the folded light curve,  
 thus reducing the number of false positives. 
  
\item
A classifier will be developed to optimise the type and subtype classification of Cepheids and RR Lyrae stars performed by the SOS Cep\&RRL pipeline.  

\item
Automated  validation and cleaning procedures will be put in place. 

\item
CMDs in absolute magnitude  and the comparison  with theoretical instability strips for Cepheids and RR Lyrae stars 
will be used  to  improve the source classification and the derivation of their intrinsic parameters (e.g. effective temperatures).%

\end{enumerate}

To conclude,
the results of  \textit{Gaia} all-sky Cepheids and RR Lyrae stars obtained with the SOS Cep\&RRL processing demonstrate the excellent quality of {\it Gaia} multi-band photometry released in DR2 and 
nicely showcase the potential of {\it Gaia} in the field of variable star studies and for  Cepheids and RR Lyrae stars in particular.
   
\begin{acknowledgements}
This work has made use of data from the ESA space mission {\it Gaia}, processed by the {\it Gaia} Data
Processing and Analysis Consortium (DPAC). Funding for the DPAC has been
provided by national institutions participating in
the {\it Gaia} Multilateral Agreement. In particular, the Italian participation in DPAC has been supported by Istituto Nazionale di
Astrofisica (INAF) and the Agenzia Spaziale Italiana (ASI) through grants
I/037/08/0,  I/058/10/0,  2014-025-R.0, and 2014-025-R.1.2015 to INAF (PI M.G.
Lattanzi), the Belgian participation by the BELgian federal Science Policy 
(BELSPO) through PRODEX grants, the Swiss participation by  the Swiss State Secretariat for Education, Research
and Innovation through the ESA Prodex program, the ``Mesures d'accompagnement'', the
``Activit\'{e}s Nationales Compl\'{e}mentaires'', the Swiss National Science
Foundation, and the Early Postdoc Mobility fellowship, the Spanish participation by the Spanish Ministry of Economy
MINECO-FEDER through grants AyA2014-55216,
AyA2011-24052 and  the Hungarian participation through the PECS programme (contracts
   C98009 and 40001066398/12/NL/KML). 
UK community participation in this work has been supported by funding
from the UK Space Agency, and from the UK Science and Technology
Research Council.
The {\it Gaia} mission website is:  \texttt{http://www.cosmos.esa.int/gaia}.
This research has made use of the International Variable Star Index (VSX) database, operated at AAVSO, Cambridge,
Massachusetts, USA (https://www.aavso.org/vsx/index.php) and of the 
SIMBAD database,
operated at CDS, Strasbourg, France. 
 We thank an anonymous referee for carefully reading our paper and for providing comments that contributed to significantly improve the readability and quality of our manuscript.
We are indebted to Dante Minniti, for sharing in advance of publication a list of  RR Lyrae candidates discovered by the VVV survey in the MW bulge and disc. 
We  gratefully acknowledge feedback from S. Cheng and S. Koposov  who identified the misclassified sources listed in Appendix C, during a project developed in part at the 2018 NYC Gaia Sprint, hosted by the Center
for Computational Astrophysics of the Flatiron Institute in New York City. We are also indebted to Hsiang-Chih Hwang from John
Hopkins University who pointed out further 4 possibly misclassified sources (one RR Lyrae star and 3 Cepheids) discovered at the Gaia DR2 Experiment Lab (2018) in Madrid and to Tim Bedding and Dan Hey who informed us of the misclassification as Cepheid of the spotted rotating star KIC 6619830 observed by Kepler.
In this study we have largely made use of TOPCAT,  Taylor, M. B. (2005), ``TOPCAT \& STIL: Starlink Table/VOTable Processing Software", in Astronomical Data Analysis Software and Systems XIV, eds. P. Shopbell et al., ASP Conf. Ser. 347, p. 29.
 
\end{acknowledgements}
%
   \bibliographystyle{aa} 
%

\begin{appendix}

\include{CEP_RR_Sample_Queries_Revised}

\include{acronyms_ceprrl-2} 

\include{misclassification_new}

\end{appendix}

\end{document}

%% file: CEP_RR_Sample_Queries_Revised.tex
\section{Examples of \textit{\textbf{Gaia}} archive queries}
\label{app:queries}


  \begin{table*}[!htbp]
    \caption{Queries to retrieve DR2 information on the Cepheids and RR Lyrae stars from the {\it Gaia} archive in the Astronomical Data Query Language (\citealt{osuna08}).}
    \label{tab:queries}
    \centering
    \begin{tabular}{p{\textwidth}}
    \footnotesize
      Query to retrieve time series of all Cepheids in the {\it Gaia} DR2.
      \begin{verbatim}
 select gaia.source_id, epoch_photometry_url from gaiadr2.gaia_source as gaia
      inner join gaiadr2.vari_cepheid as cep on gaia.source_id=cep.source_id
      \end{verbatim} \\

      Query to retrive time series of all RR Lyrae stars in the {\it Gaia} DR2.
      \begin{verbatim}
 select gaia.source_id, epoch_photometry_url from gaiadr2.gaia_source as gaia
      inner join gaiadr2.vari_rrlyrae as rrl on gaia.source_id=rrl.source_id
      \end{verbatim} \\

      Query to retrieve the number of processed observations and SOS Cep\&RRL-computed parameters of all Cepheids in the {\it Gaia} DR2.
      \begin{verbatim}
select cep.*,tsr.num_selected_g_fov,tsr.num_selected_bp,tsr.num_selected_rp from gaiadr2.vari_cepheid cep 
     inner join gaiadr2.vari_time_series_statistics tsr on cep.source_id=tsr.source_id   
\end{verbatim} \\

      Query to retrieve the number of processed observations and SOS Cep\&RRL-computed parameters of all RR Lyrae in the {\it Gaia} DR2.
      \begin{verbatim}
      select rrl.*,tsr.num_selected_g_fov,tsr.num_selected_bp,tsr.num_selected_rp from gaiadr2.vari_rrlyrae rrl 
           inner join gaiadr2.vari_time_series_statistics tsr on rrl.source_id=tsr.source_id    
     \end{verbatim}
    \end{tabular}
  \end{table*}

%% file: acronyms_ceprrl-2.tex
\section{Acronyms}
\label{app:acro}

  \begin{table*}[!htbp]
      \caption{List of acronyms used in this paper.}
\begin{tabular}{ll}
\hline
\hline
\noalign{\smallskip}
{\bf Acronym} & {\bf Description}  \\
\hline 
\noalign{\smallskip}
ACEP & Anomalous Cepheid \\ 
ALL\_SKY&The celestial region excluding  the LMC and SMC regions \\
Amp($G$)& Amplitude of the light variation in the $G$ band\\
Amp($G_{BP}$)& Amplitude of the light variation in the $G_{BP}$ band\\
Amp($G_{RP}$)& Amplitude of the light variation in the $G_{RP}$ band\\
BLHER &  BL Herculis class of variables  \\
CMD & Colour Magnitude Diagram \\
DCEP &  Classical Cepheid (Population I) \\
dSph& Dwarf spheroidal galaxy \\ 
DR&Data Release \\
F&Fundamental mode of pulsation \\ 
FO&First overtone mode of pulsation \\
$G$ & Gaia photometric $G$-band \\ 
$G_{BP}$ &Gaia photometric $G_{BP}$ band \\
$G_{RP}$&Gaia photometric $G_{RP}$ band \\ 
GC& Globular cluster \\ 
LMC&Large Magellanic Cloud \\
MW&Milky Way \\
PA&Period--Amplitude \\
PL&Period--Luminosity \\
PW&Period--Wesenheit \\
RRab&RR Lyrae star of ab type \\
RRc&RR Lyrae star of c type \\
RRd& double-mode RR Lyrae star \\
RVTAU &  RV Tauri class of  variables  \\ 
SMC&Small Magellanic Cloud \\
SOS& Specific Object Study \\
T2CEP & Type II Cepheid (Population II)\\
WVIR &  W Virginis class of variables  \\
 \hline  
\noalign{\smallskip}
\end{tabular} 
  \end{table*}

%% file: misclassification_new.tex
\section{Misclassified sources}
\label{app:misclass}
{\bf 
  \begin{table*}[!htbp]
      \caption{List of sources likely misclassified as RR Lyrae stars (courtesy of S. Cheng, S. Koposov and H-C. Hwang).}
\begin{tabular}{rc}
\hline
\hline
\noalign{\smallskip}
\textit{Gaia} sourceid ~~~~~~~~~~ & Notes \\
\hline
\noalign{\smallskip}
  1754525270341133312 &  weird folded light curve \\
  32205593226566016 & galaxy \\
  361431775815909376 & galaxy \\   
  361451975047400448\ & galaxy \\   
  585136968493947264 & galaxy \\   
 \hline  
\noalign{\smallskip}
\end{tabular}
\tablefoot{This table is available in its entirety in the electronic edition of the journal, only a portion is shown here for guidance regarding its form and content.}\\
 \end{table*}
 
   \begin{table*}[!htbp]
      \caption{List of sources likely misclassified as Cepheids (courtesy of H-C. Hwang, T. Bedding and D. Hey).}
\begin{tabular}{rcc}
\hline
\hline
\noalign{\smallskip}
\textit{Gaia} sourceid~~~~~~~~~~ & Literature sourceid & Notes \\
\hline
\noalign{\smallskip}
 2077108036182676224 &  KIC 6619830 & Spotted rotating star \\
 2641587994382309632 & & Fainter and redder than typical Cepheids \\
5424247204666300032 & & Fainter and redder than typical Cepheids \\
5804085561048983552 & & Fainter and redder than typical Cepheids \\
 \hline  
\noalign{\smallskip}
\end{tabular}
\tablefoot{A more complete listing of likely misclassified Cepheids is published in {\bf Ripepi et al. (2018).}}\\
 \end{table*}
 }